\documentclass[twocolumn]{aastex63}

\usepackage{xcolor,colortbl}

\shorttitle{Peptide-like Bond Molecules in HMC}
\shortauthors{Gorai et al.}

\begin{document}

\title{Identification of pre-biotic molecules containing Peptide-like bond in a hot molecular core, G10.47+0.03}
\email{ankan.das@gmail.com}
\email{prasanta.astro@gmail.com}
\author[0000-0003-1602-6849]{Prasanta Gorai}
\author[0000-0002-5224-3026]{Bratati Bhat}
\author[0000-0001-5720-6294]{Milan Sil}
\author[0000-0002-7657-1243]{Suman K. Mondal}
\author[0000-0003-1745-9718]{Rana Ghosh}
\author[0000-0002-0193-1136]{Sandip K. Chakrabarti}
\author[0000-0003-4615-602X]{Ankan Das}
\affiliation{Indian Centre for Space Physics, 43 Chalantika, Garia Station Road, Kolkata 700084, India}

\begin{abstract}
After hydrogen, oxygen, and carbon, nitrogen is one of the most chemically active species in the interstellar medium (ISM). Nitrogen bearing 
molecules have great importance as they are actively involved in the formation of biomolecules. Therefore, it is essential to look for nitrogen-bearing 
species in various astrophysical sources, 
specifically around high-mass star-forming regions where the evolutionary history 
is comparatively poorly understood. In this paper, we report observation of three potential pre-biotic molecules, namely, isocyanic acid (HNCO), 
formamide ($\rm{NH_2CHO}$), and methyl isocyanate ($\rm{CH_3NCO}$), which contain peptide-like bonds (-NH-C(=O)-) in a hot molecular core, 
G10.47+0.03 (hereafter, G10). Along with the identification of these three complex nitrogen-bearing species, we speculate their spatial distribution in the source and
discuss their possible formation pathways under such conditions.
Rotational diagram method under LTE condition has been employed to estimate the excitation temperature and the
column density of the observed species. Markov Chain Monte Carlo method was used to obtain the best suited physical parameters of G10 as well
as line properties of some species. We also 
determined the hydrogen column density and the optical depth for different continuum observed in various frequency ranges. Finally, based on these 
observational results, we have constructed a chemical model to explain the observational findings. We found that HNCO, $\rm{NH_2CHO}$, and $\rm{CH_3NCO}$ 
are chemically linked with each other.
\end{abstract}

\section{Introduction}
Various inter-disciplinary studies are involved in the search of the origin of life on Earth. Whether life evolved
{\it ab-initio} here on the Earth or came from another part of the space is debatable, 
but it is accepted that our ancestors
(may be the unicellular species) should have formed from the raw materials present at that time somewhere
in the universe. When, where, and how the first life came is not straight forward to answer. However, at the current era,
it is necessary to try to explain how the building blocks of life (simple $\rightarrow$ complex molecule $\rightarrow$ bio-molecule)
could be indigenously produced in the universe.

{ Around $200$} molecular species have been identified in the ISM or circumstellar shells
\url{(https://www.astro.uni-koeln.de/cdms/molecules)}. Among them several species are
marked as the precursor to biomolecules. Study of the pre-biotic molecules is always fascinating 
as they involved in the formation of amino acids, proteins, and the basic building blocks of life
\citep{chak00a,chak00b,das08,garr13,chak15,maju15,das19}. Protein synthesis occurs through peptide bond formation \citep{gold10}. 
CN is the first observed nitrogen-bearing species in space \citep{mcke40}. Since then, various nitrogen-bearing species were 
identified in numerous astronomical objects. Hot core regions are the unique laboratory of complex organic molecules 
(COMs). Forest of molecular lines has been identified in several hot molecular cores (HMCs) \citep[e.g.,][]{bell16,garr17}.  Here, we will focus 
on the observation done towards a hot molecular core, G10, which is located at a distance of 8.6 kpc \citep{sana14} having 
luminosity $\rm{5\times 10^{5}\ L_{\odot}}$ \citep{cesa10}.

Among the pre-biotic molecules, methanimine (CH$_2$NH) and methylamine ($\rm{CH_3NH_2}$) are the simple imine and 
amine respectively which play a significant role in the synthesis of the simplest amino acid, glycine ($\rm{NH_2CH_2COOH}$) \citep{altw17,sil18}.
These molecules have been identified in G10, which strengthens the possibility of the presence of glycine in this source \citep{ohis17}.

Isocyanic acid (HNCO) is the simple molecule which has four biogenic elements (C, N, O, and H) making a peptide bond, -NH-C(=O)-. HNCO was observed
long ago towards the high-mass star-forming region, Sgr B2 \citep{snyd72}.  Presently, it has been observed in various astronomical objects such 
as translucent molecular cloud \citep{turn99}, dense core \citep{marc18}, and low-mass protostar, IRAS 16293-2422 \citep{biss08}. It was also previously detected in G10 \citep{wyro99}.

Formamide (NH$_2$CHO) is the simplest possible amide and a potential pre-biotic molecule which contains a 
peptide bond that can link with the amino acids and form proteins. NH$_2$CHO is also a precursor of genetic and metabolic molecules \citep{sala12}. 
This molecule is one of the key species for the formation of nucleobases and nucleobase analogs. NH$_2$CHO was observed for the first time towards high-mass star-forming region, Sgr B2 \citep{rubi71}. Subsequently, it was
identified in other hot cores, such as, Orion KL, G327.3-0.6, G34-3+0.15, NGC 6334 \citep{turn91,boge19}, solar-type low-mass protostar IRAS 16293-2422
\citep{kaha13}, and in the shock of the prestellar core, L1157-B1 \citep{code17}. NH$_2$CHO was previously detected in G10 using millimeter and
sub-millimeter wavelength facility with Sub-millimeter Array (SMA) observation \citep{rolf11}.

Methyl isocyanate (CH$_3$NCO) is another potential pre-biotic molecules, which also
has a peptide-like bond. It has recently been
observed in a high-mass star-forming region, Sgr B2 \citep{cern16} and low-mass star-forming region,
IRAS 16293-2422 \citep{ligt17,mart17}. Here, for the first time, we are reporting the identification of CH$_3$NCO in G10.
{ HNCO has firmly been identified in G10, but for NH$_2$CHO, no clear peak was present in the observed spectra of \cite{rolf11}.}
{ Recently, HNCO and NH$_2$CHO both have been identified, but $\rm{CH_3NCO}$ has been tentatively identified
in the 67P/Churyumov-Gerasimenko comet by Double Focusing Mass Spectrometer (DFMS) of the ROSINA experiment on ESA's
Rosetta mission \citep{altw17}.}

In this paper, we present a combined study of observational analysis and chemical modeling of the peptide-like bond molecules.
We report identifications of HNCO, $\rm{NH_2CHO}$, and $\rm{CH_3NCO}$ in G10. To understand 
the formation of these three species, we prepare a chemical model which mimics the observed results. 
We have organized this paper as follows. In Section \ref{sec:obs-data-analysis}, we describe observational details and data analysis procedures.
Observational results are presented in Section \ref{sec:obs-results}. Chemical model and results are described in Section \ref{sec:model-results}. 
Finally, in Section \ref{sec:conclusions}, we make concluding remarks.

\section{Observations, data analysis and line identification \label{sec:obs-data-analysis}
}
In this paper, we have used ALMA cycle 4 archival data of G10.47+0.03 observation ($\#$2016.1.00929.S.). 
The phase center of the observation is located at $\alpha$(J2000)=18$^h$08$^m$38.232$^s$ and $\delta$(J2000)=-19$^0$51$'$50.4$^{''}$. Observations were performed with 
ALMA Band 4 covering four spectral ranges; (i) 129.50-131.44 GHz, (ii) 147.50-149.43 GHz, (iii) 153.00-154.93 GHz, and (iv) 158.49-160.43 GHz. In this observation, 
the flux calibrator was J1733-1304, the phase calibrator was J1832-2039 and the bandpass calibrator was J1924-2914. The systematic velocity of this source 
was $67-68$ Km s$^{-1}$ \citep{rolf11}. Observational summary is given in Table \ref{table:observ_sum}. All the analysis, such as spectral and line analysis, 
were done using CASA 4.7.2 software \citep{mcmu07}. We have implemented a first-order baseline fit by using  `uvcontsub' command
available in the CASA program. We have divided each spectral window into two data cubes: continuum and line emission for the analysis.
{ We did not apply the self-calibration and ALMA missing flux correction.}
The line identification of all the observed species presented in this paper was carried out using CASSIS\footnote{\url{http://cassis.irap.omp.eu}} 
software together with the Cologne Database for Molecular Spectroscopy  \citep[CDMS,][]{mull01,mull05}\footnote{\url{https://www.astro.uni-koeln.de/cdms}} 
and Jet Propulsion Laboratory \cite[JPL,][]{pick98}\footnote{\url{http://spec.jpl.nasa.gov}}. To firmly identify a molecular transition
corresponding to the observed spectra, we checked line blending, V$_{LSR}$ velocity, upper state energy (E$_{up}$), and Einstein coefficient. 
After assigning a molecular species to the observed spectral feature, we used LTE modeling to confirm or reject the identification.

\begin{deluxetable*}{lccccccc}
\tabletypesize{\footnotesize}
\tablewidth{0pt}
\small
\tablecaption{Summary of the observation.\label{table:observ_sum}}
\tablehead{\colhead{Source name}&\colhead{Observation date}&\colhead{On-source time}&\colhead{Number} &\colhead{Frequency range}&\colhead{Channel spacing}&\multicolumn{2}{c}{Baseline}\\
\colhead{}&\colhead{yyyy-dd-mm}&\colhead{hh:mm}&\colhead{of antennas}&\colhead{GHz}&\colhead{kHz}&\colhead{Maximum (m)}&\colhead{Minimum (m)}}

\startdata
G10.47+0.03&2017-05-03&001:53&39&129.50-131.44&244, 976&310&15\\ 
&2017-28-01&00:33.6&40&147.50-149.43&244, 976&272&15\\
&2017-06-03&01:03.5&41&153.00-154.93&244, 976&331&15\\
&2017-07-03&00:28.72&39&158.49-160.63&244, 976&331&15\\
\enddata
\end{deluxetable*}

\section{Observational results \label{sec:obs-results}}

\subsection{Continuum images}
\begin{figure*}
\begin{minipage}{0.53\textwidth}
\includegraphics[width=\textwidth]{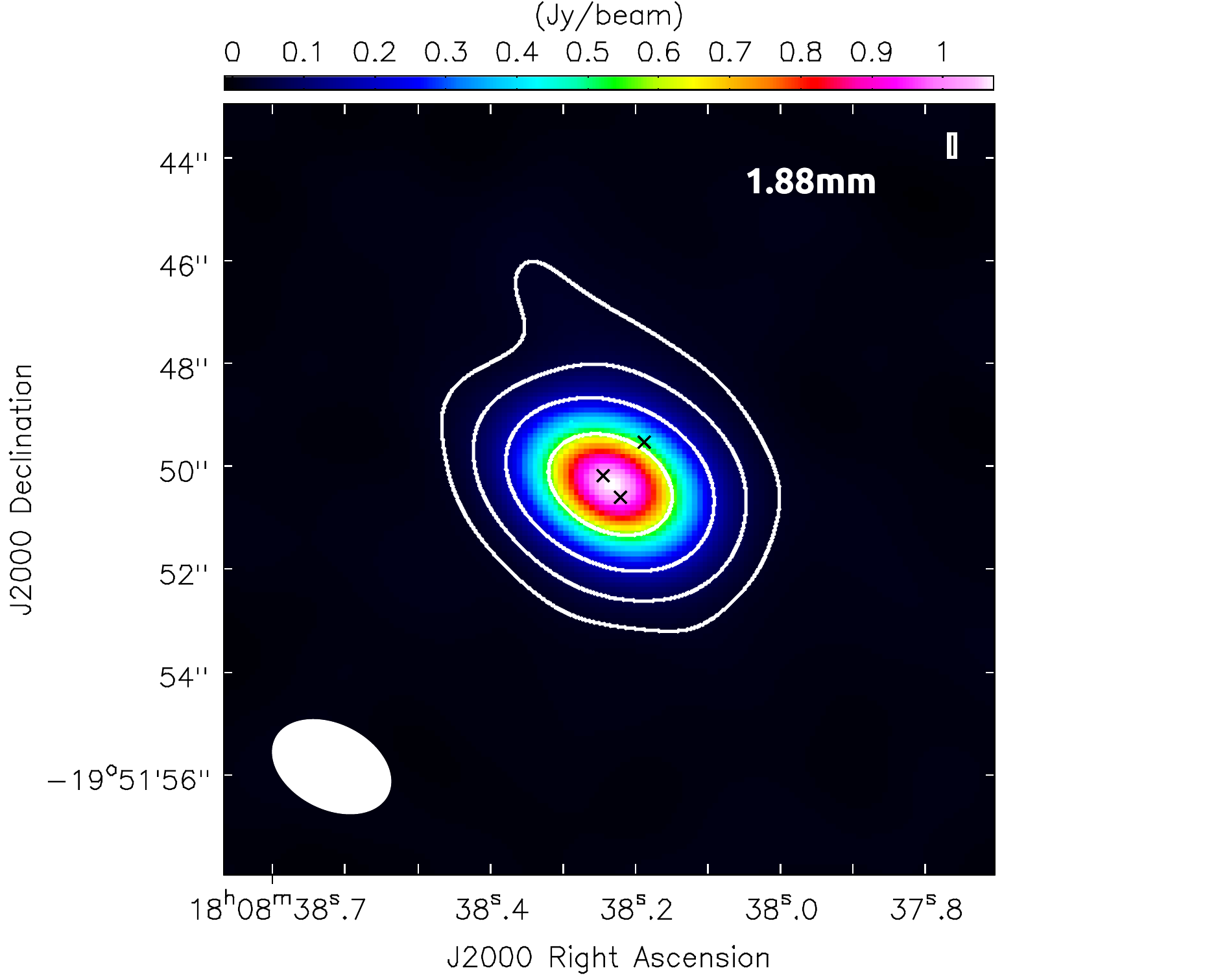}
\end{minipage}
\begin{minipage}{0.53\textwidth}
\includegraphics[width=\textwidth]{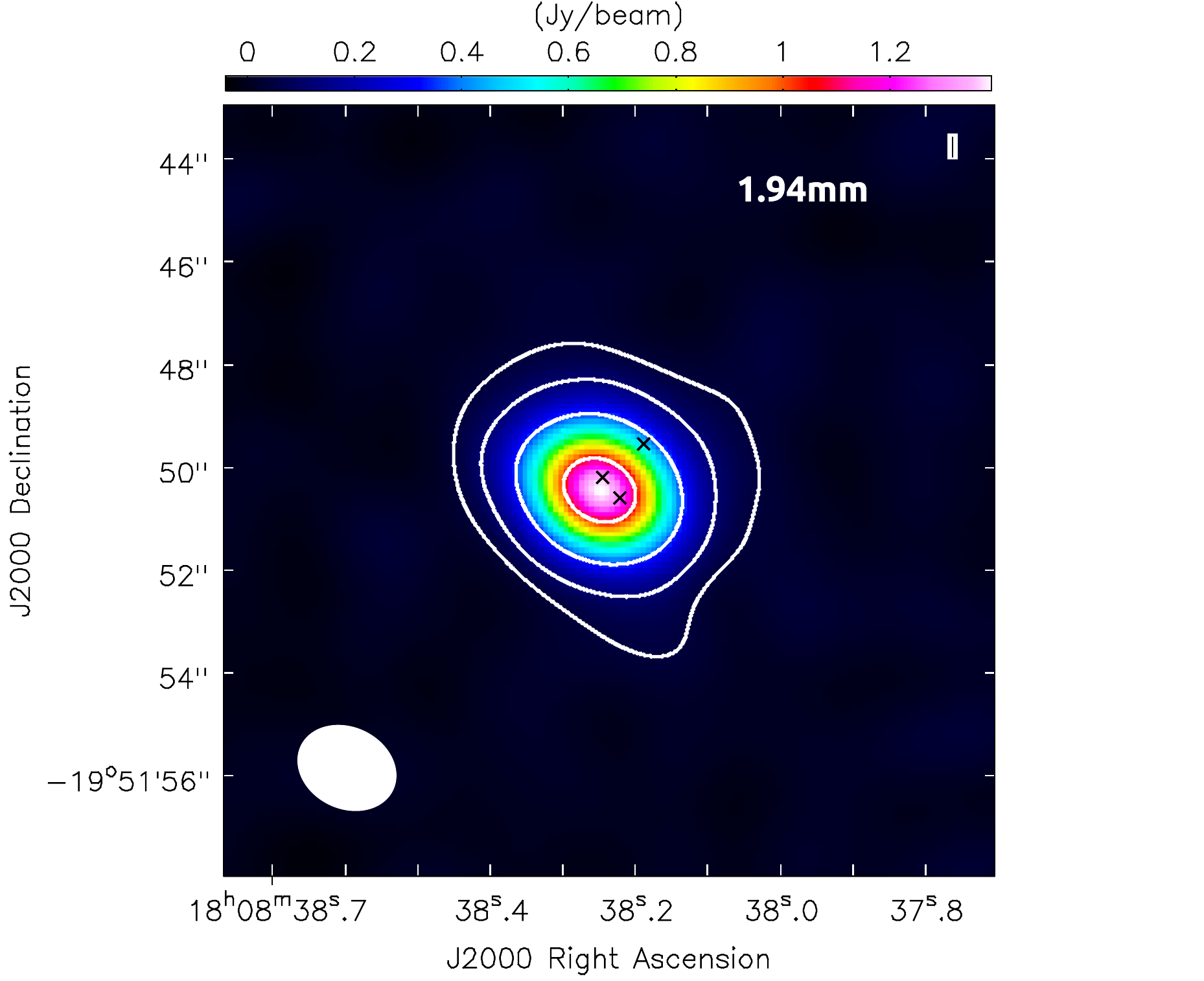}
\end{minipage}
\begin{minipage}{0.53\textwidth}
\includegraphics[width=\textwidth]{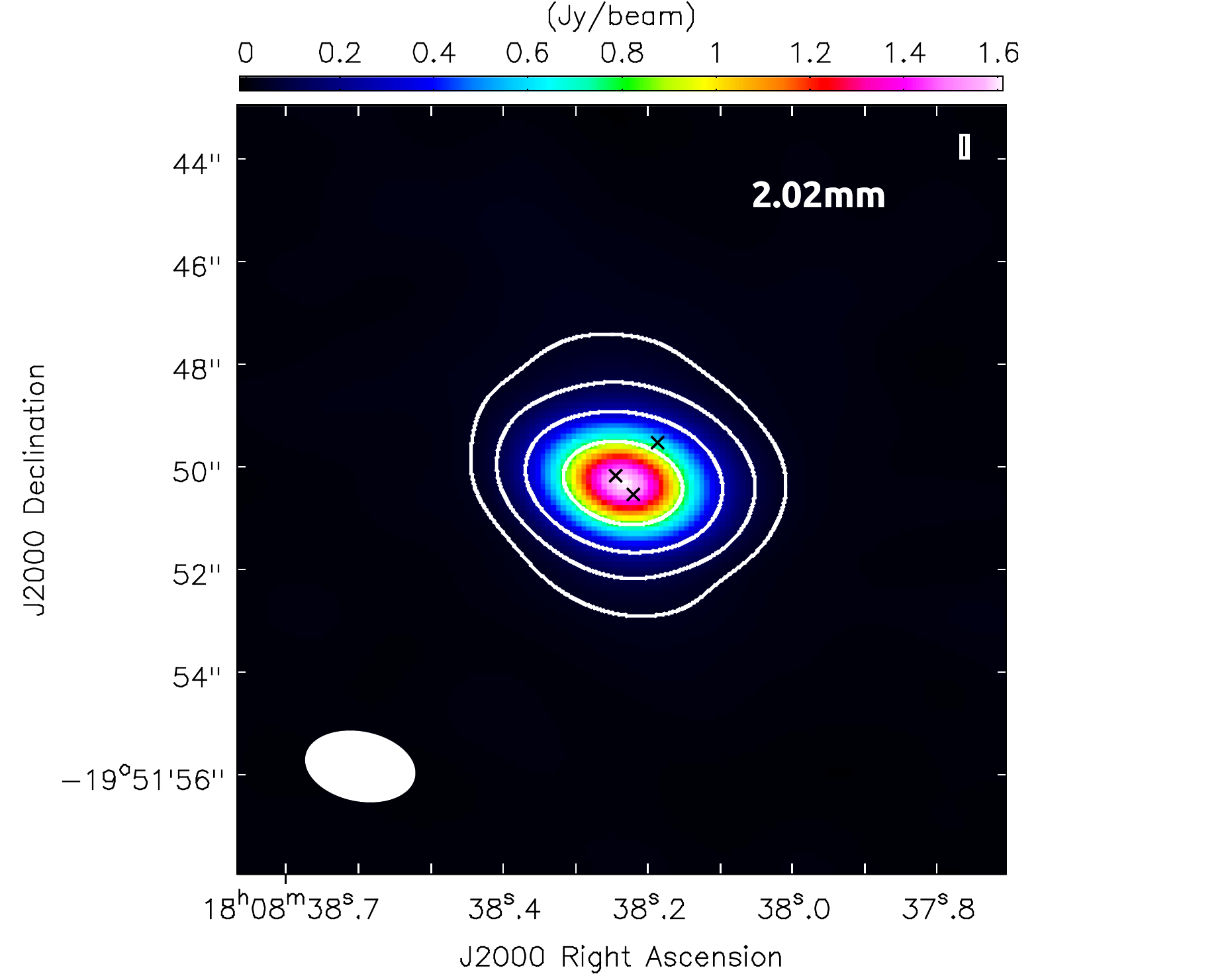}
\end{minipage}
\begin{minipage}{0.53\textwidth}
\includegraphics[width=\textwidth]{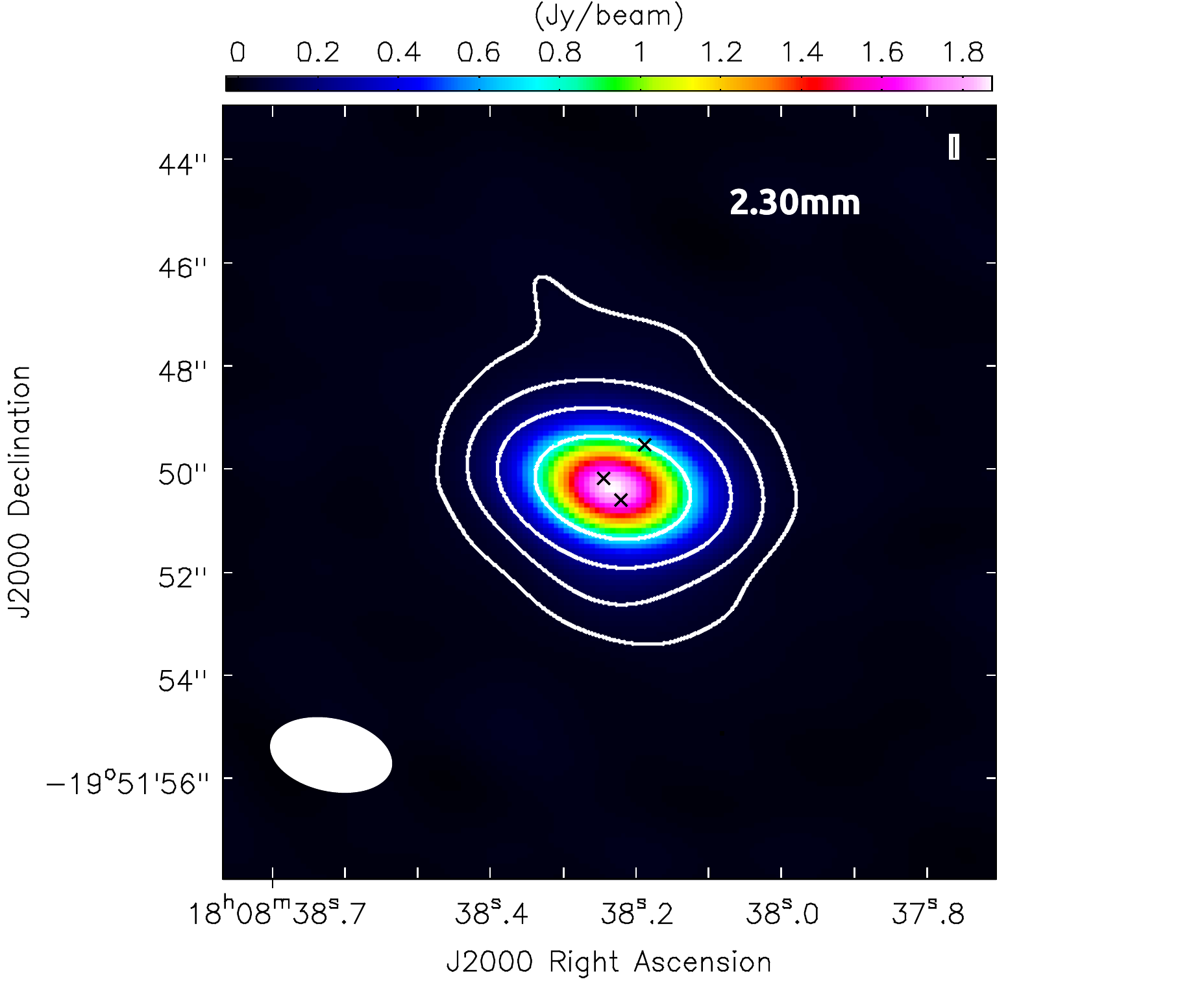}
\end{minipage}
\caption{{ Images of continuum emission observed towards G10 with ALMA at (i) $130.5$ GHz {($\sigma$ = 7 mJy/beam)}, (ii) $148.51$ GHz ($\sigma$ = 
13 mJy/beam), (iii) $153.96$ GHz ($\sigma$ = 10 mJy/beam), and (iv) $159.45$ GHz ($\sigma$ = 9 mJy/beam). Contours are drawn at 3$\sigma$, 9$\sigma$, 
27$\sigma$, and 81$\sigma$ for all the continuum maps. The observed beam is shown in the lower left corner of each figure. Black crosses in the continuum 
images indicate the HII regions B1, B2, and A in anticlockwise direction starting from the left black cross situated in the white and pink contour of 
continuum images \citep{cesa10}.}}
\label{figure:continuum}
\end{figure*}

{ \cite{cesa10} observed G10 with a very large array (VLA) and identified three distinct HII regions A, B1, and B2 inside the HMC.
The RMS noises of these observation were 269 $\mu$Jy/beam,  $73$ $\mu$Jy/beam, 773 $\mu$Jy/beam, and 227 $\mu$Jy/beam for 6 cm, $3.6$ cm, 2 cm, and $1.3$ cm
continuum and corresponding synthesized beam and position angle are 0$^{''}$.73$\times$0$^{''}$.42 and -11$^{\circ}$.2; 0$^{''}$.37$\times$0$^{''}$.19 and
-15$^{\circ}$.5; 0$^{''}$.74$\times$0$^{''}$.39 and 16$^{\circ}$.2; and 0$^{''}$.15$\times$0$^{''}$.092 and 6$^{\circ}$.8
respectively. They referred B1 and B2 as hypercompact (HC) HII regions and A as ultracompact (UC) HII regions. \cite{rolf11} observed this source with SMA. They observed continuum at three different frequency regions, 201/211 GHz, 345/355 GHz, 681/691 GHz. However, the beam size of  201/211 GHz, 681/691 GHz
frequency ranges were not sufficient to resolve the continuum, but the extension can be seen at 345/355 GHz. Here, we also observed continuum maps of G10 at
four different frequencies ($130.5$ GHz, $148.51$ GHz, $153.96$ GHz, and $159.45$ GHz), which are presented in Figure \ref{figure:continuum}. Our observed
beam sizes are also not sufficient to resolve the continuum.  The observed parameters of continuum images such as frequency, position, synthesized beam
size, position angle, peak flux, integrated flux, and deconvolved beam size (FWHM) are provided in Table \ref{table:properties}.
We obtained the peak flux, integrated flux, and the deconvolved beam size by using two-dimensional Gaussian fitting of the continuum images.}

\begin{deluxetable*}{lcccccccc}
\tabletypesize{\footnotesize}
\tablewidth{0pt}
\small
\tablecaption{Summary of the continuum images.\label{table:properties}}
\tablehead{\colhead{Frequency}&\multicolumn{2}{c}{Position (ICRS 2000)}&\colhead{Synthesized beam} &\colhead{Position angle}&\colhead{Peak flux }&\colhead{Integrated flux}&\colhead{FWHM}&\colhead{ RMS}\\
\colhead{(GHz)}&\colhead{$\alpha$, hms}&\colhead{$\delta$, ${\circ}$ $'$ $''$}&\colhead{$''\times ''$}&\colhead{in degree}&\colhead{(Jy/beam)}&\colhead{(Jy)}
&\colhead{$''$}&\colhead{(mJy/beam)}}
\startdata
130.50&18:08:38.23& -19.51.50.34&2.44$\times$1.64&63.16&1.06(0.007)&1.367(0.015)&1.04(0.057)&{$7$}\\
148.51&18:08:38.24& -19.51.50.42&1.98$\times$1.57&64.28&1.37(0.012)&2.031(0.026)&1.22(0.041)&{$13$}\\
153.96&18:08:38.24& -19.51.50.32&2.03$\times$1.47&73.50&1.59(0.001)&2.138(0.020)&1.00(0.031)&{$10$}\\
159.45&18:08:38.23& -19.51.50.35&2.38$\times$1.39&77.84&1.86(0.014)&2.477(0.030)&1.01(0.047)&{$9$}\\
\enddata
\end{deluxetable*}

\begin{deluxetable}{cccc}
\tabletypesize{\footnotesize}
\tablewidth{0pt}
\small
\tablecaption{Hydrogen column density and optical depth.\label{table:hcolden}}
\tablehead{\colhead{Wavelength}&\colhead{Hydrogen column density}&\colhead{Optical depth}\\
\colhead{(mm)}&\colhead{(cm$^{-2}$)}&\colhead{($\tau$)}}
\startdata
2.30&{ $\rm{9.93\times10^{24}}$ ($\rm{5.52\times10^{24}}$)$^{*}$} & { 0.098}\\
2.02&{ $\rm{1.32\times10^{25}}$ ($\rm{5.84\times10^{24}}$)$^{*}$} & { 0.133}\\
1.94&{ $\rm{1.48\times10^{25}}$ ($\rm{6.30\times10^{24}}$)$^{*}$} & { 0.151}\\
1.88&{ $\rm{1.62\times10^{25}}$ ($\rm{6.46\times10^{24}}$)$^{*}$} & { 0.166}\\
\hline
{ Average Value} &{ $\rm{1.35\times10^{25}}$ ($\rm{6.04\times10^{24}}$)$^{*}$} & { 0.135}\\
\enddata

{ $^{*}$ These are estimated by considering new set of mass absorption coefficient values, which are obtained following \cite{moto19}
as described in Section \ref{sec:hcolden}.}
\end{deluxetable}

\subsection{Rotation diagram analysis}
\label{sec:rotdia}
{ In this work, we have detected multiple lines of HNCO, NH$_2$CHO, and CH$_3$NCO and carried out rotation diagram analysis to obtain the 
temperature and column density of the observed species.} Assuming the observed transitions of these species are optically thin and are in Local Thermodynamic 
Equilibrium (LTE), we performed rotational diagram analysis. For optically thin lines, column density can be expressed as \citep{gold99},
\begin{equation}
\frac{N_u^{thin}}{g_u}=\frac{3k_B\int{T_{mb}dV}}{8\pi^{3}\nu S\mu^{2}},
\end{equation}
where, g$_u$ is the degeneracy of the upper state, k$_B$ is the Boltzmann constant, $\rm{\int T_{mb}dV}$ is the integrated intensity,
$\nu$ is the rest frequency, $\mu$ is the electric dipole moment, and S is the transition line strength. Under LTE conditions, the total
column density can be written as,
\begin{equation}
\frac{N_u^{thin}}{g_u}=\frac{N_{total}}{Q(T_{rot})}\exp(-E_u/k_BT_{rot}),
\end{equation}
where, $T_{rot}$ is the rotational temperature, E$_u$ is the upper state energy, $\rm{Q(T_{rot})}$ is the partition function at rotational
temperature. Equation 2 can be rearranged as,
\begin{equation}
log\Bigg(\frac{N_u^{thin}}{g_u}\Bigg)=-\Bigg(\frac{log\ e}{T_{rot}}\Bigg)\Bigg(\frac{E_u}{k_B}\Bigg)+log\Bigg(\frac{N_{total}}{Q(T_{rot})}\Bigg).
\end{equation}

Above equation shows that there is a linear relationship between the upper state energy and column density at the upper level. 
Using this equation, we can extract both column density and rotational temperature. 
{
Line parameters of the observed transitions are estimated with a single Gaussian fit. Observed and Gaussian fitted spectra of various transitions of 
HNCO, NH$_2$CHO, and CH$_3$NCO are provided in the Appendix (see Figures \ref{Gfit-hnco}-\ref{Gfit-ch3nco}).
All the line parameters of  observed molecules such as molecular transitions (quantum numbers) along with their rest frequency ($\nu$), upper state energy (E$_u$), line width 
($\Delta V$), line intensity (S$\mu^{2}$), integrated intensity ($\rm{\int T_{mb}dV}$) 
are presented in Table \ref{table:line-parameters}.} 

\begin{figure}
\begin{minipage}{0.42\textwidth}
\includegraphics[width=\textwidth]{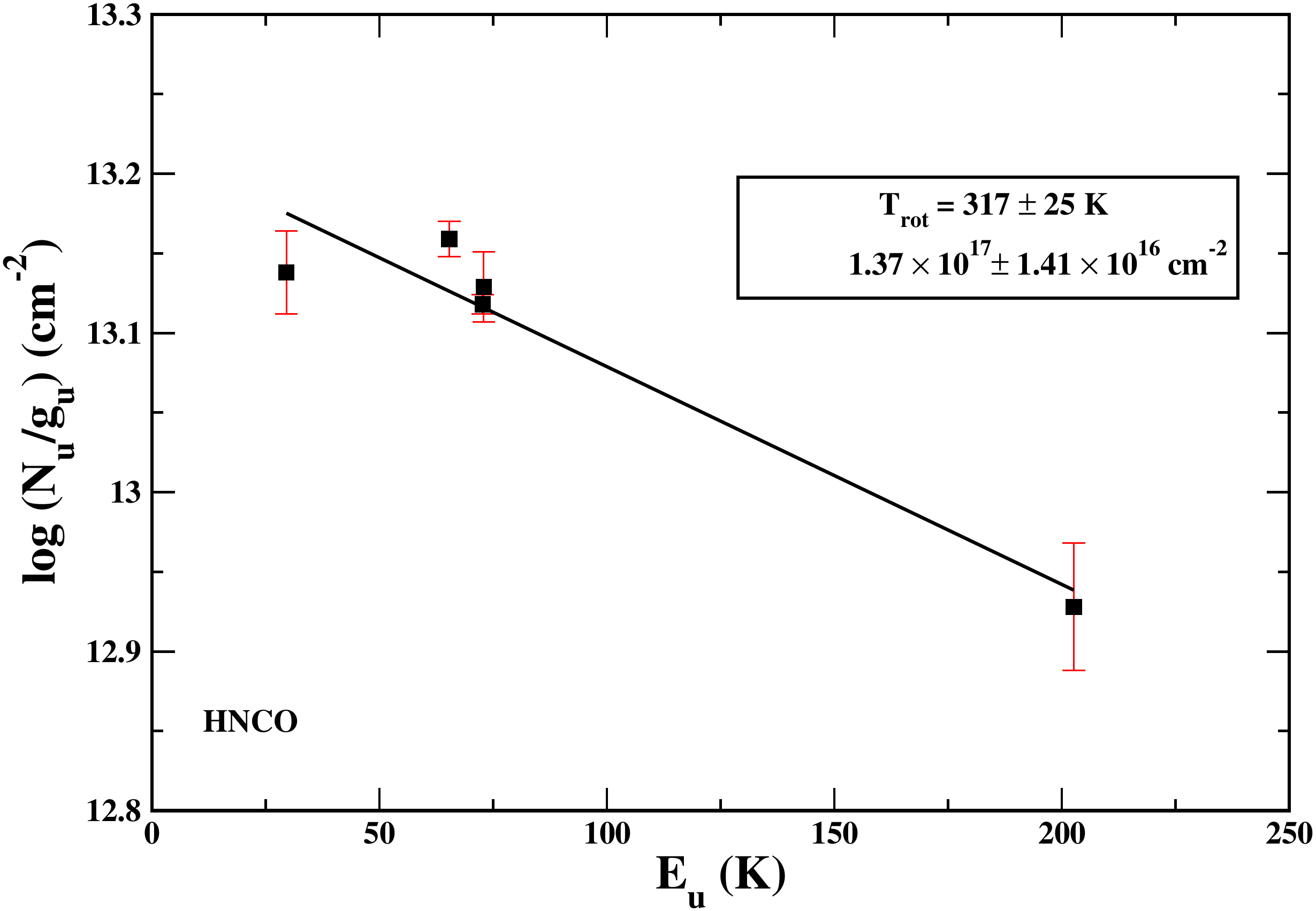}
\end{minipage}
\vskip 0.5cm
\begin{minipage}{0.42\textwidth}
\includegraphics[width=\textwidth]{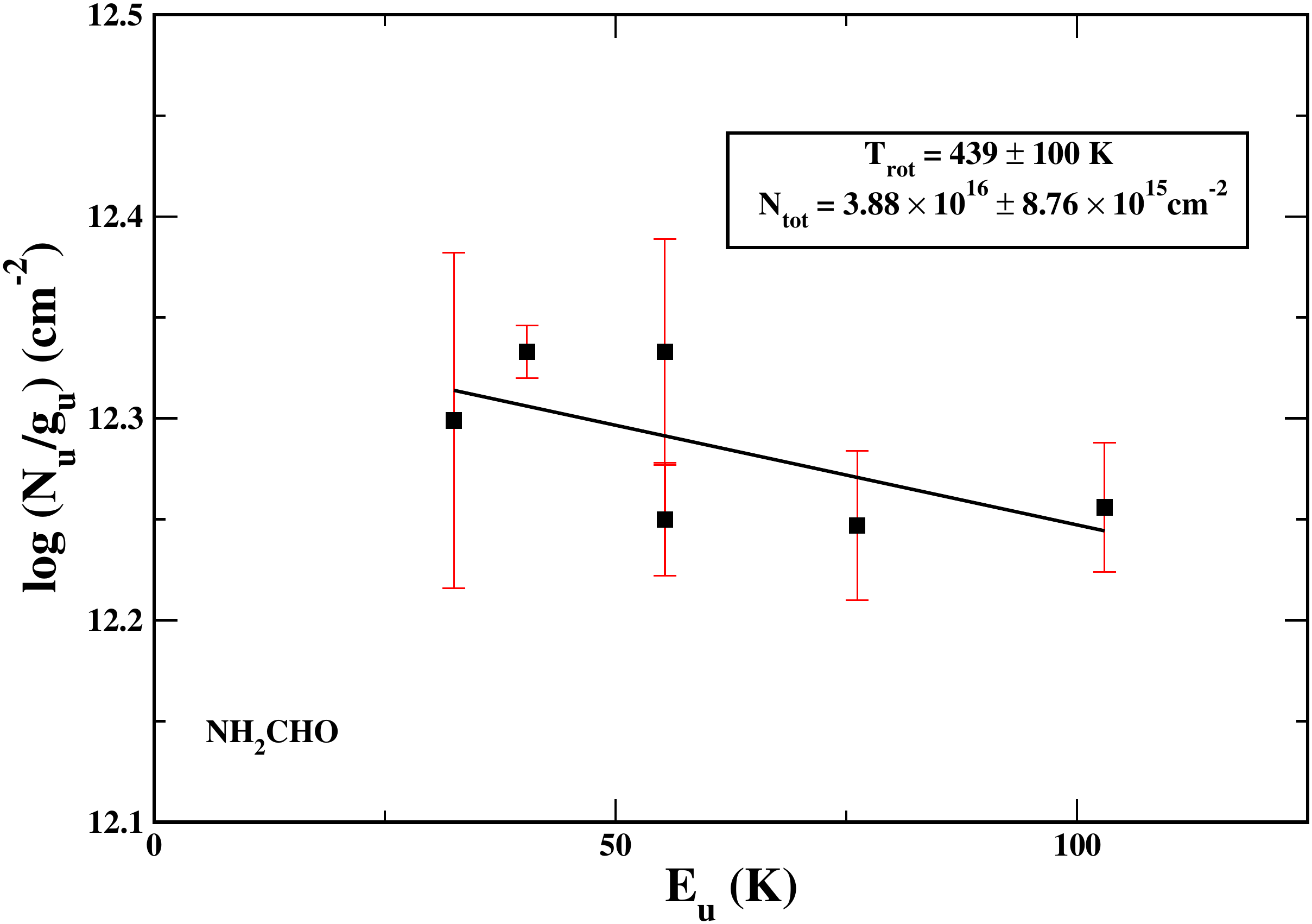}
\end{minipage}
\vskip 0.5cm
\begin{minipage}{0.42\textwidth}
\includegraphics[width=\textwidth]{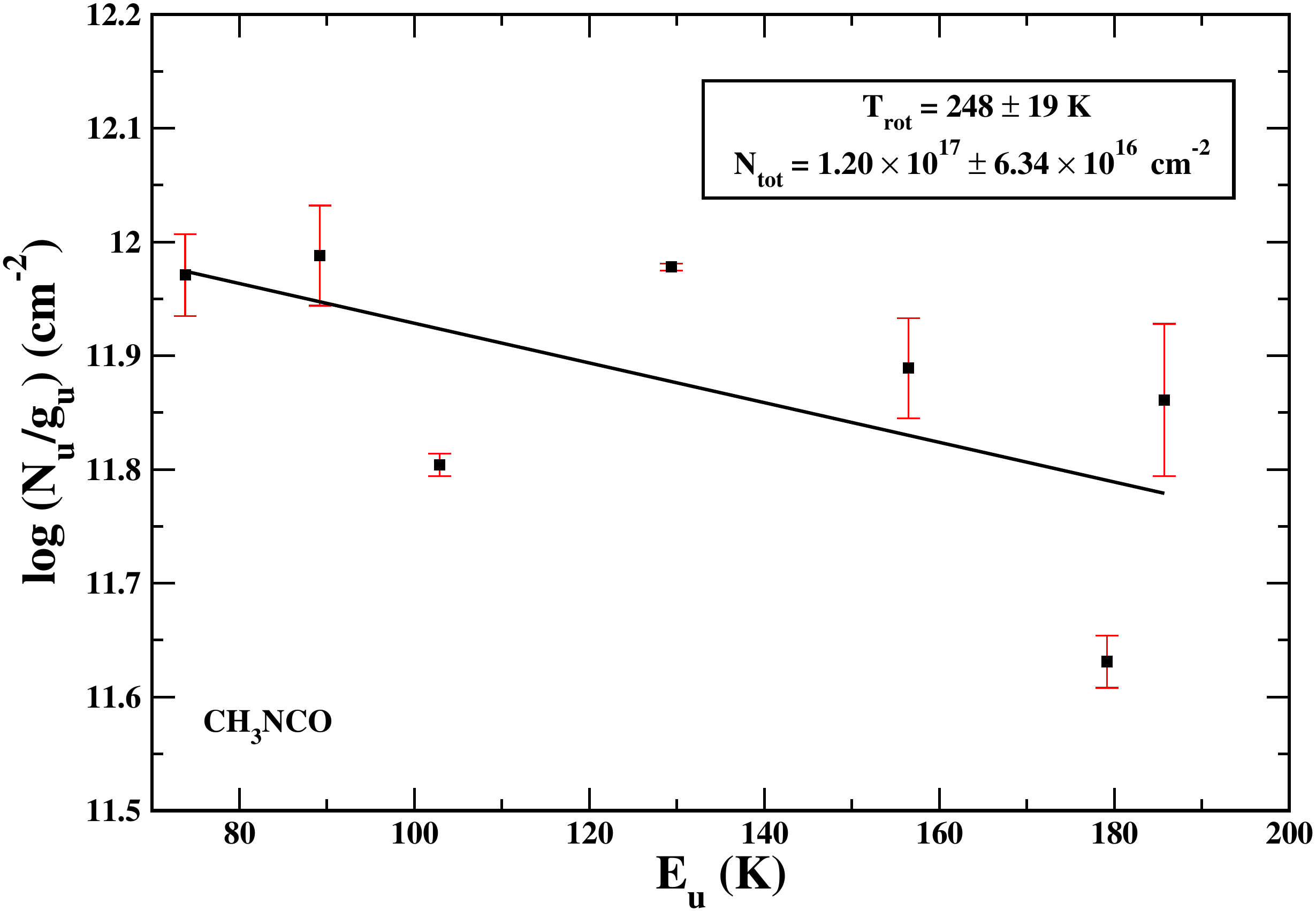}
\end{minipage}
\caption{{ Rotational diagram of HNCO, NH$_2$CHO, and CH$_3$NCO. Black filled squares are the data points and red lines represent the 
error bar. Best fit rotational temperature and column density are mentioned inside the small box for corresponding figure.}}
\label{figure:rotdia}
\end{figure}

We have identified multiple hyperfine transitions of HNCO and $\rm{NH_2CHO}$. However, with the present spectral resolution
it is not possible to resolve these transitions. 
{ Since there were various hyperfine transitions corresponds to a single observed spectral profile,
we have split the observed intensity flux according to their S$\mu^{2}$ values. Then we have used the most probable (maximum intensity) hyperfine transitions
in rotational diagram analysis. Selected transitions are then used to obtain the rotational temperature and column density from the rotational diagram.
We have detected many transitions of CH$_3$NCO but few of them are blended with some other nearby molecular transitions.
In Table \ref{table:line-parameters}, we have provided all the observed transitions. However, the integrated intensity is estimated only for
non-blended transitions of CH$_3$NCO, which are further used in the rotational diagram analysis of it. Rotational diagrams of HNCO, NH$_2$CHO,
and CH$_3$NCO are presented in Fig. \ref{figure:rotdia}.}

\subsection{Hydrogen Column density estimation}
\label{sec:hcolden}
Flux density of the dust continuum ($S_\nu$) for the optically thin condition can be written as,
\begin{equation}
 S_\nu =\Omega \tau_\nu B_\nu (T_d),	
\end{equation}
where, $\Omega$ is solid angle of the synthesized beam, $\tau_\nu$ is optical depth, $T_d$ is dust temperature, and $\rm{B_\nu(T_d)}$ is the
Planck function \citep{whit92}. Optical depth can be expressed as,
\begin{equation}
 \tau_\nu =\rho_d\kappa_\nu L,
\end{equation}
 where, $\rho_d$ is mass density of dust, $\kappa_{\nu}$ is the mass absorption coefficient, and L is the path length.
Using the dust-to-gas mass ratio ($Z$), the mass density of the dust can be written as,
\begin{equation}
\rho_d = Z\mu_H\rho_{H_2}=Z\mu_HN_{H_2}2m_H/L,
\end{equation}
where, $\rho_H$ is the mass density of hydrogen,  N$_H$ is the column density of hydrogen, m$_H$ is the  hydrogen mass
and $\mu_H$ is the mean atomic mass per hydrogen. Here, we used $Z=0.01$, $\mu_H=1.41$ \citep{cox00}, and dust temperature 200 K.
Measured peak flux density of the dust continuum of the source at different frequencies are noted in Table \ref{table:properties}. From 
the above equations, the column density of molecular hydrogen can be written as,
\begin{equation}
N_{H_2} = \frac{S_\nu /\Omega}{2\kappa_\nu B_\nu(T_d)Z\mu_H m_H}.
\end{equation}
According to the { extrapolation} of the data presented in \cite{osse94}, the mass absorption coefficient per gram of
dust at $130.50$, $148.51$, $153.96$, and $159.45$ GHz ($2.30$, $2.02$, $1.94$, and $1.88$ mm respectively) is of $\sim 0.20$ cm$^2$/g for the thin ice condition.
{ If we adopt the formula $\rm{k_\nu = 0.90(\nu/230 GHz)^{\beta}\ cm^{2}\ g^{-1}}$ \citep{moto19} in estimating the 
mass absorption coefficient, where $\rm{k_{230} = 0.90 \ cm^{2}\ g^{-1}}$ is the 
emissivity of the dust grains at a gas density of $\rm{10^{6}\ cm^{-3}}$ covered by a thin ice mantle at 230 GHz. Dust spectral index $\beta$ is 
used of $\sim$1.6 \citep{frie05}.  Following the above mentioned formula, the obtained value of mass absorption coefficient is 0.36, 0.45, 0.47, and
0.50 for the frequency 130.5 GHz, 148.5 GHz, 153.96 GHz, and 159.45 GHz respectively.} We estimated the hydrogen column density and optical depth of 
dust for the four frequency regions. Estimated hydrogen 
column density and optical depth values are given in Table \ref{table:hcolden}. We take a mean value to find out the resultant column density of the source. By 
taking the average of these four continuum values, we obtained column density of {$\sim 1.35 \times10^{25}$ cm$^{-2}$}. { 
Observed average hydrogen column density is of $\sim$2 times lower (see Table \ref{table:hcolden}) if we consider the mass absorption coefficients values as 
estimated following \cite{moto19}. Optical depth of the dust is estimated to be $0.135$. Achieved optical depth suggests that the source 
is optically thin in this frequency range and with present angular resolution of the observation.}

\subsection{Results of observed species }

\begin{figure*}[t]
\begin{center}
\includegraphics[width=15.6cm]{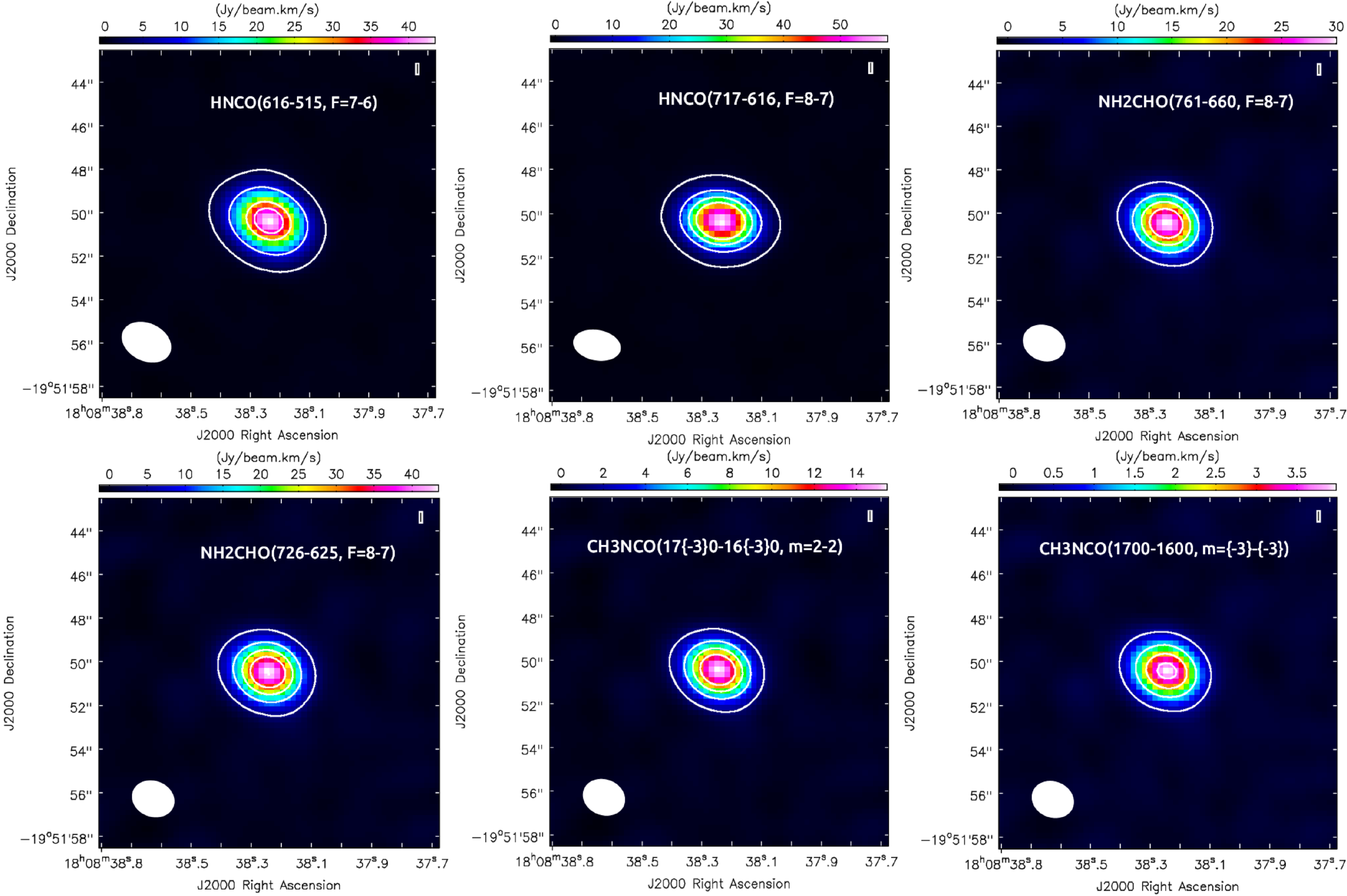}
\caption{ { Moment maps of HNCO, NH$_2$CHO, and CH$_3$NCO for various transitions. The contours
are from 3$\sigma$ to 43$\sigma$ in steps of 10$\sigma$ for HNCO. For $\rm{NH_2CHO}$ and
$\rm{CH_3NCO}$ the contours are from 3$\sigma$ to 18$\sigma$ in steps of 5$\sigma$. The RMS noise levels are
140, 130, 282, 412, 146, and 40 mJy beam$^{-1}$ Km s$^{-1}$ for HNCO ($6_{16}-5_{15}$, $7_{17}-6_{16}$), NH$_2$CHO 
($7_{61}-6_{60}$, $7_{26}-6_{25}$) and CH$_3$NCO ($17_{-30}-16_{-30}$, $17_{00}-16_{00}$) respectively.}}
\label{figure:moment}
\end{center}
\end{figure*}

\subsubsection{Isocyanic acid, HNCO}
We have observed numerous hyperfine transitions of HNCO. All the line parameters of perceived transitions are summarized in
Table \ref{table:line-parameters}. Spatial distribution of the observed HNCO transitions is shown in Fig. \ref{figure:moment}.
Here, we have depicted the spatial distribution of two transitions of HNCO with two different upper state energies.
{ To determine the emitting region of various molecular transitions, we have used two-dimensional Gaussian fittings of 
the first-order moment map image.} $\rm{6_{1,6}-5_{1,5}}$ emission of HNCO is found to be compact ($\rm{\theta \sim \ 1.11^{''}}$)
than $\rm{7_{1,7}-6_{1,6}}$ transition ($\rm{\theta \sim \ 1.25^{''}}$). For 7$_{0, 7}$-6$_{0, 6}$ and 7$_{2, 6}$-6$_{2, 5}$ transitions 
we found slightly extended region ($\rm{\theta \sim \ 1.38^{''}}$). However,
the morphological structures of spatial distribution of all transitions
are similar. Obtained rotational temperature, column density, and  
fractional abundances are given in Table
\ref{table:abunobs}. { From the rotational diagram analysis, we obtained a rotational temperature of about $317$ K,
and column density of $\rm{\sim 1.37 \times10^{17}\ cm^{-2}}$. \cite{gibb03} estimated the rotational temperature and column density of 
HNCO  to be $328$ K and 6.76$\rm{\times10^{16}\ cm^{-2}}$ respectively in G10. Here, we obtained a column density which is about 
two times higher and almost similar rotational temperature as reported in \cite{gibb03}}.

\subsubsection{Formamide, NH$_2$CHO}
We have identified several hyperfine transitions of NH$_2$CHO. All the line parameters of the observed NH$_2$CHO 
transitions are presented in Table \ref{table:line-parameters}.
Spatial distribution of the observed NH$_2$CHO transitions are depicted in Fig. \ref{figure:moment}. 
Spatial distribution of $\rm{7_{2,6}-6_{2,5}}$ ($E_u=40.40$ K) and $\rm{7_{6,1}-6_{6,0}}$ ($E_u=135.74$ K) transitions show 
similar nature.  We obtained a higher rotational temperature of $\sim 439$ K.
We obtained the column density of NH$_2$CHO of $\sim3.88 \times10^{16}$ cm$^{-2}$ which is in good agreement with previous study \citep{rolf11}.
Spatial distribution of NH$_2$CHO is also found to be similar as HNCO. Emitting region ($\theta$) of 
NH$_2$CHO transitions varies between $\rm{1.18^{''}}$ and $1.55^{''}$.

\subsubsection{Methyl isocyanate, CH$_3$NCO}
In CH$_3$NCO, there is an internal rotation of the methyl group (CH$_3$, described by quantum number $m$) and
low-frequency CNC bending motion. Estimated energies for the sub-states m = 1, 2 relative to the ground state (m = 0)
is $8.4$ and $36.8$ cm$^{-1}$ respectively.  For $m=3$ sub-state this estimated relative energy is $79.7$, and $80.3$ cm$^{-1}$ for the
nearly two degenerate $m=3$ sub-state and $140.6$ cm$^{-1}$ for $m = 4$ sub-state.
The next higher vibrational state was found to be the first excited
state of the CNC bending mode, $V_b = 1$ at $182.2$ cm$^{-1}$ \citep{cern16}. Observed transitions of CH$_3$NCO and their
line parameters are summarized in Table \ref{table:line-parameters}. Several transitions of methyl isocyanate have observed
in this work with $m=0, 1, 2, 3$. Spatial distribution of CH$_3$NCO transitions
for $m= 2$ and $3$ are depicted in Figure \ref{figure:moment}. Interestingly, it is observed that emission with
$m=0$ transitions of CH$_3$NCO are compact ($\rm{\theta \sim 1.0^{''}}$) with the continuum emission 
whereas the transitions with a higher value of m, the emitting regions are comparatively extended
(for m = 1, $\rm{\theta \sim 1.15^{''}}$; m = 2, $\rm{\theta \sim 1.19^{''}}$). However, these transitions are marginally
resolved and it is not possible to draw any conclusion regarding the spatial distribution of these molecules in this
source. A high angular and spatial resolution observation can shed some light on this issue more elaborately. 
The obtained rotational temperature of CH$_3$NCO is $248$ K.
For another high-mass star-forming region, \cite{cern16} observed CH$_3$NCO in the warm gas of Sgr B2 and obtained rotational 
temperature of $\sim$ 200 K  { and column density of $\sim\rm{(3-5)\times10^{17}\ cm^{-2}}$. We estimated the column density and fractional abundance CH$_3$NCO as $\rm{1.20\times10^{16}}$ 
cm$^{-2}$ and $\rm{8.88\times10^{-09}}$ respectively. Our observed column density of CH$_3$NCO in this source to be 2-4 times lower as 
compared to that in Sgr B2 observation \citep{cern16}}.

\begin{table*}
\centering
\tiny
\caption{Summary of the line parameters of observed molecules towards G10. \label{table:line-parameters}}
\begin{tabular}{|c|c|c|c|>{}c|c|>{}c|c|}
\hline
\hline
Species&(${\rm J^{'}_{K_a^{'}K_c^{'}}}$-${\rm J^{''}_{K_a^{''}K_c^{''}}}$)&Frequency (GHz)&E$_u$ (K)&FWHM (Km s$^{-1}$)&S$\mu^{2}$(Debye$^{2}$)&
${\int{T_{mb}}dv (K.Km/s)}$&Remarks\\
\hline\hline
\hline
&6$_{1, 6}$-5$_{1, 5}$, F=7-6&131.394262&65.35&3.85&0.00291&228.88$\pm$5.72&\\
HNCO&7$_{1, 7}$-6$_{1, 6}$, F=8-7&154.414770&72.92&1.93&0.35919&289.88$\pm$12.91&\\
&7$_{0, 7}$-6$_{0, 6}$, F=8-7&153.865080&29.54&2.27&0.36664&301.17$\pm$17.68&\\
&7$_{2, 6}$-6$_{2, 5}$, F=8-7&153.818870&202.63&1.99&0.33673&170.14$\pm$14.40&\\
&7$_{1, 7}$-6$_{1, 6}$, F=8-7&153.291840&72.70&2.16&0.00184&280.76$\pm$3.66&\\
\hline
&7$_{2, 6}$-6$_{2, 5}$, F= 8-7&148.223354&40.40&3.79&95.26&212.49$\pm$6.33&\\
&7$_{6, 1}$-6$_{6, 0}$, F=8-7&148.556276&135.74&1.71&20.55&74.04$\pm$6.22&\\
&7$_{5, 3}$-6$_{5, 2}$, F=8-7&148.567249&102.99&1.81&43.92&95.13$\pm$6.96&\\
NH$_2$CHO&7$_{4, 3}$-6$_{4, 2}$, F=8-7&148.599727&76.19&2.11&52.15&128.08$\pm$11.02&\\
&7$_{3, 5}$-6$_{3, 4}$, F=8-7&148.667591&55.34&4.23&73.20&189.77$\pm$12.34&\\
&7$_{3, 4}$-6$_{3, 3}$, F=8-7&148.709316&55.35&3.93&63.22&161.56$\pm$10.54&\\
&7$_{1, 6}$-6$_{1, 5}$, F=8-7&153.432351&32.47&3.82&101.58&217.18$\pm$41.43&\\
\hline
&15$_{0, 15}$-14$_{0, 14}$, m=0-0&129.957471&49.91&--&123.94&--&Blended\\
&15$_{3, 12}$-14$_{3, 11}$, m=0-0&129.669703&103.61&--&120.32&--&Blended\\
&15$_{2, 14}$-14$_{2, 13}$, m=0-0&130.146799&73.80&11.63&121.88&103.58$\pm$8.52&\\
&15$_{-3, 0}$- 14$_{-3, 0}$, m=1-1&130.300215&115.63&9.61&&50.89$\pm$5.36&\\
&15$_{2, 13}$-14$_{2, 12}$, m=0-0&130.228419&73.82&--&121.89&--&Blended\\
&15$_{2, 0}$-14$_{2, 0}$, m=1-1&130.541066&85.84&--&121.08&--&Blended\\
&15$_{-1, 0}$-14$_{-1 0}$, m=2-2&130.583038&108.84&--&122.79&--&Blended\\
&15$_{0, 0}$-14$_{0, 0}$, m=2-2&130.653851&102.88&9.39&123.19&71.61$\pm$1.57&\\
&14$_{3, 0}$-13$_{3,0}$, m=1-1&130.661691&109.38&--&95.64&--&Blended\\
&15$_{1, 0}$-14$_{1, 0}$, m=2-2 &130.88332&108.84&--&124.79&--&Blended\\
CH$_3$NCO&17$_{3, 0}$ -16$_{3, 0}$, m=(-3)-(-3)&148.833657&232.81&--&122.25&--&Blended\\
&17$_{-3, 0}$ -16$_{-3, 0}$, m=2-2 &148.437799&170.22&9.92&134.84&200.34$\pm$9.84&\\
&17$_{1, 0}$-16$_{1 0}$, m=2-2&148.376092&85.19&--&138.51&--&Blended\\
&17$_{1, 0}$-16$_{1, 0}$, m=3-3  &148.326896&184.87&--&141.48&--&Blended\\
&17$_{1, 0}$-16$_{1, 0}$, m=(-3)-(-3)&148.280088&185.73&12.11&141.57&106.45$\pm$16.28&\\
&17$_{0, 0}$-16$_{0, 0}$,  m=(-3)-(-3) &148.262442&179.18&10.03&139.37&61.71$\pm$3.20&\\
&16$_{3, 0}$-15$_{3, 0}$, m=1-1&148.075687&122.28&--&112.35&--&Blended\\
&17$_{0, 0}$-16$_{0, 0}$, m=2-2 &148.061901&116.62&--&139.59&--&Blended\\
&17$_{-3, 0}$ -16$_{-3,0}$, m=1-1&147.673312&129.36&10.53&136.04&133.62$\pm$0.98&\\
&17$_{2, 15}$-16$_{2, 14}$, m=0-0 &147.603962&87.56&--&138.69&--&Blended\\
&18$_{1, 0}$-17$_{1, 1}$, m=1-1&154.742833&89.20&11.62&150.95&158.88$\pm$15.98&\\
&18$_{1, 18}$-17$_{1, 17}$, m=0-0&154.636867&76.47&--&148.31&--&Blended\\
\hline

\end{tabular}
\end{table*}

\begin{table}
{
\centering
\caption{Estimated rotational temperatures, column densities, and fractional abundances
of the observed species.}
\begin{tabular}{cccc}
\hline
\hline
Species& Rotational& Column & Fractional \\
&temperature& density &abundance\\
&(K)&(cm$^{-2}$)&\\
\hline
HNCO&317 $\pm$ 25&$\rm{1.37\times10^{17}}$&$\rm{1.02\times10^{-08}}$\\
NH$_2$CHO&439 $\pm$ 100&$\rm{3.88\times10^{16}}$&$\rm{2.87\times10^{-09}}$\\
CH$_3$NCO&248 $\pm$ 19&$\rm{1.20\times10^{17}}$&$\rm{8.88\times10^{-09}}$\\
\hline
\end{tabular}
\vskip 0.25cm
{ Notes.} Assuming the mean value of $\rm{N_{H_2}=1.35\times10^{25}}$ cm$^{-2}$ as estimated
in Table \ref{table:hcolden}.}
\label{table:abunobs}
\end{table}

\subsection{LTE fitting using MCMC}
We have used Markov Chain Monte Carlo (MCMC) method to fit the observed line profiles of $\rm{HNCO, NH_2CHO}$, and $\rm{CH_3NCO}$ towards the hot core G10. 
We have assumed that the source is under the LTE condition. We have extracted the best fitted physical parameters (column density, excitation temperature,
FWHM, optical depth, and source velocity) from the fitting. 
{ We have used the python scripting interface available in CASSIS for our model calculation to find out the best-fitted physical
parameters for the astronomical source. To determine the best-fitted set that can fit the observational result,  we have used the $\chi^2$
minimization process by considering the N number of spectra. This python script computes the $\chi^2$ between the observed and simulated data
and finds the minimal value of $\chi^2$ following the relation:
$$
{\rm {\chi_i}^2=\sum_{j=1}^{N_i} \frac{(I_{obs,ij}-I_{model,ij})^2}{rms^2_i+(cal_i\times I_{obs,ij})^2}},
$$
where, ${\rm I_{obs,ij}}$ and ${\rm I_{model,ij}}$  are observed and modeled intensity in the channel j of transition i respectively,
rms$_i$ is the rms of the spectrum i, and cal$_i$ is the calibration error. The reduced $\chi^2$ is computed using the following 
relation:
$$
{\rm \chi^2_{red}=\frac{1}{ \sum_{i}^{N_{spec}} N_i} \sum_{i=1}^{N_{spec}} \chi^2_i}.
$$
In the MCMC calculation, the initial physical values are chosen randomly between the minimum (${\rm X_{max}}$) to the maximum
(${\rm X_{min}}$) range set by the user
during their modeling. The step
of MCMC computation ($\theta_l$) depends on the iteration number $l$ and other parameters ($\alpha$ and $v$), where, 
$\theta_{l+1}=\theta_{l}+\alpha (v-0.05)$ 
(v is
a random number between 0 to 1). Here $\alpha$ is defined as,
$$
{\rm \alpha=\frac{k(X_{max}-X_{min})}{k^\prime}},
$$
where, k is defined as 
$$
k=r_c \hskip 3.5cm \rm{when \ l>c},
$$
$$
k=\frac{(r_c-1)}{c}l+1  \hskip 2cm  \rm{when \ l<c},
$$ 
where, c and r$_c$ are the parameter cutoff and ratio at cutoff respectively which are set by the
user during the modeling. 
k$^\prime$ is defined as reduced
physical parameter which is set to a value during computation. 
$\alpha$ determines the amplitude of the steps, which starts with a bigger step at the initial
stage of
the computation to find a good $\chi^2$ and shorter steps at the end of the computation to extract the value of the potential best $\chi^2$.}

\begin{figure}
\begin{minipage}{0.23\textwidth}
\includegraphics[width=\textwidth]{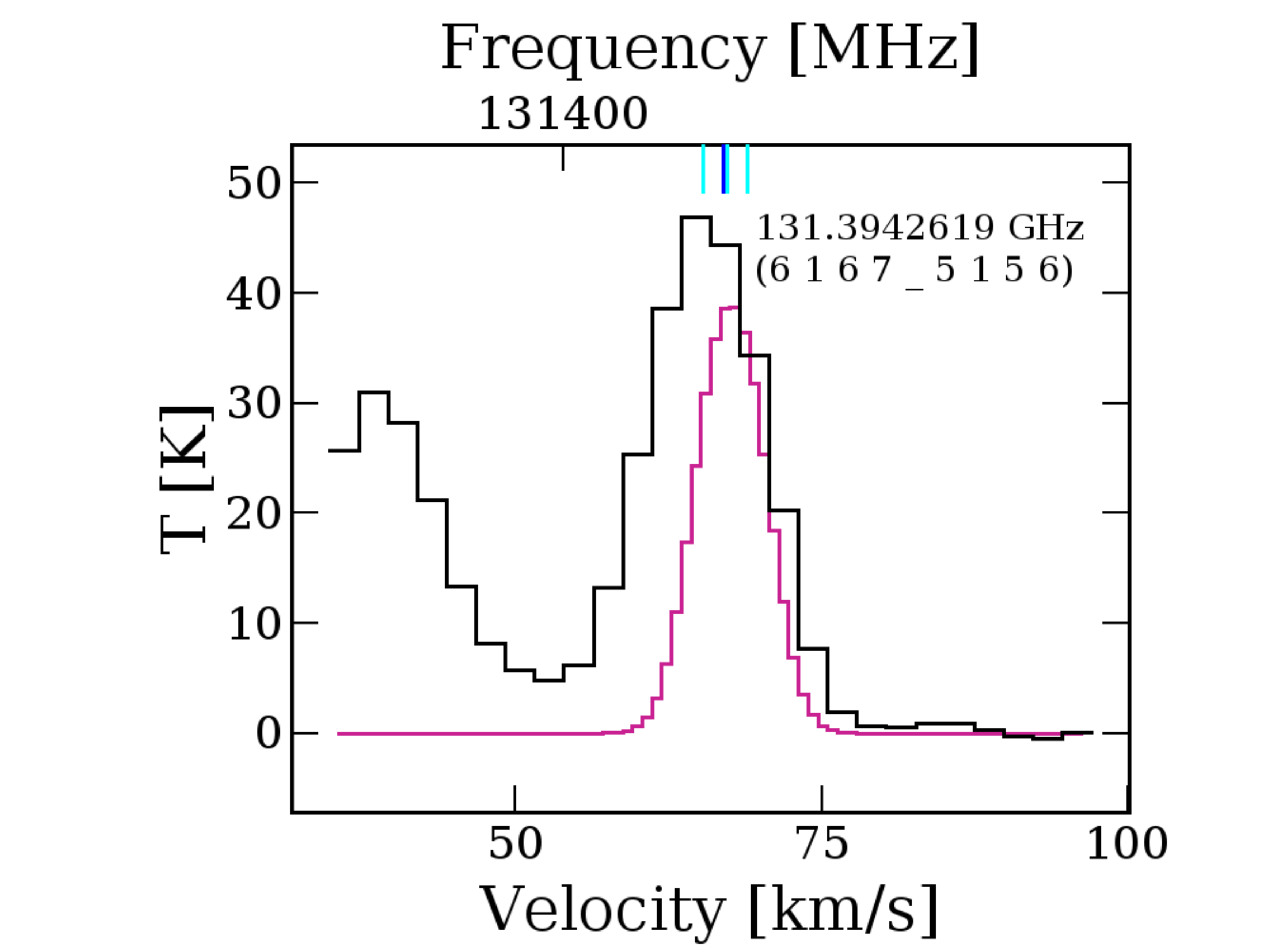}
\end{minipage}
\begin{minipage}{0.23\textwidth}
\includegraphics[width=\textwidth]{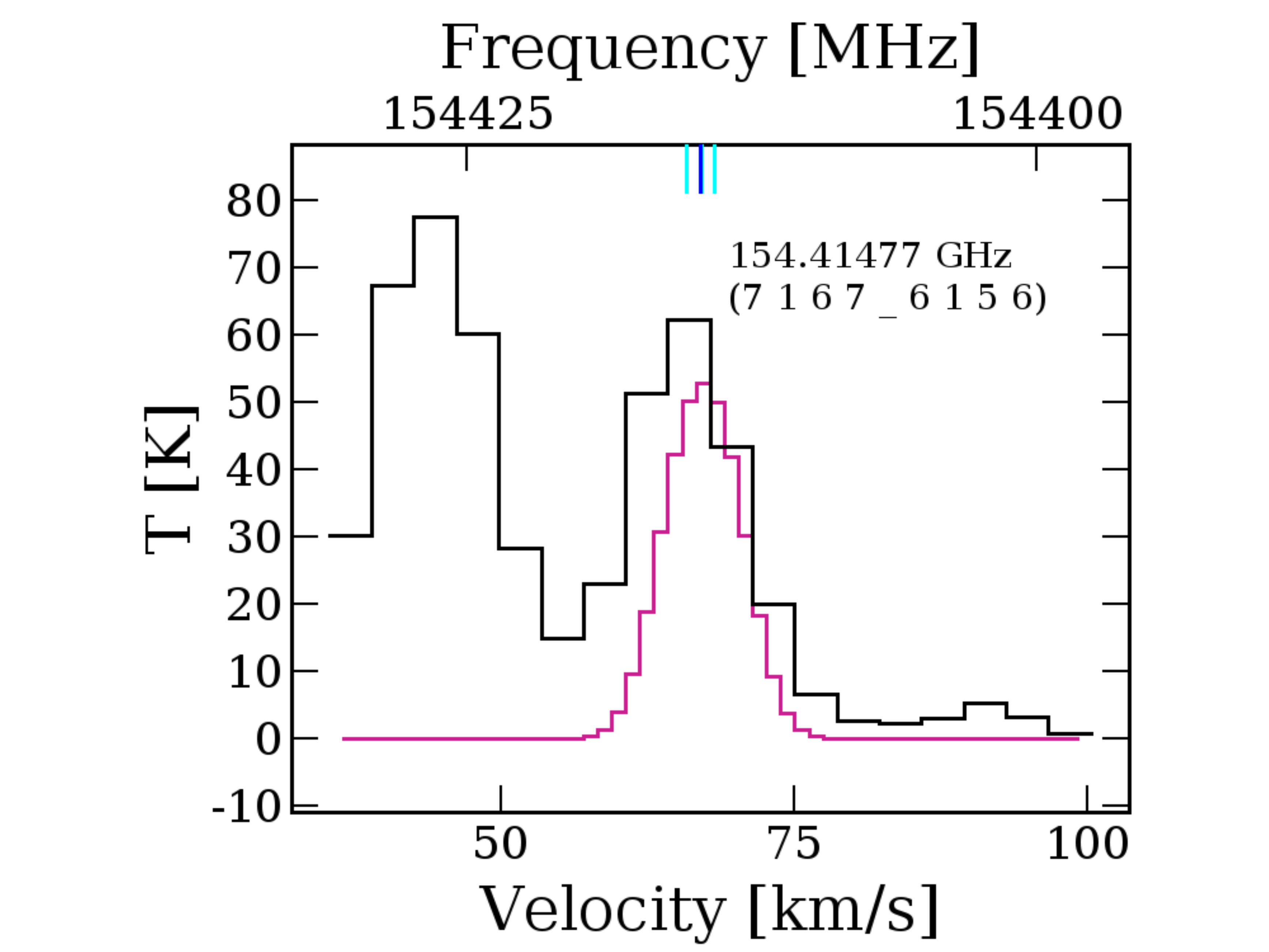}
\end{minipage}
 \begin{minipage}{0.23\textwidth}
 \includegraphics[width=\textwidth]{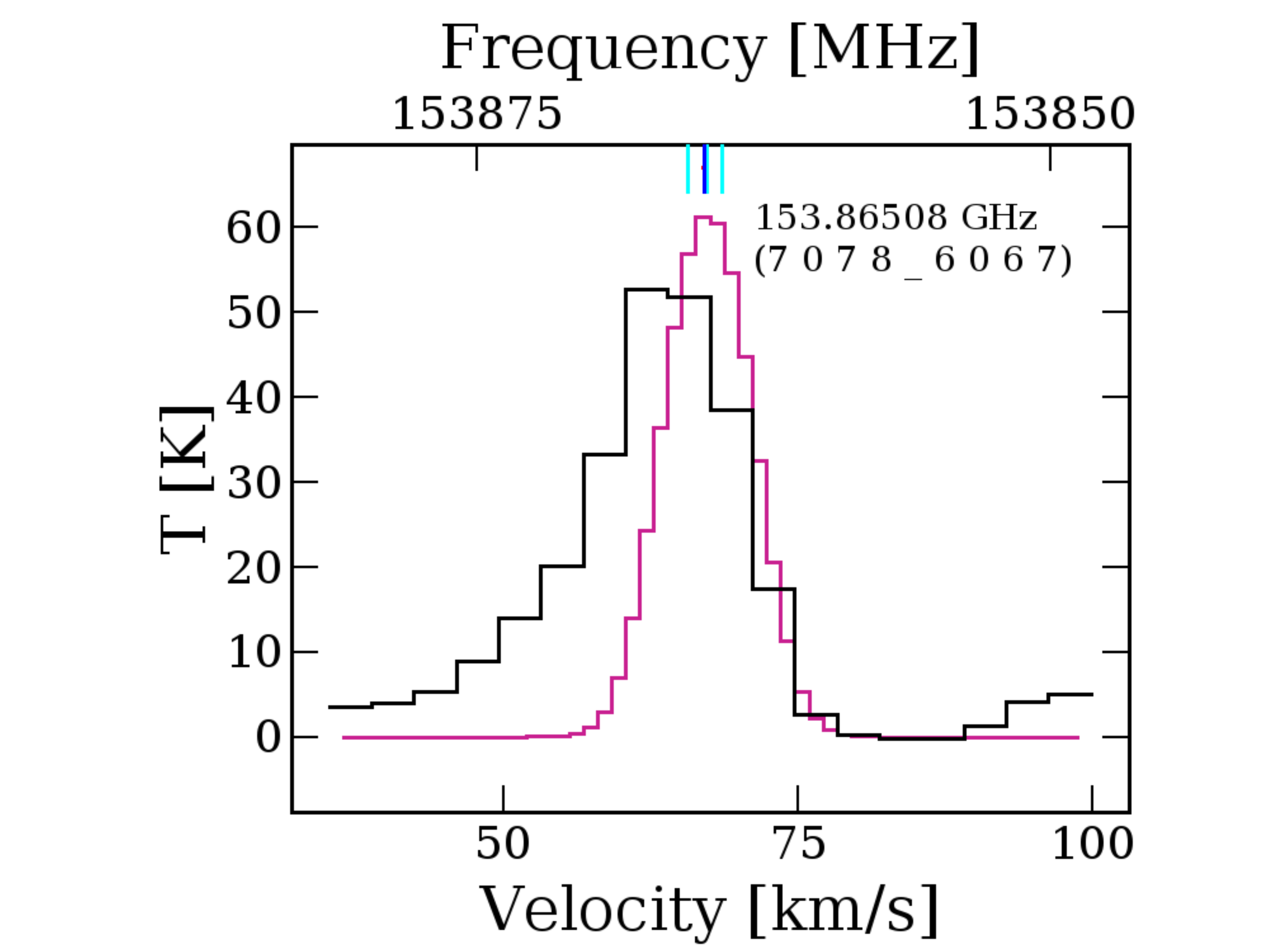}
 \end{minipage}
\begin{minipage}{0.23\textwidth}
\includegraphics[width=\textwidth]{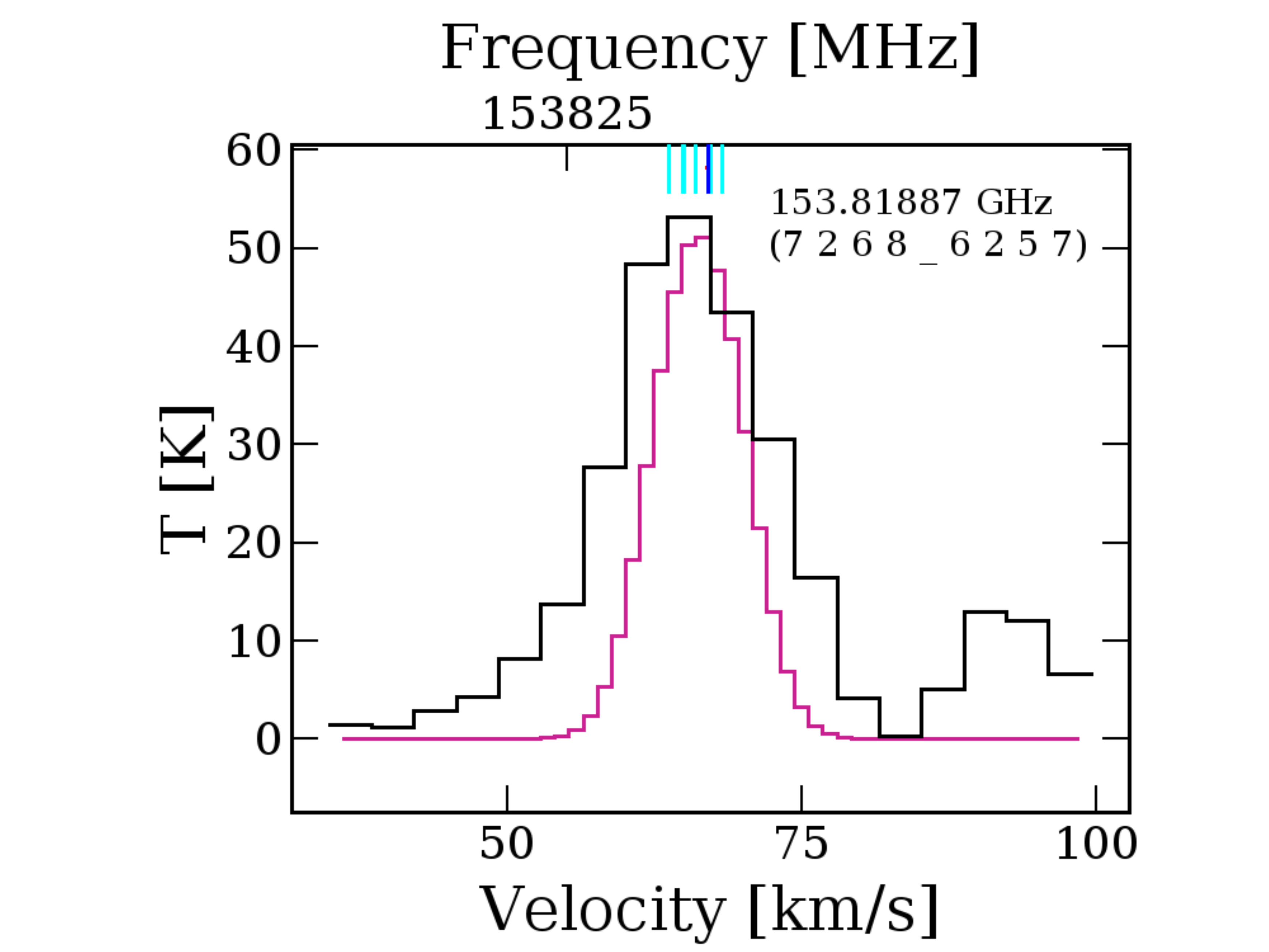}
\end{minipage}
\begin{minipage}{0.23\textwidth}
\includegraphics[width=\textwidth]{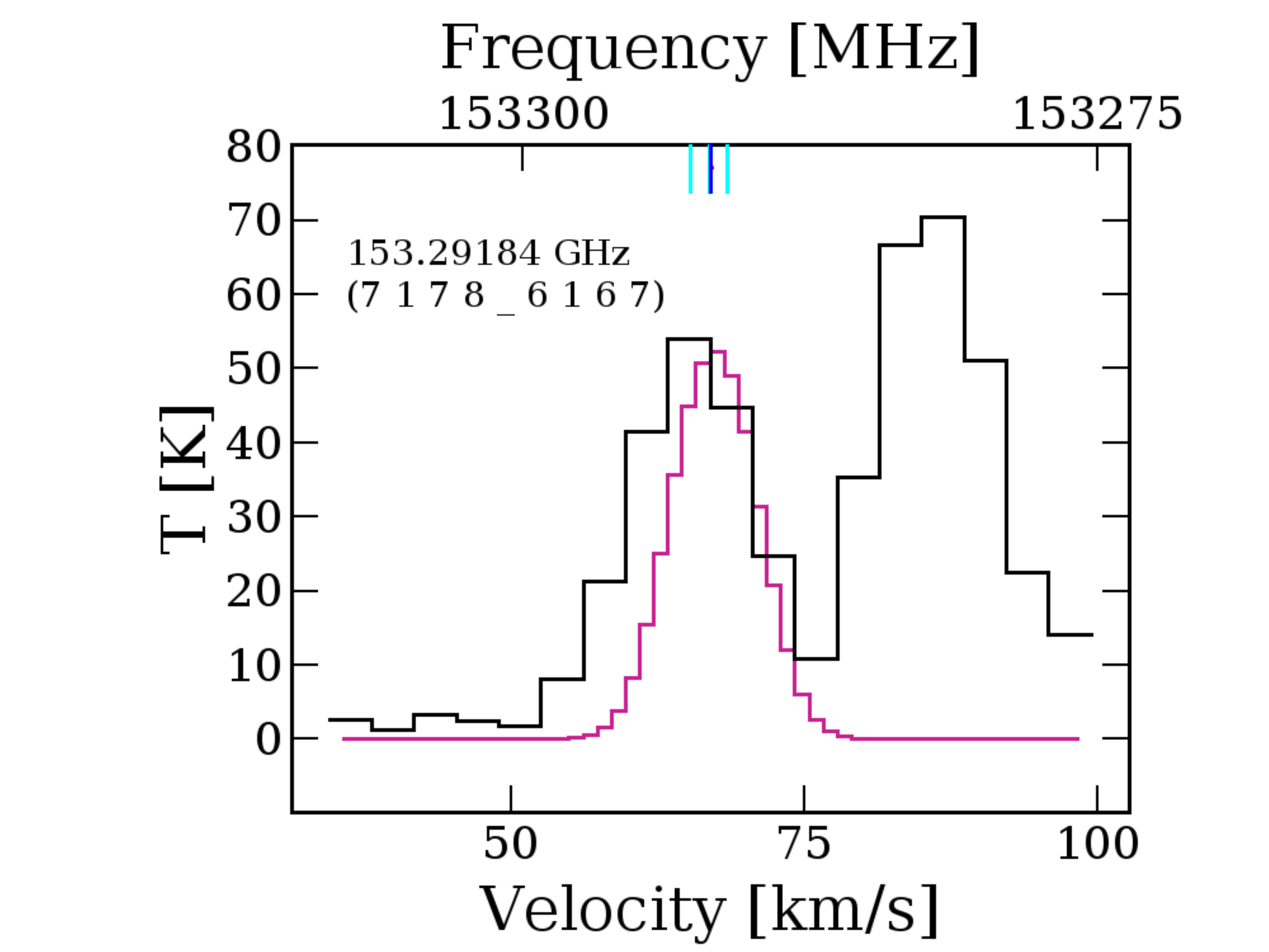}
\end{minipage}
\caption{LTE fitting of observed lines of HNCO towards G10.  Black line represents the observed spectra and pink line 
is the fitted profile.}
\label{fig:HNCO-rot}
\end{figure}

\begin{figure}
\begin{minipage}{0.23\textwidth}
\includegraphics[width=\textwidth]{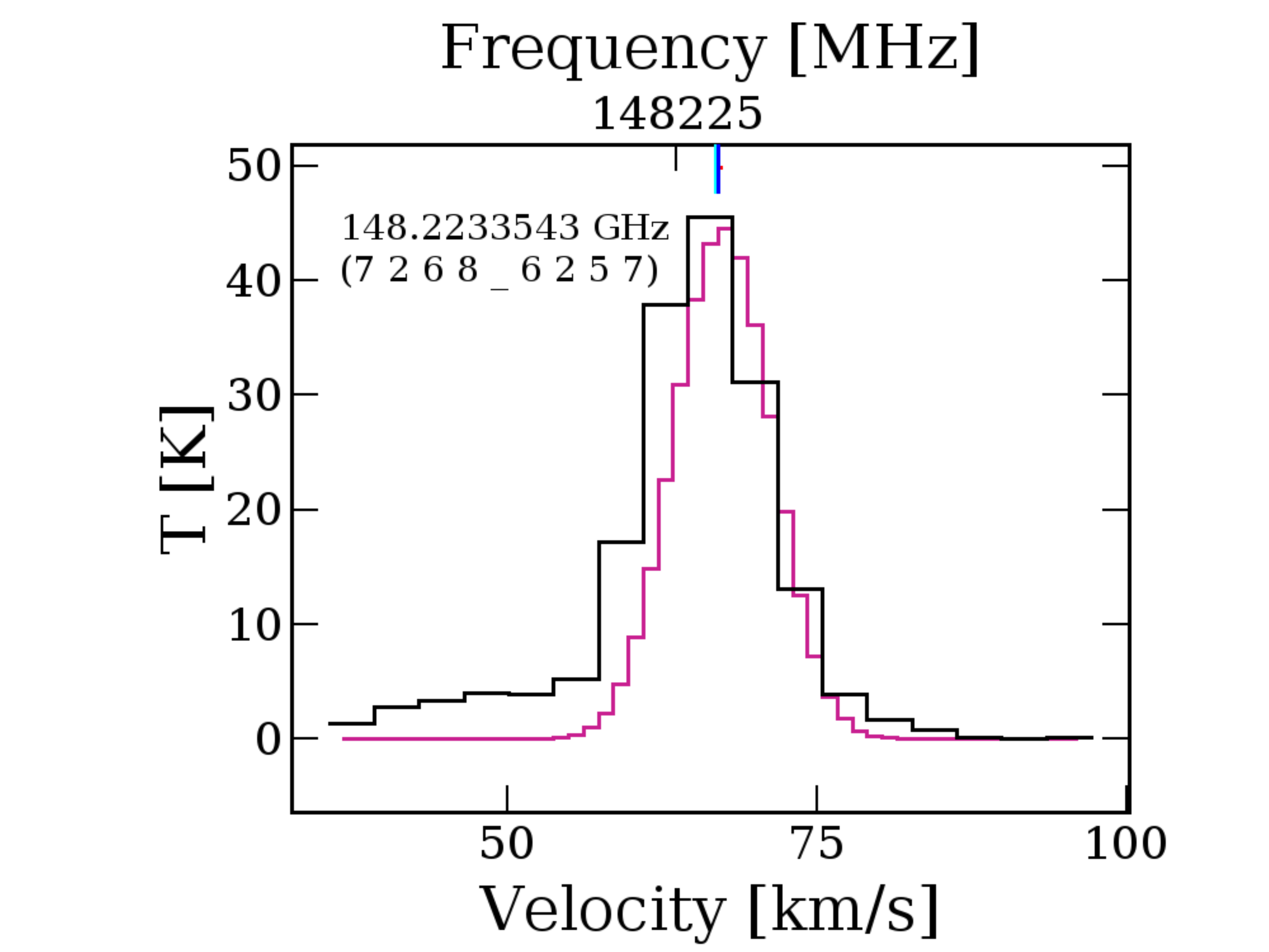}
\end{minipage}
\begin{minipage}{0.23\textwidth}
\includegraphics[width=\textwidth]{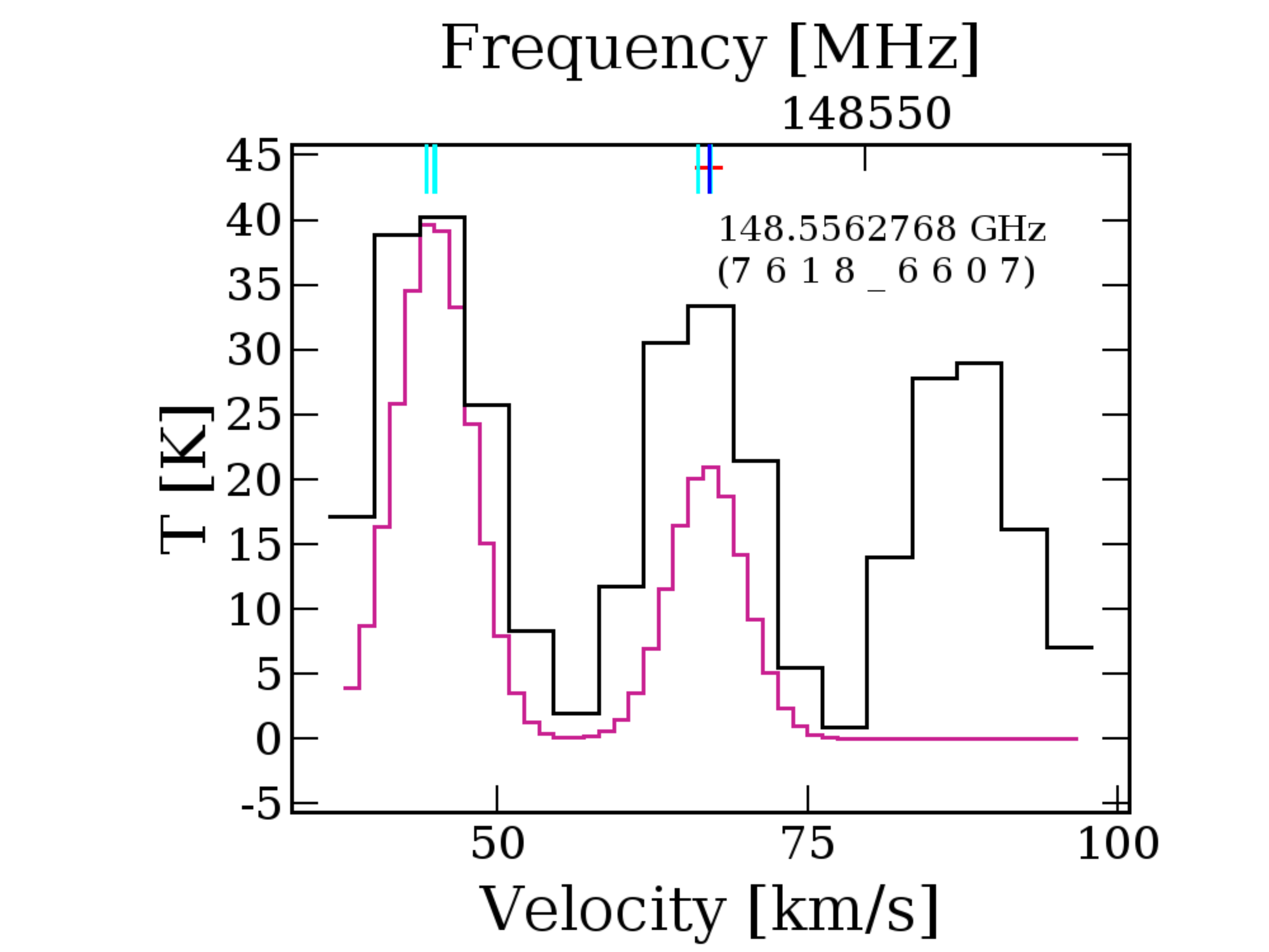}
\end{minipage}
 \begin{minipage}{0.23\textwidth}
 \includegraphics[width=\textwidth]{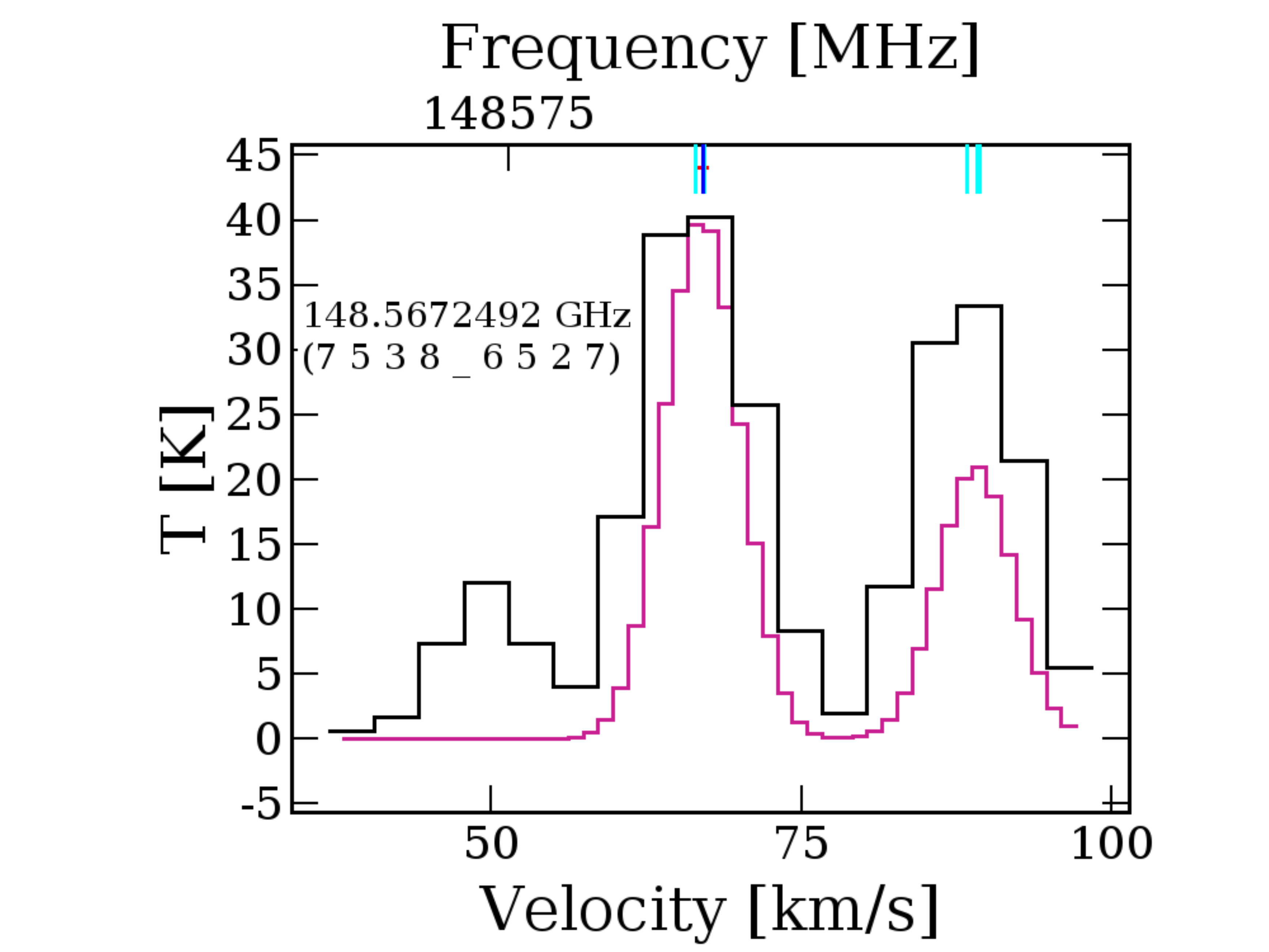}
 \end{minipage}
\begin{minipage}{0.23\textwidth}
\includegraphics[width=\textwidth]{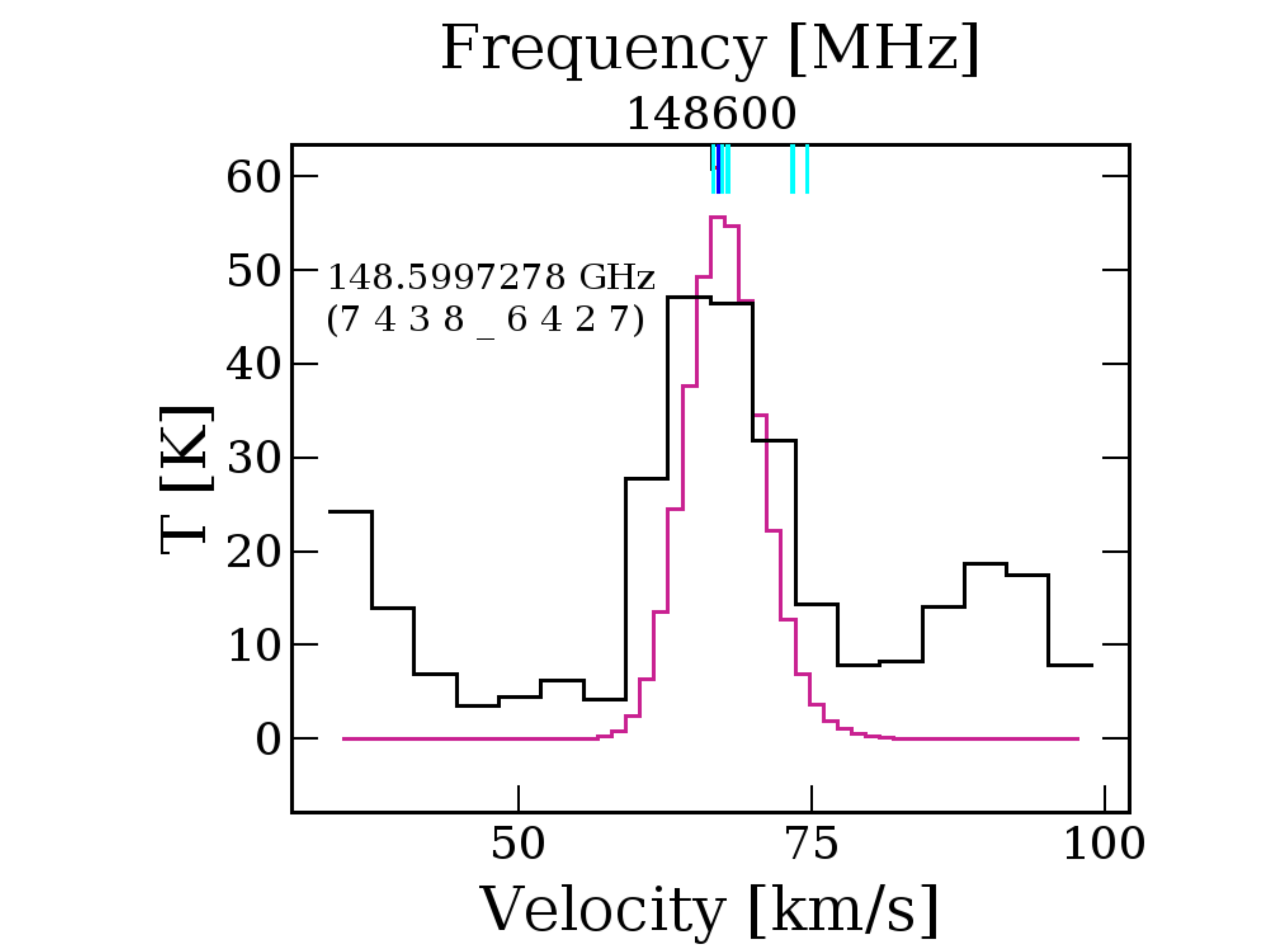}
\end{minipage}
\begin{minipage}{0.23\textwidth}
\includegraphics[width=\textwidth]{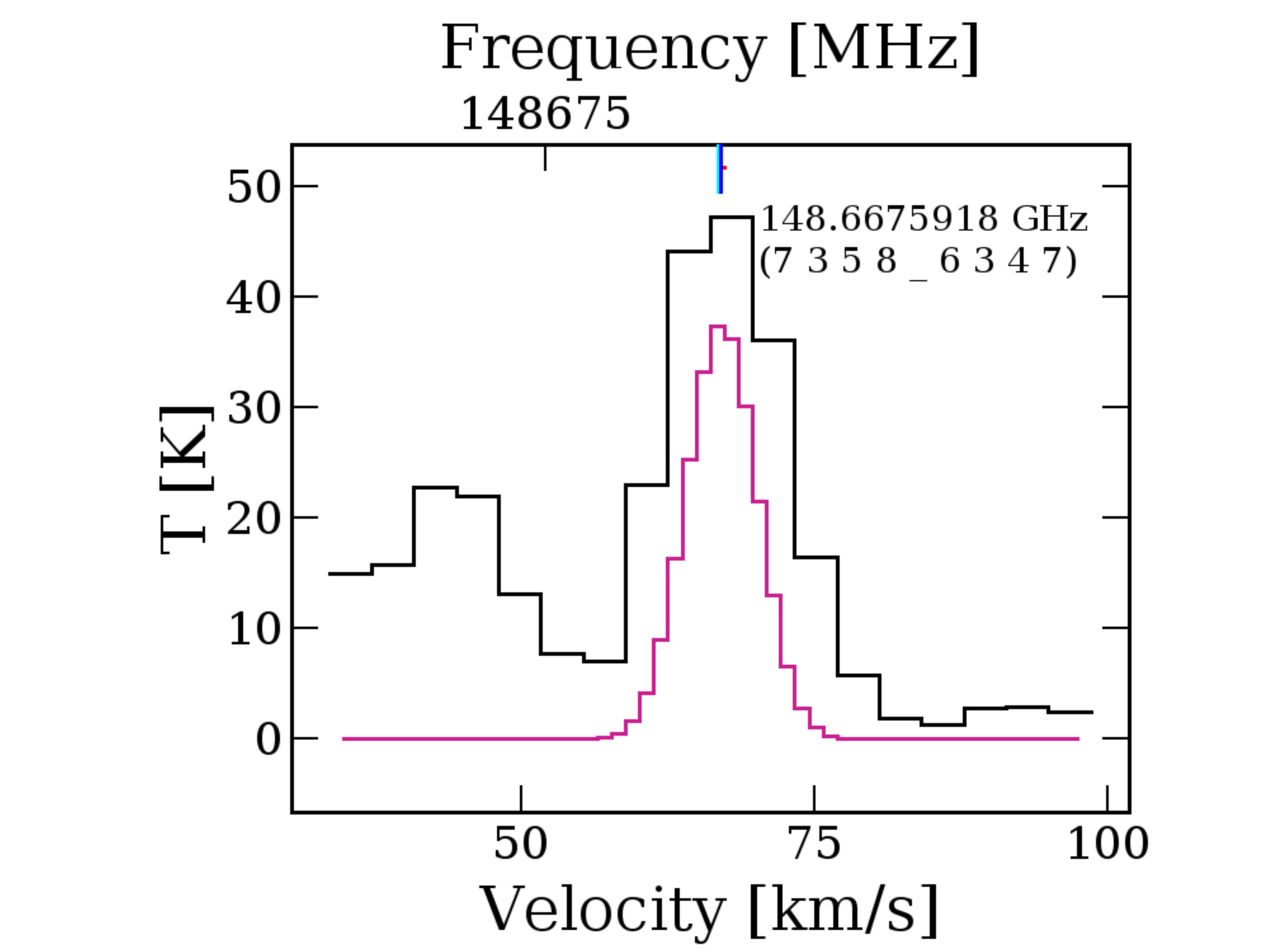}
\end{minipage}
\begin{minipage}{0.23\textwidth}
\includegraphics[width=\textwidth]{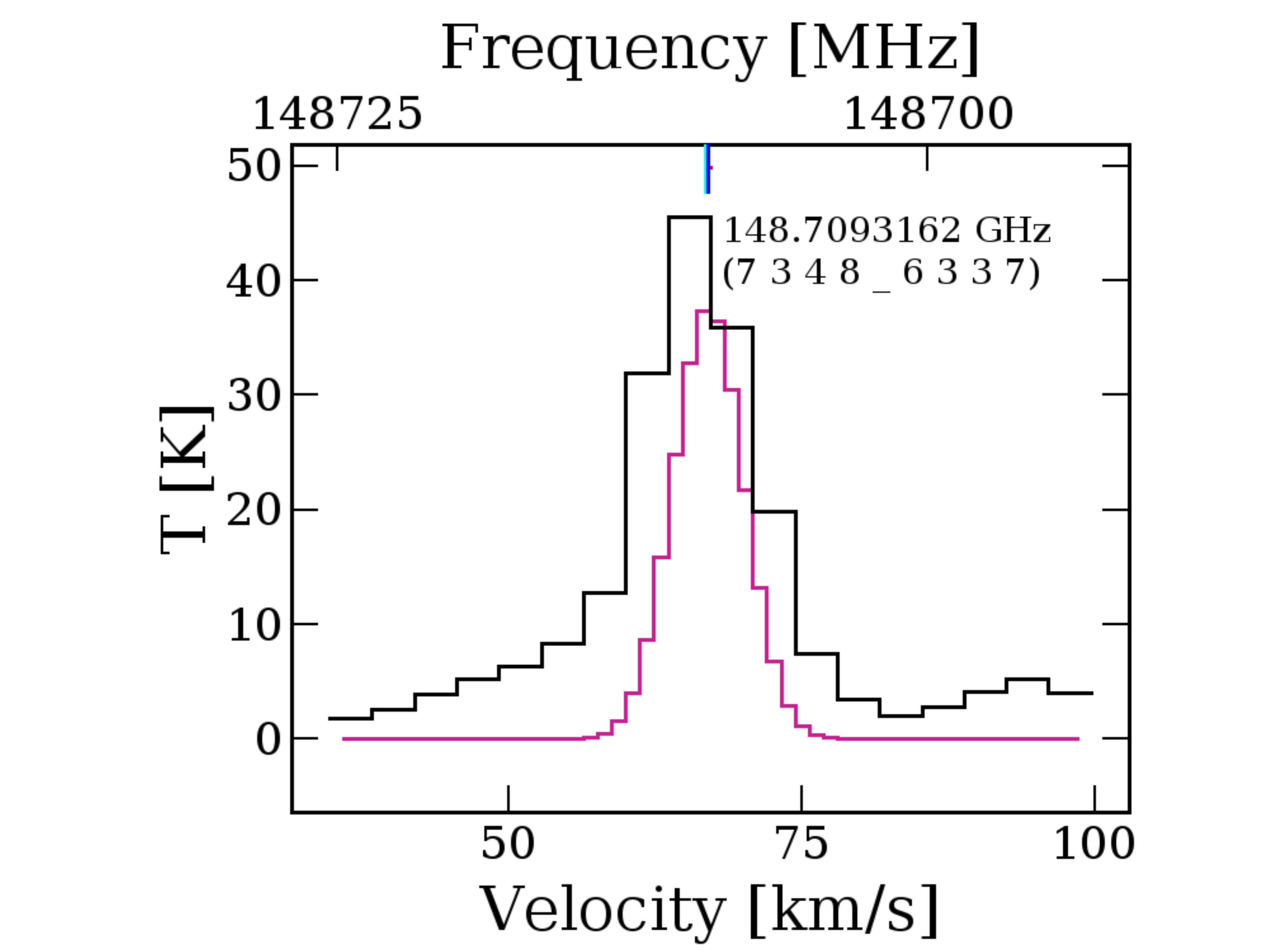}
\end{minipage}
\begin{minipage}{0.23\textwidth}
\includegraphics[width=\textwidth]{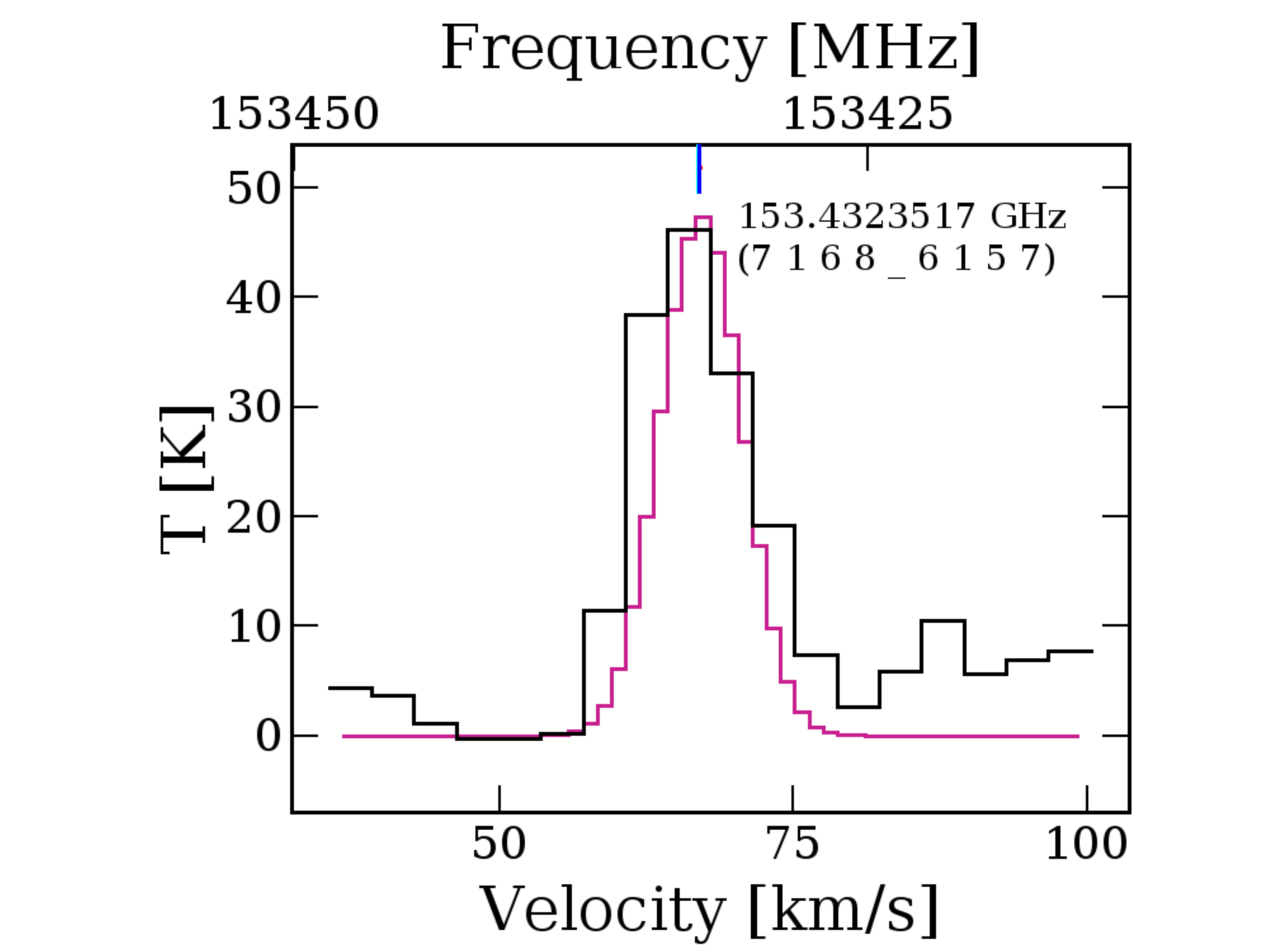}
\end{minipage}
\caption{LTE fitting of observed lines of $\rm{NH_2CHO}$ towards G10. Black line represents the observed spectra and pink line 
is the fitted profile.}
\label{fig:NH2CHO-rot}
\end{figure}

\begin{figure*}[t]
\begin{minipage}{0.23\textwidth}
\includegraphics[width=\textwidth]{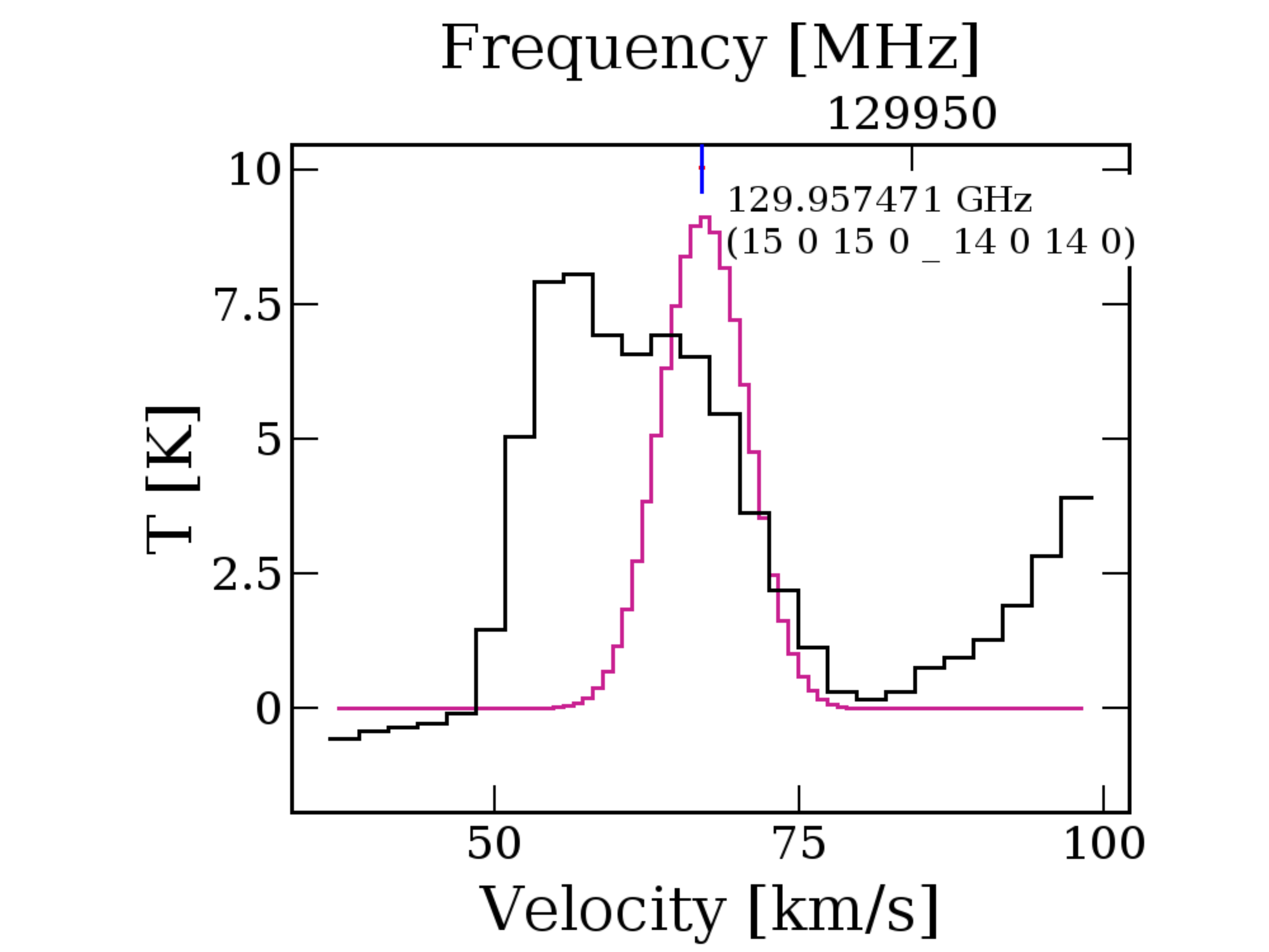}
\end{minipage}
\begin{minipage}{0.23\textwidth}
\includegraphics[width=\textwidth]{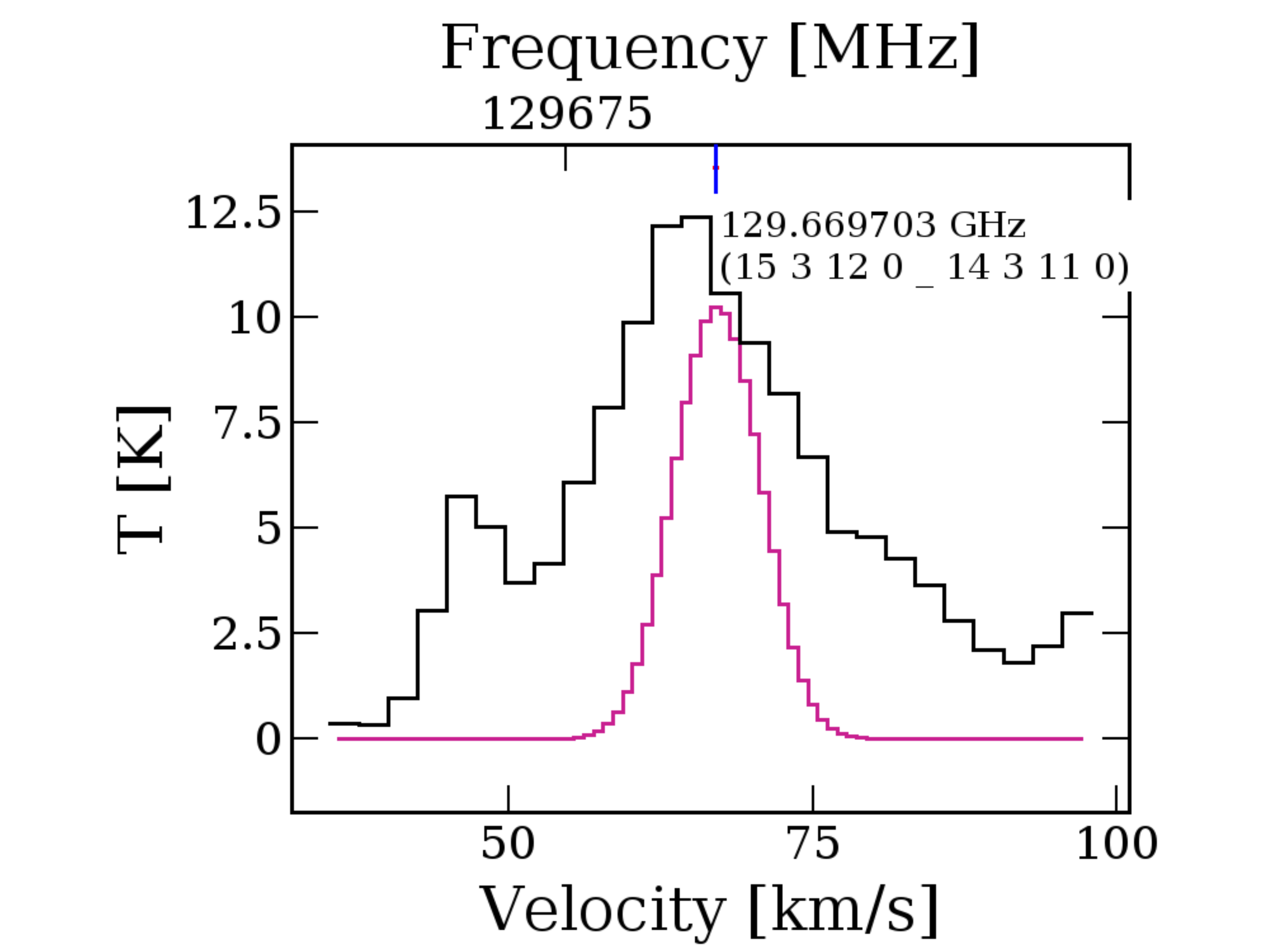}
\end{minipage}
 \begin{minipage}{0.23\textwidth}
 \includegraphics[width=\textwidth]{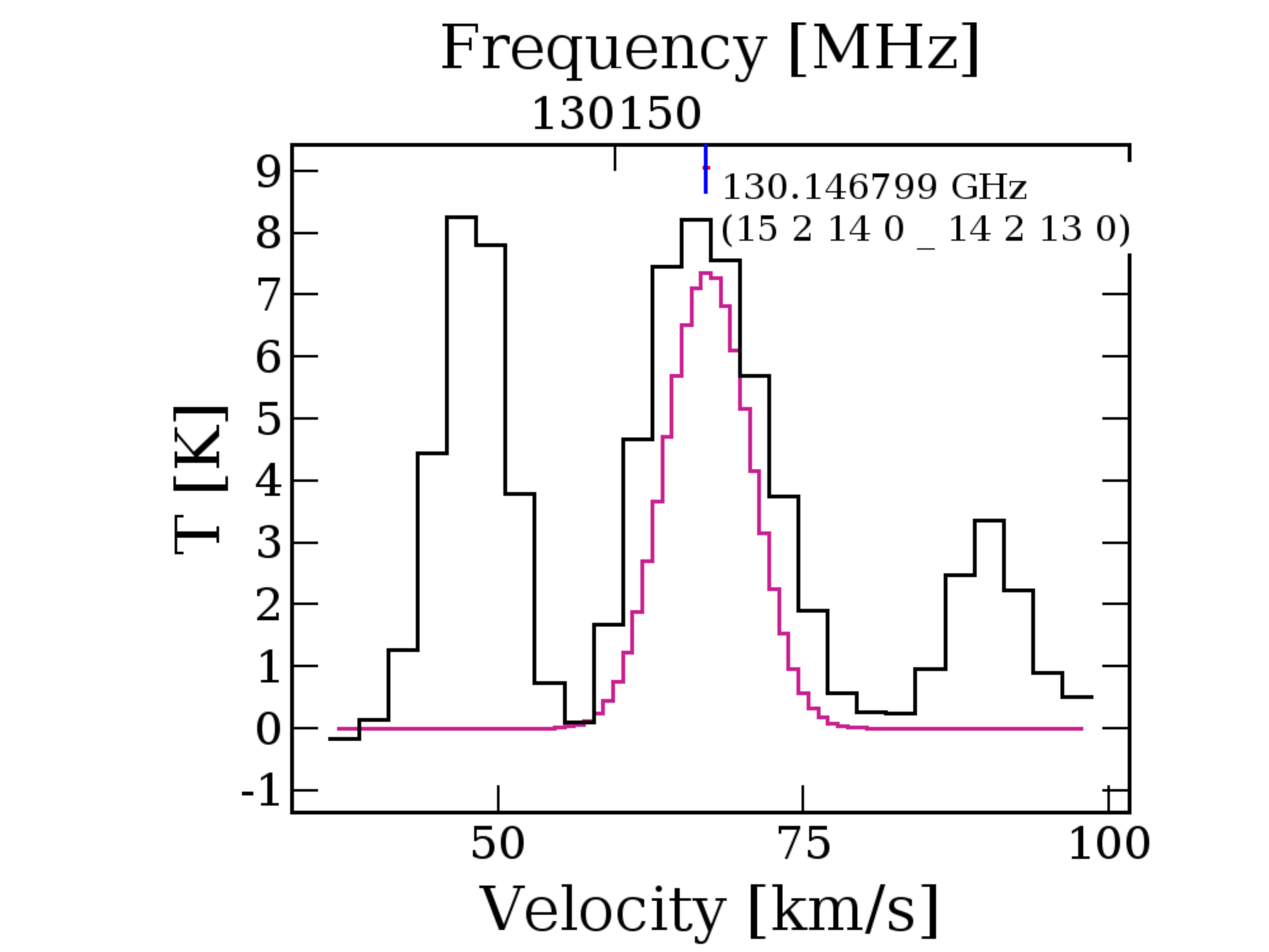}
 \end{minipage}
\begin{minipage}{0.23\textwidth}
\includegraphics[width=\textwidth]{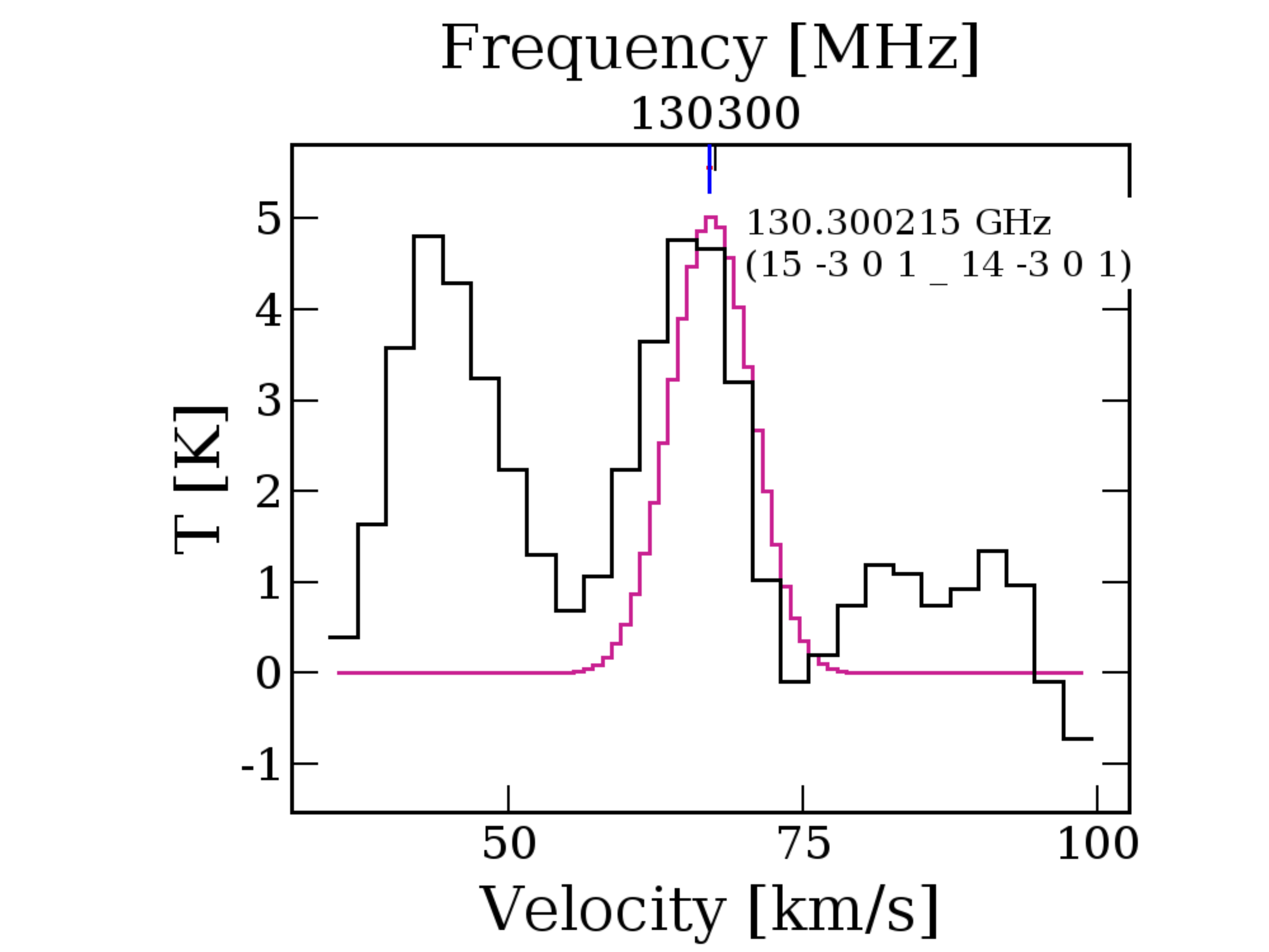}
\end{minipage}
\begin{minipage}{0.23\textwidth}
\includegraphics[width=\textwidth]{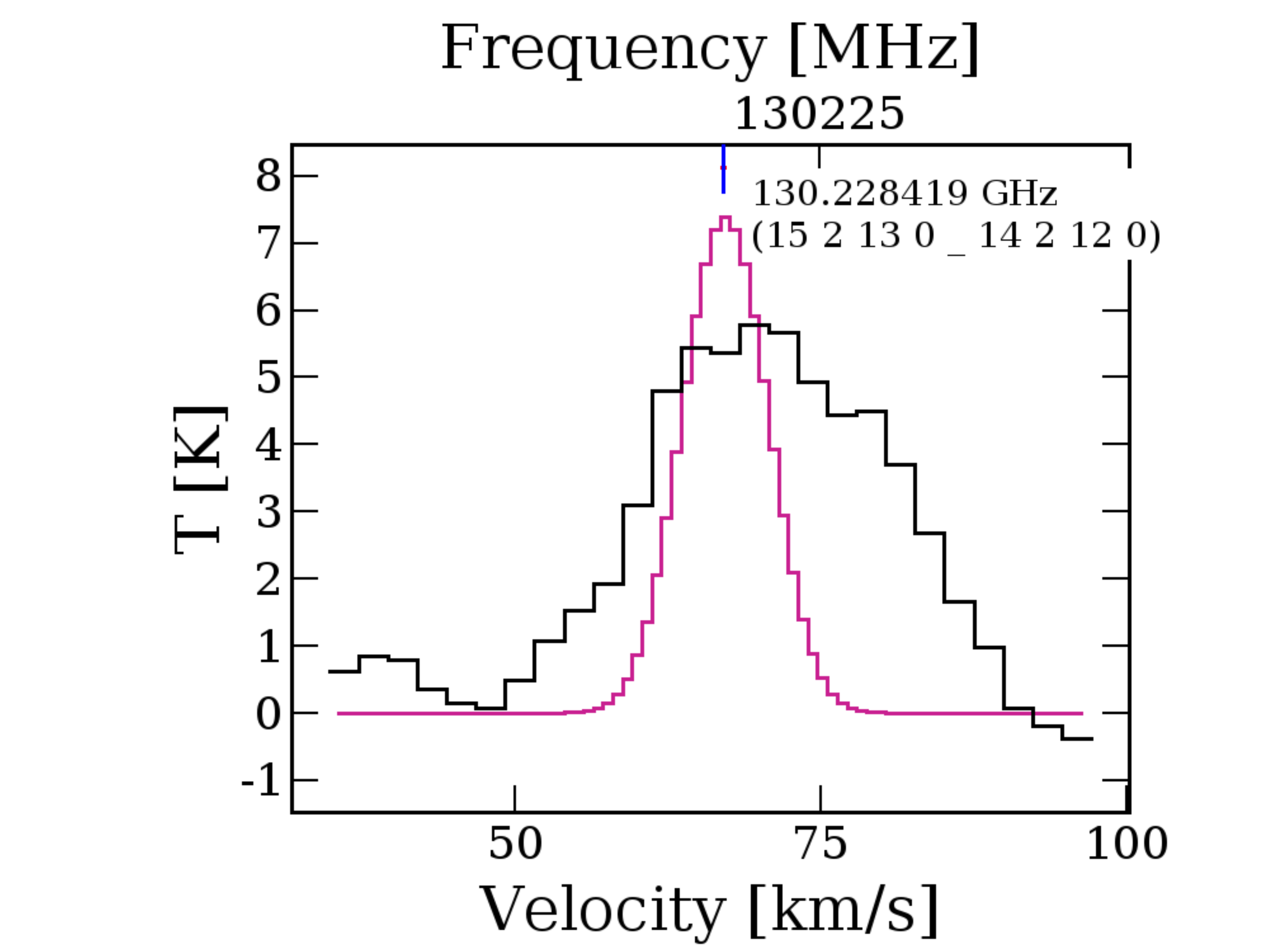}
\end{minipage}
\begin{minipage}{0.23\textwidth}
\includegraphics[width=\textwidth]{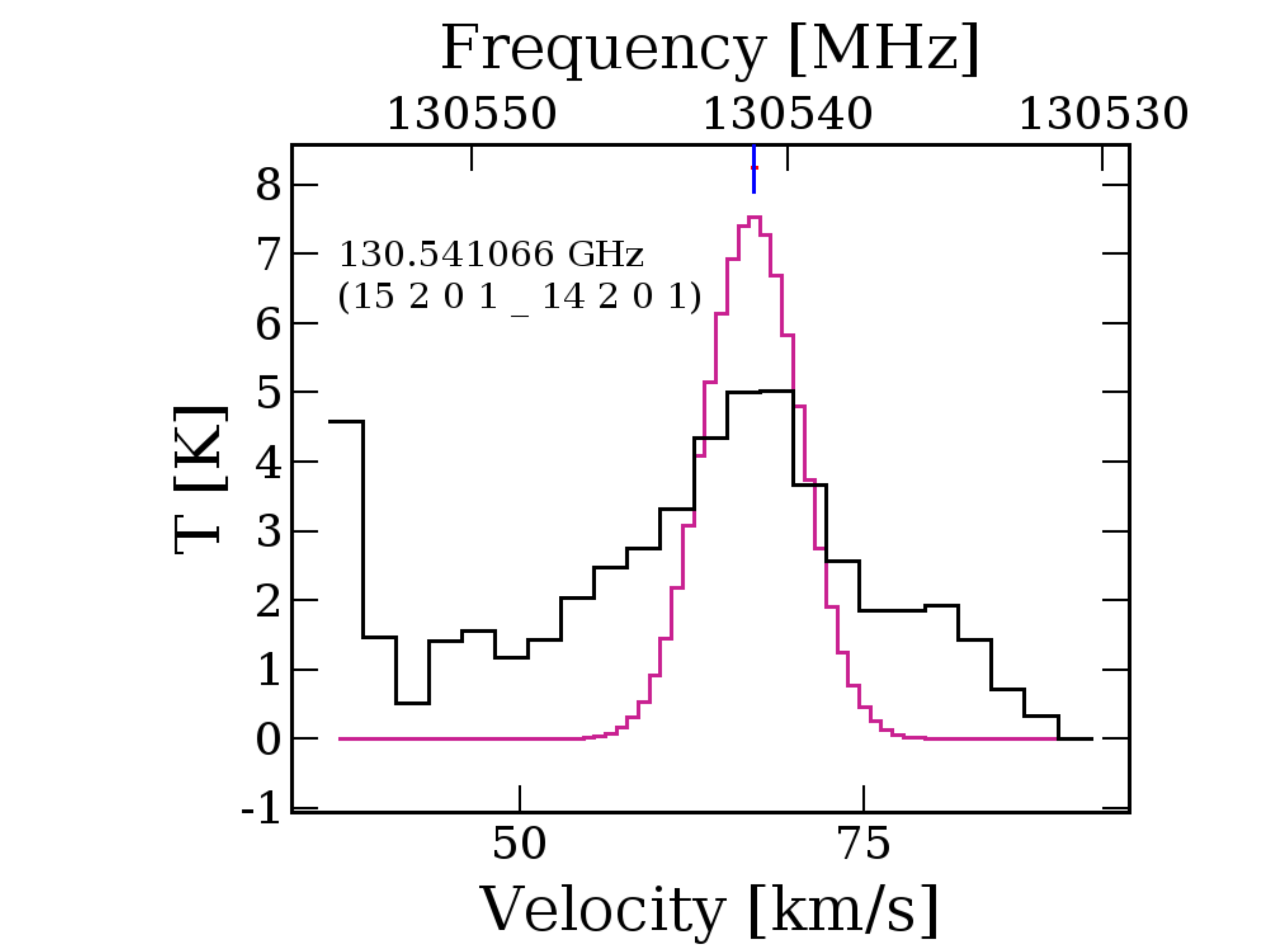}
\end{minipage}
\begin{minipage}{0.23\textwidth}
\includegraphics[width=\textwidth]{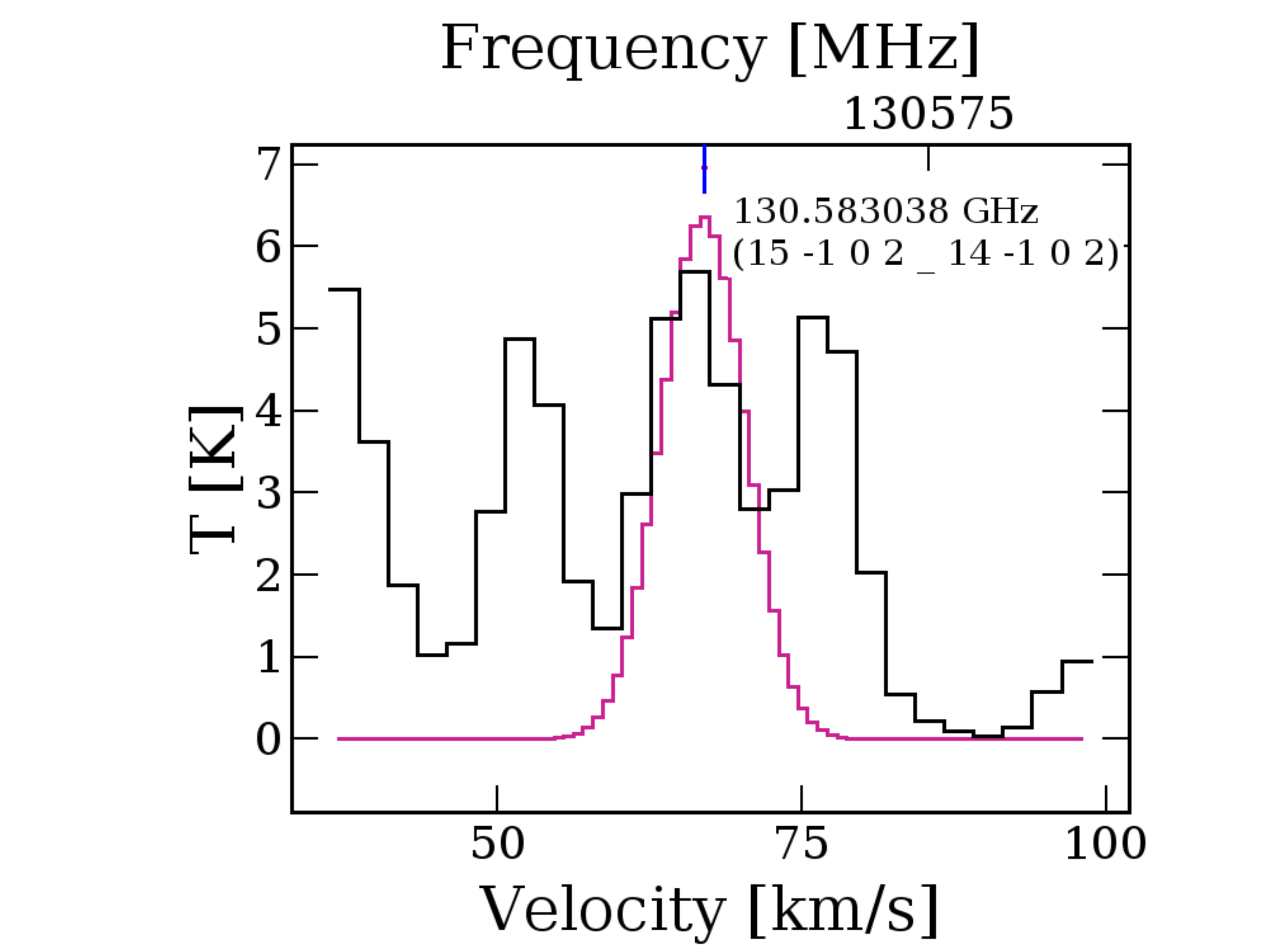}
\end{minipage}
\begin{minipage}{0.23\textwidth}
\includegraphics[width=\textwidth]{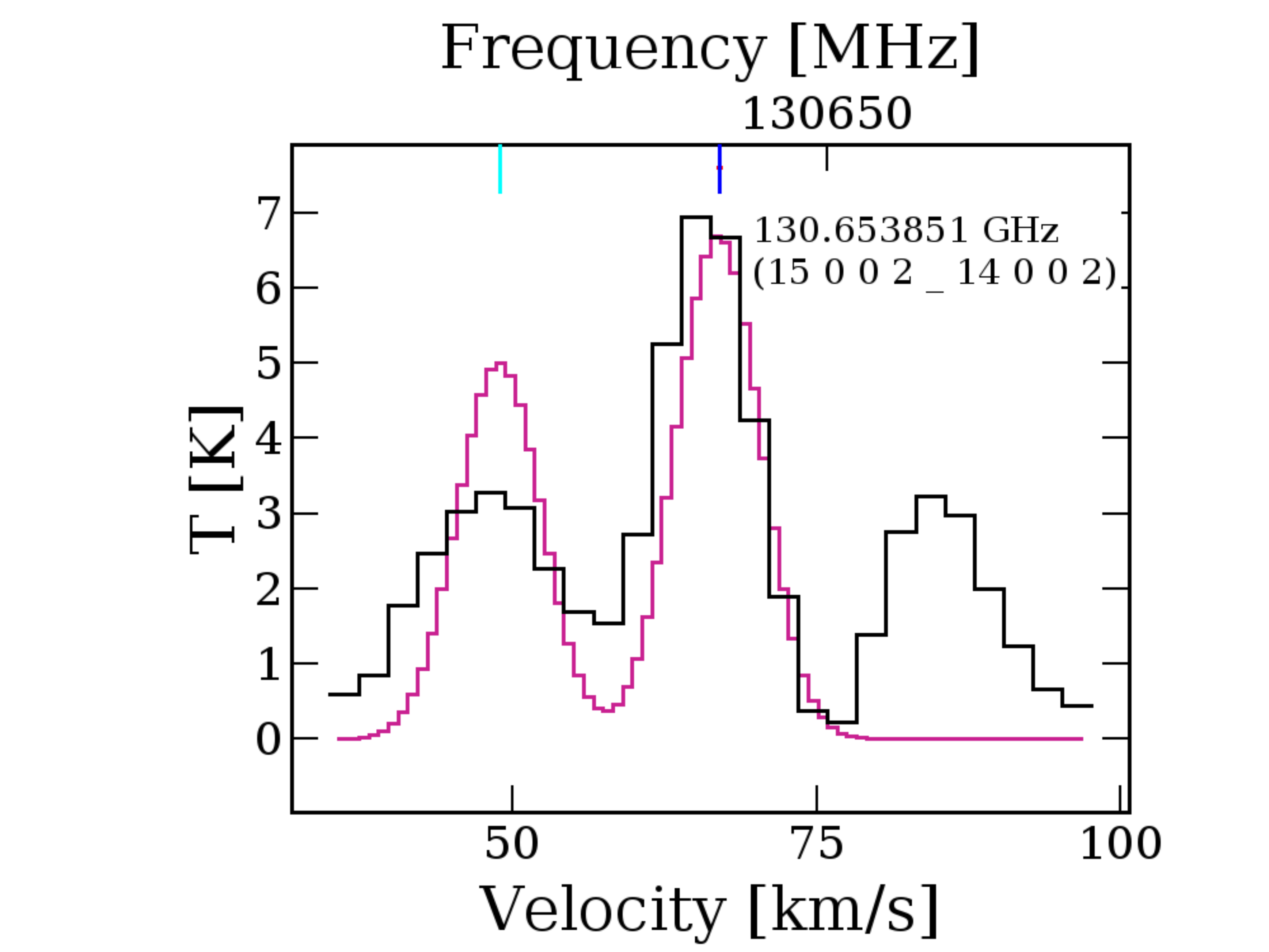}
\end{minipage}
\begin{minipage}{0.23\textwidth}
\includegraphics[width=\textwidth]{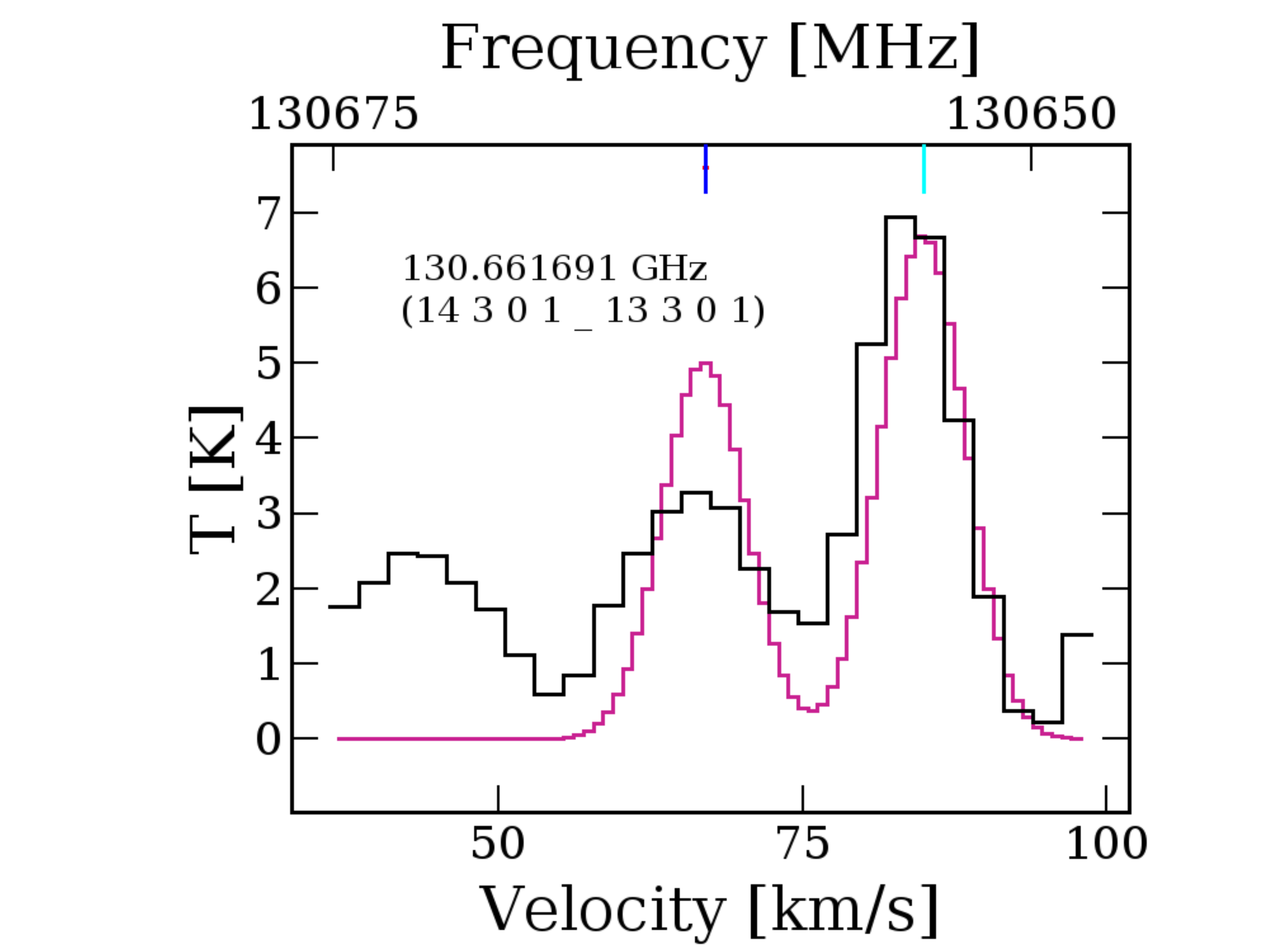}
\end{minipage}
\begin{minipage}{0.23\textwidth}
\includegraphics[width=\textwidth]{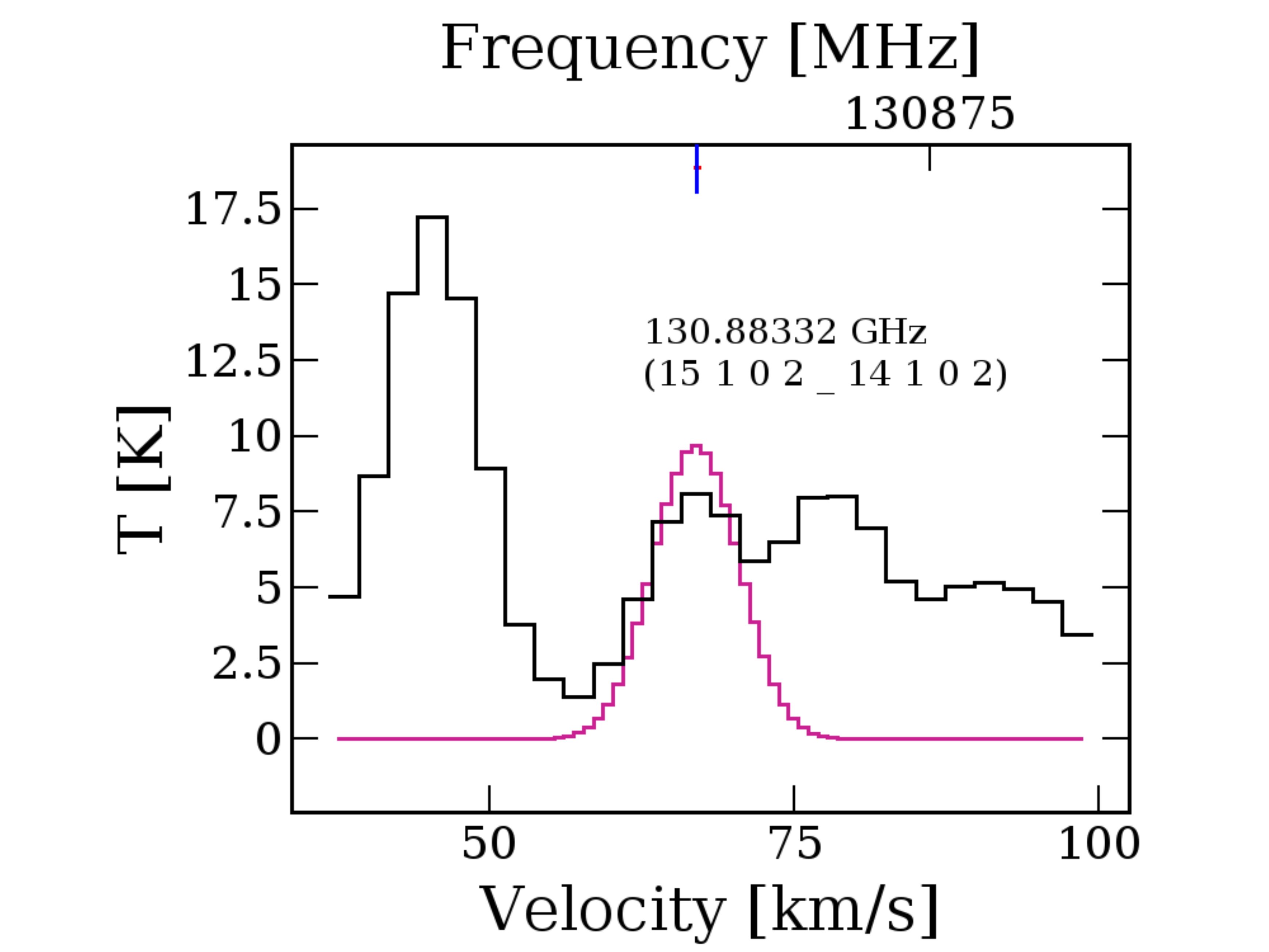}
\end{minipage}
\begin{minipage}{0.23\textwidth}
\includegraphics[width=\textwidth]{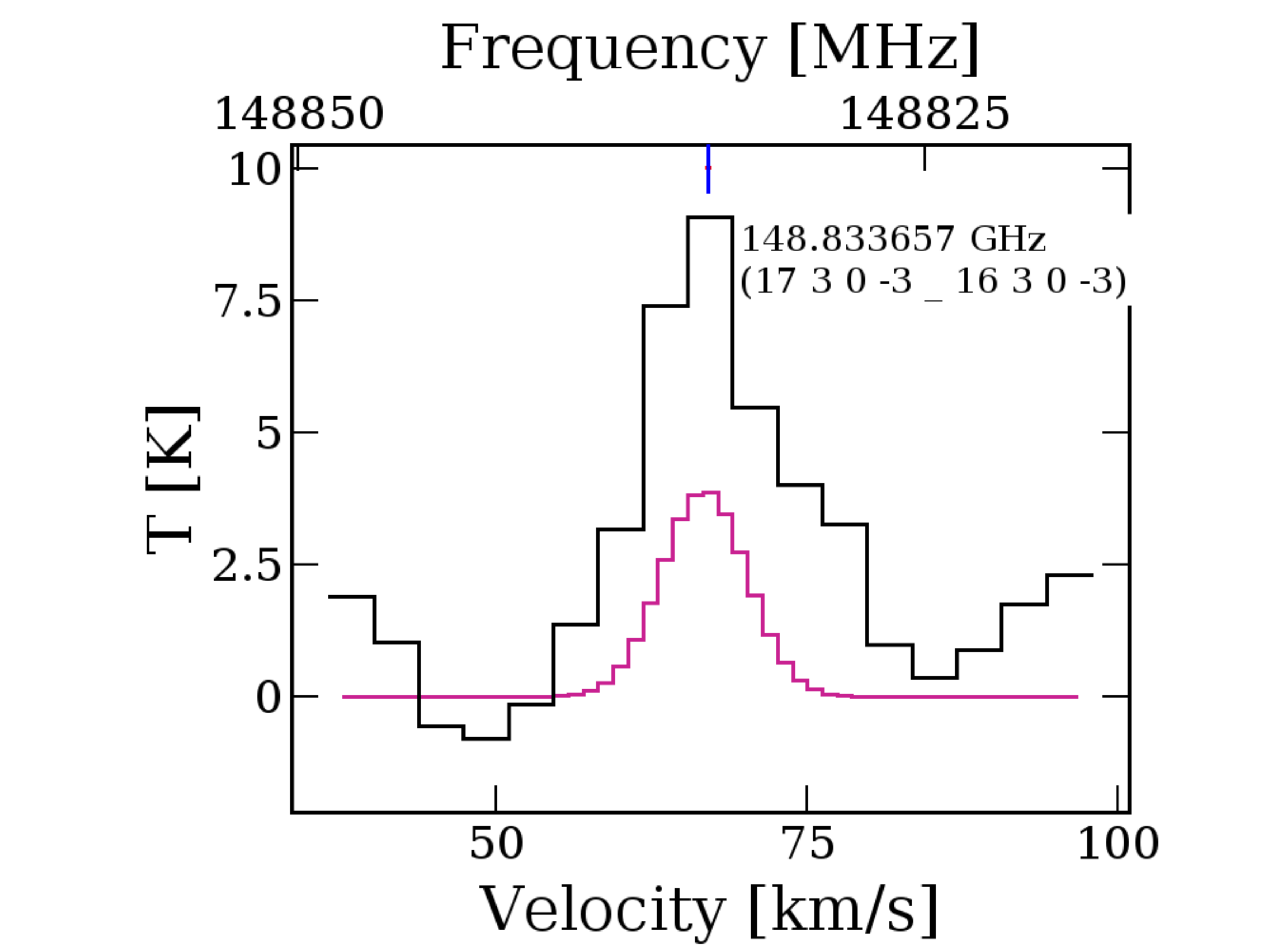}
\end{minipage}
\begin{minipage}{0.23\textwidth}
\includegraphics[width=\textwidth]{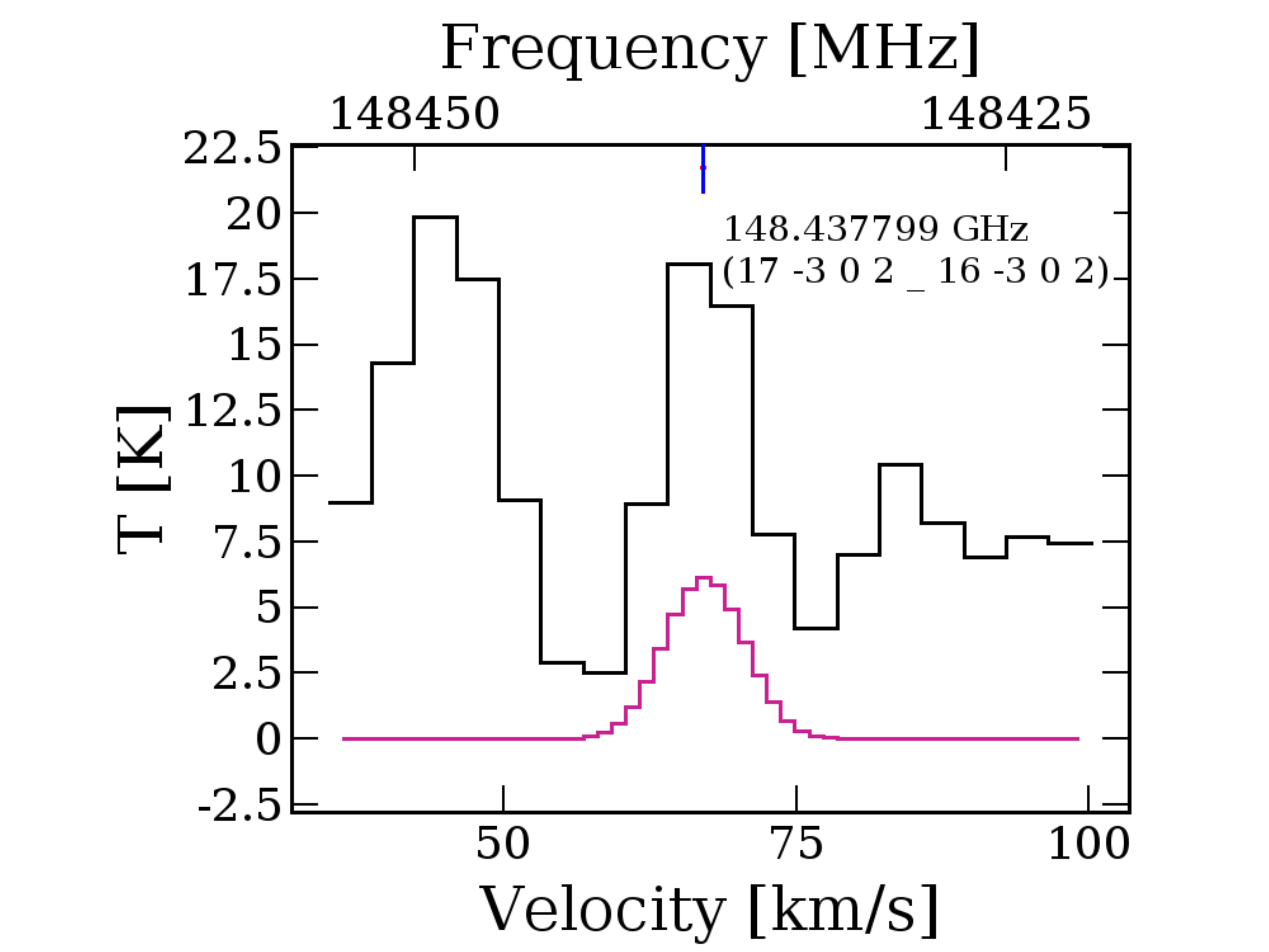}
\end{minipage}
\begin{minipage}{0.23\textwidth}
\includegraphics[width=\textwidth]{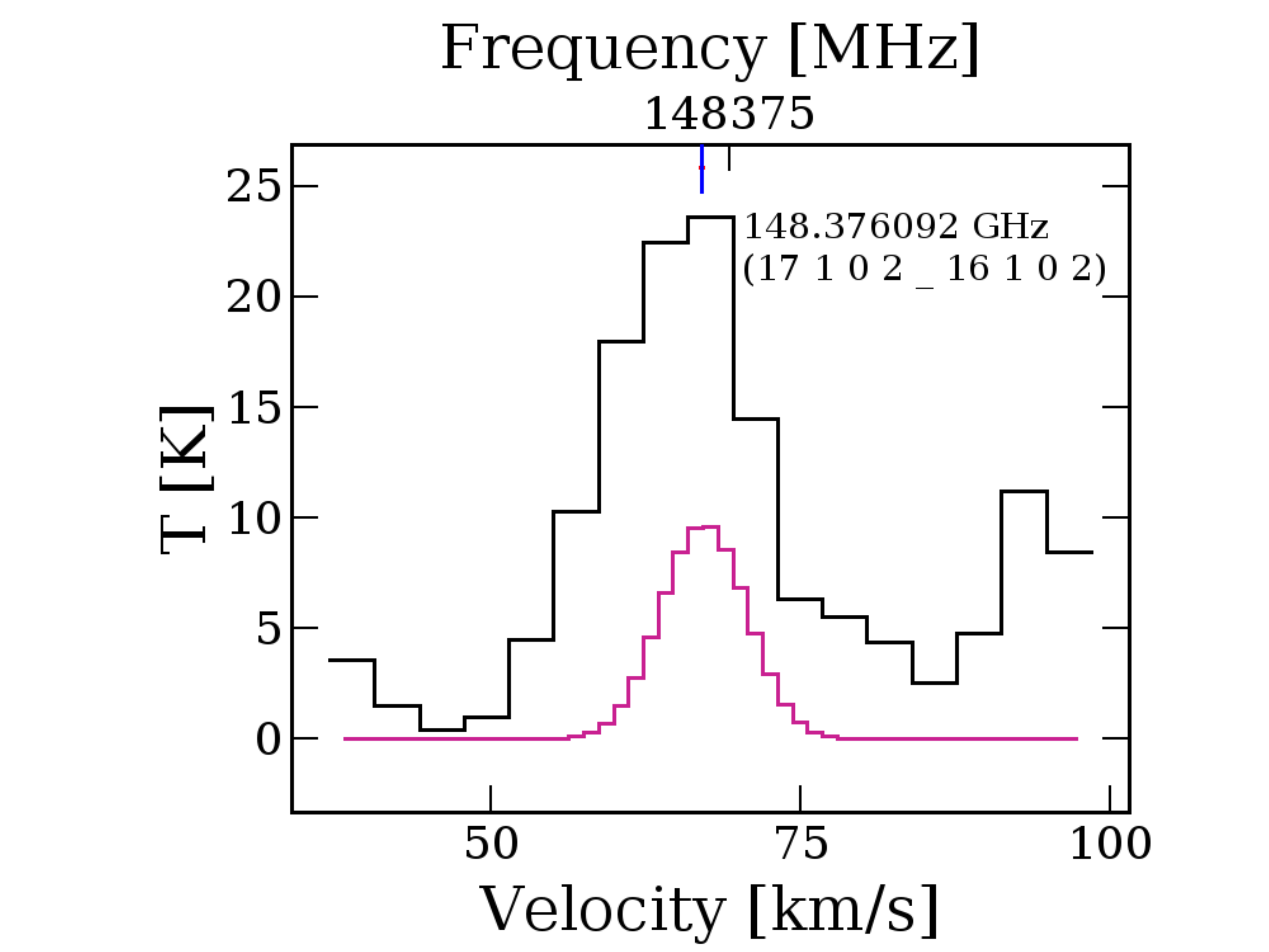}
\end{minipage}
\begin{minipage}{0.23\textwidth}
\includegraphics[width=\textwidth]{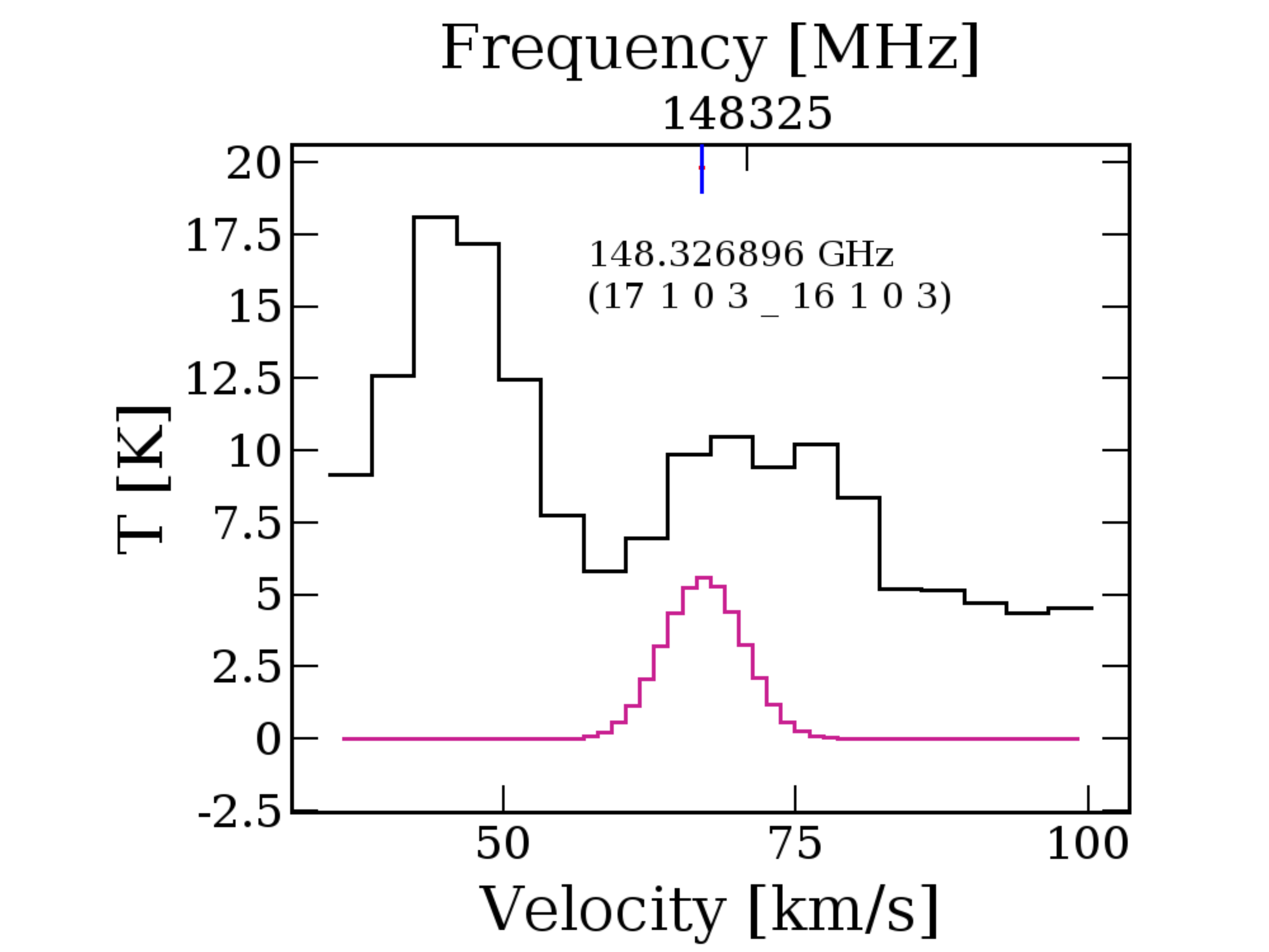}
\end{minipage}
\begin{minipage}{0.23\textwidth}
\includegraphics[width=\textwidth]{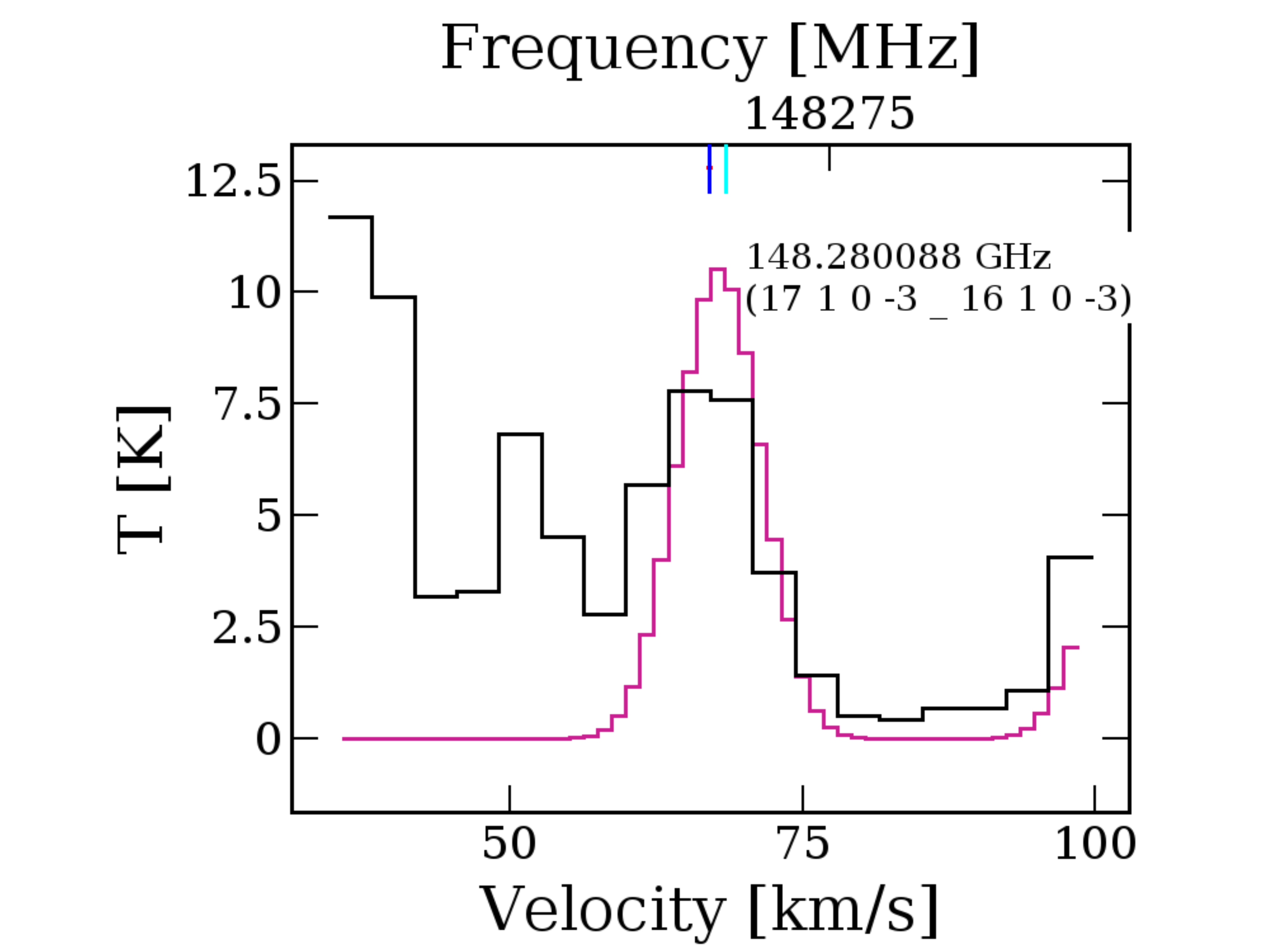}
\end{minipage}
\begin{minipage}{0.23\textwidth}
\includegraphics[width=\textwidth]{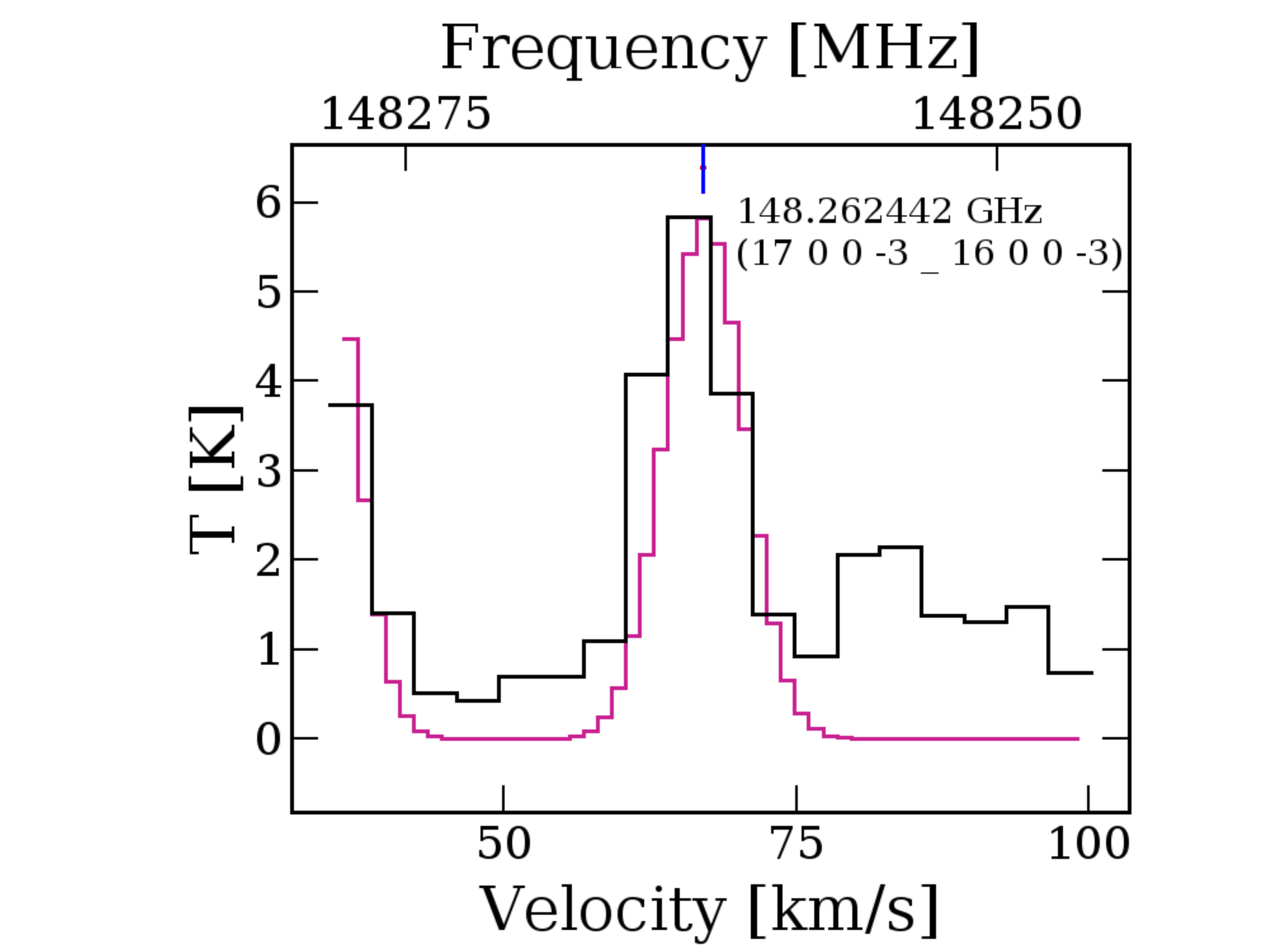}
\end{minipage}
\begin{minipage}{0.23\textwidth}
\includegraphics[width=\textwidth]{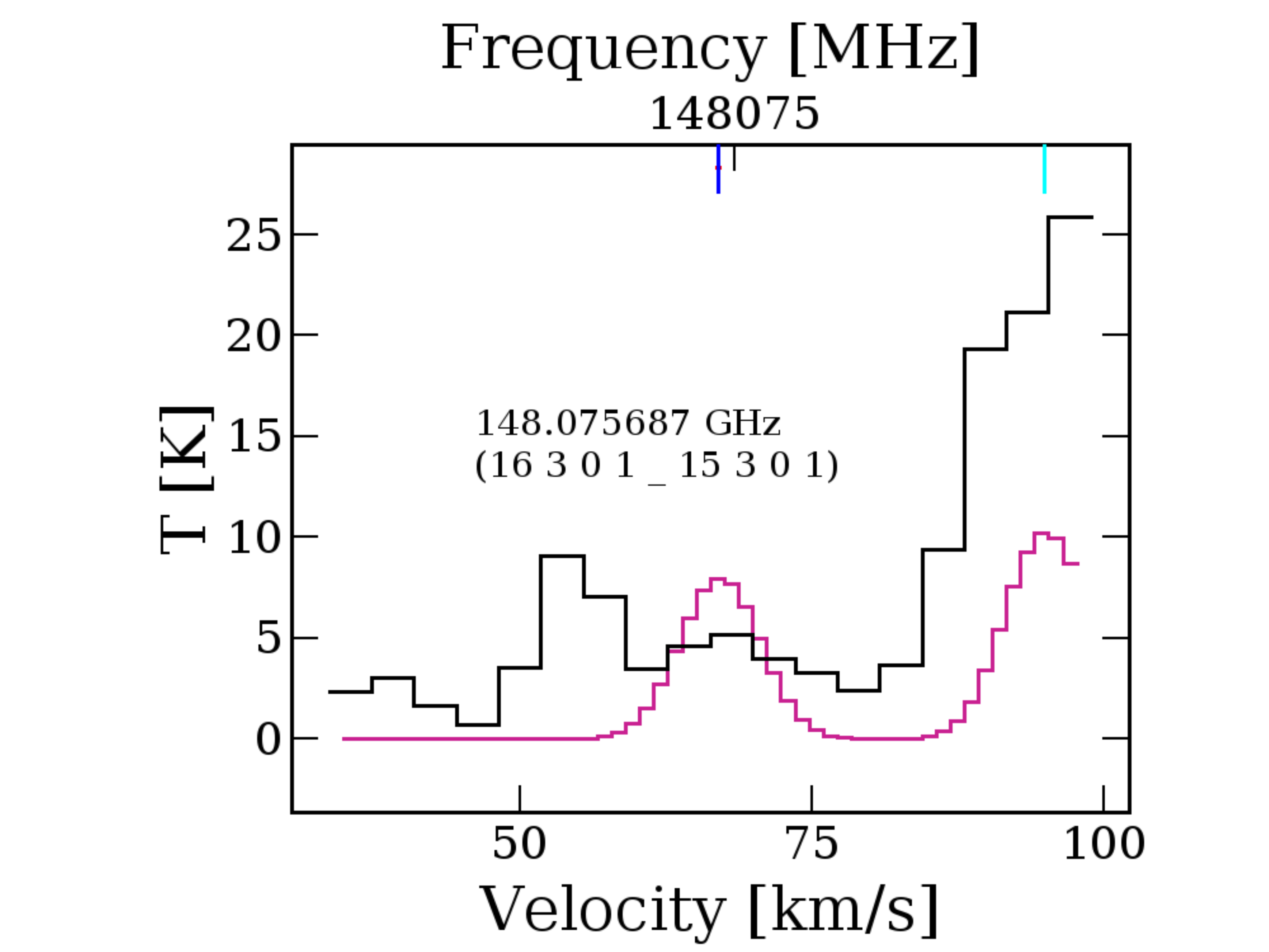}
\end{minipage}
\begin{minipage}{0.23\textwidth}
\includegraphics[width=\textwidth]{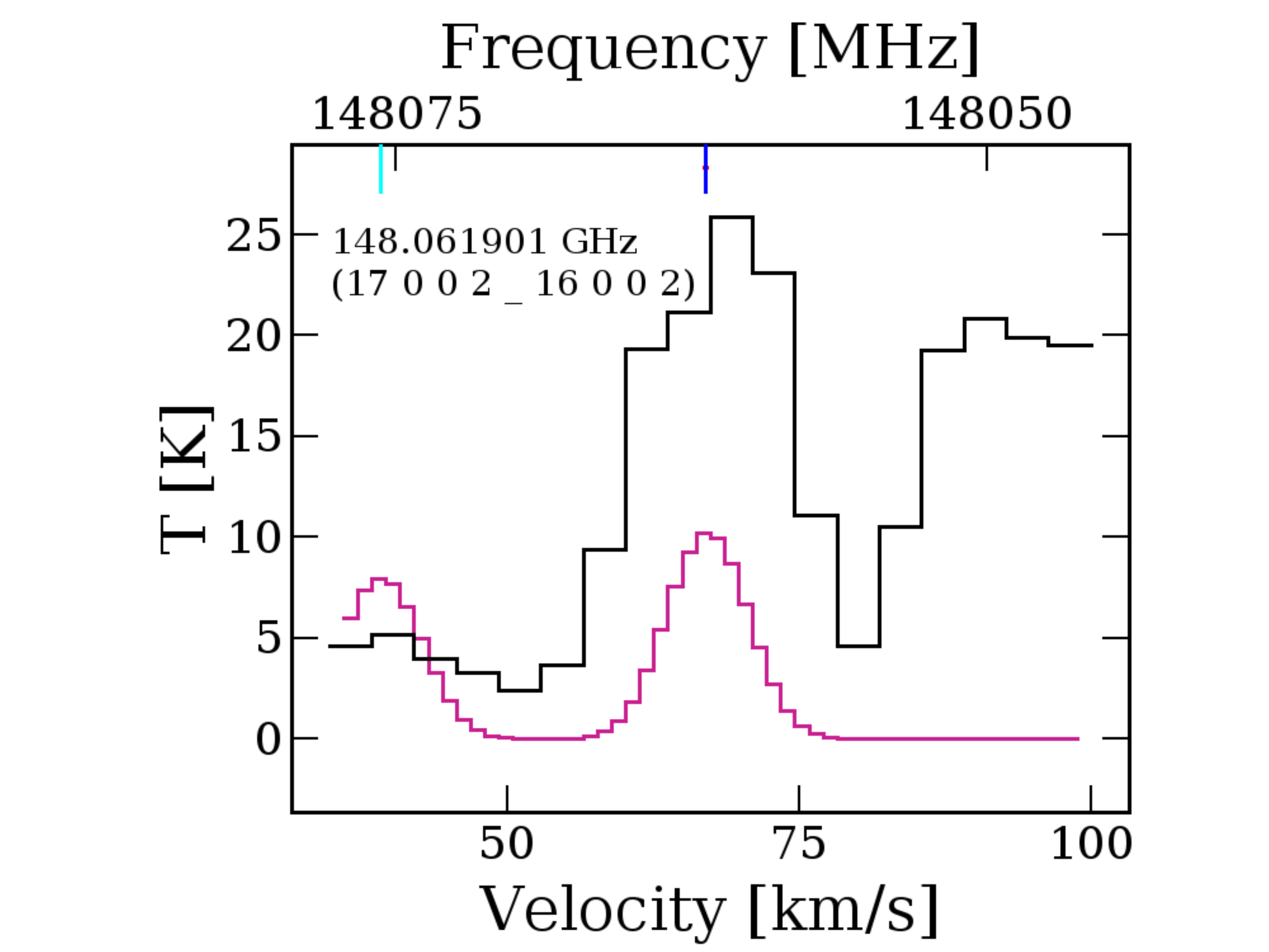}
\end{minipage}
\begin{minipage}{0.23\textwidth}
\includegraphics[width=\textwidth]{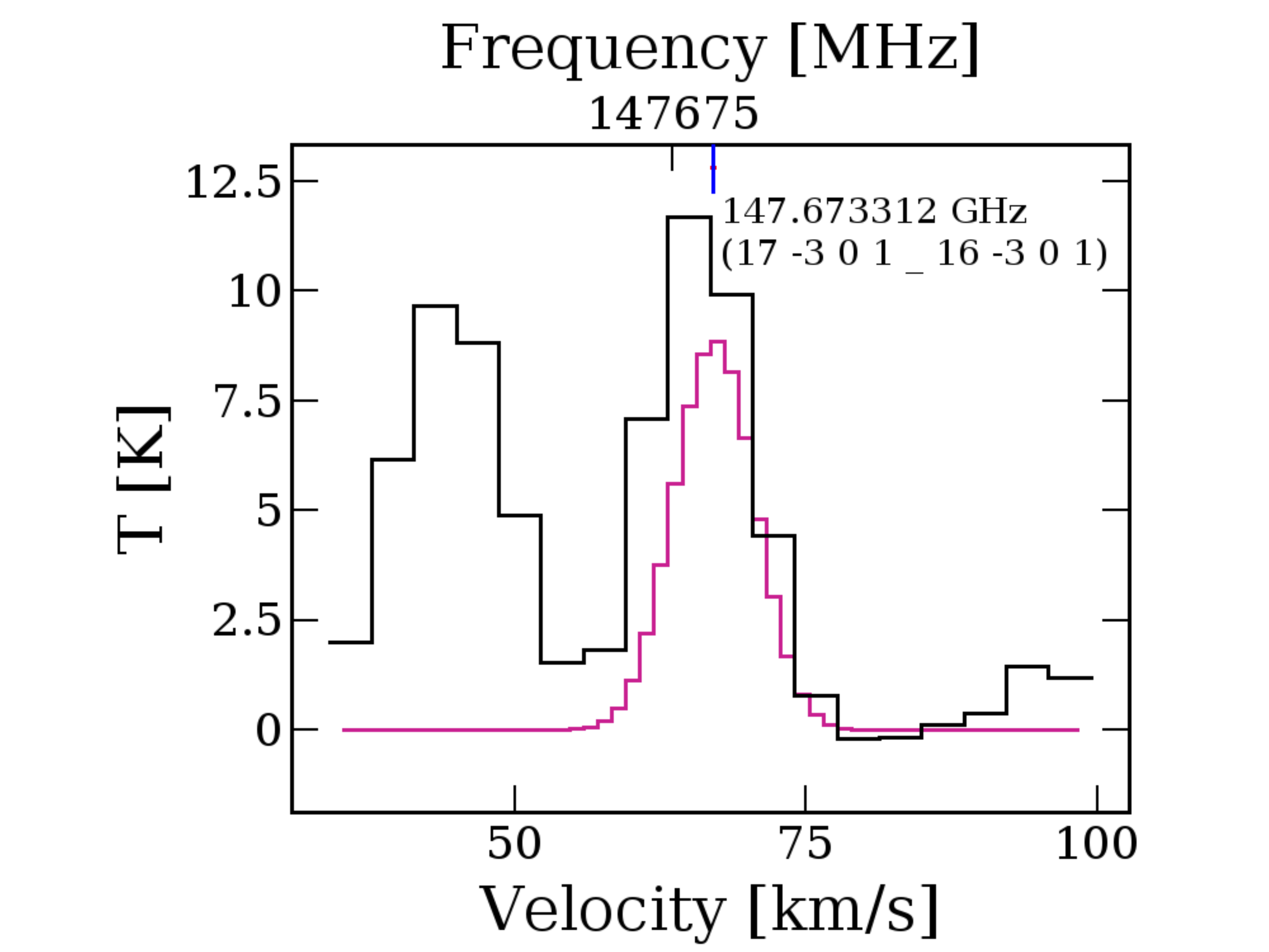}
\end{minipage}
\begin{minipage}{0.23\textwidth}
\includegraphics[width=\textwidth]{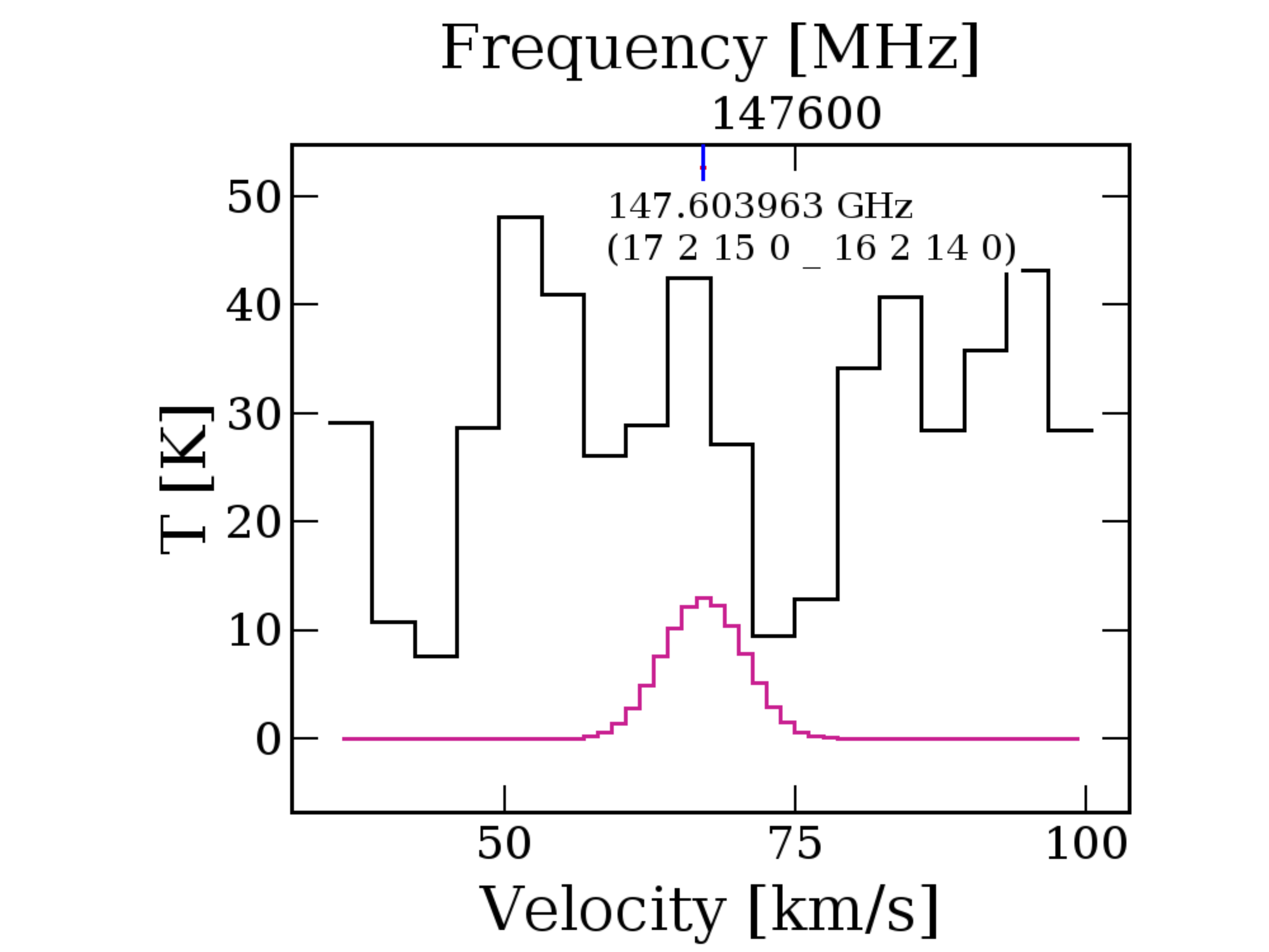}
\end{minipage}
\begin{minipage}{0.23\textwidth}
\includegraphics[width=\textwidth]{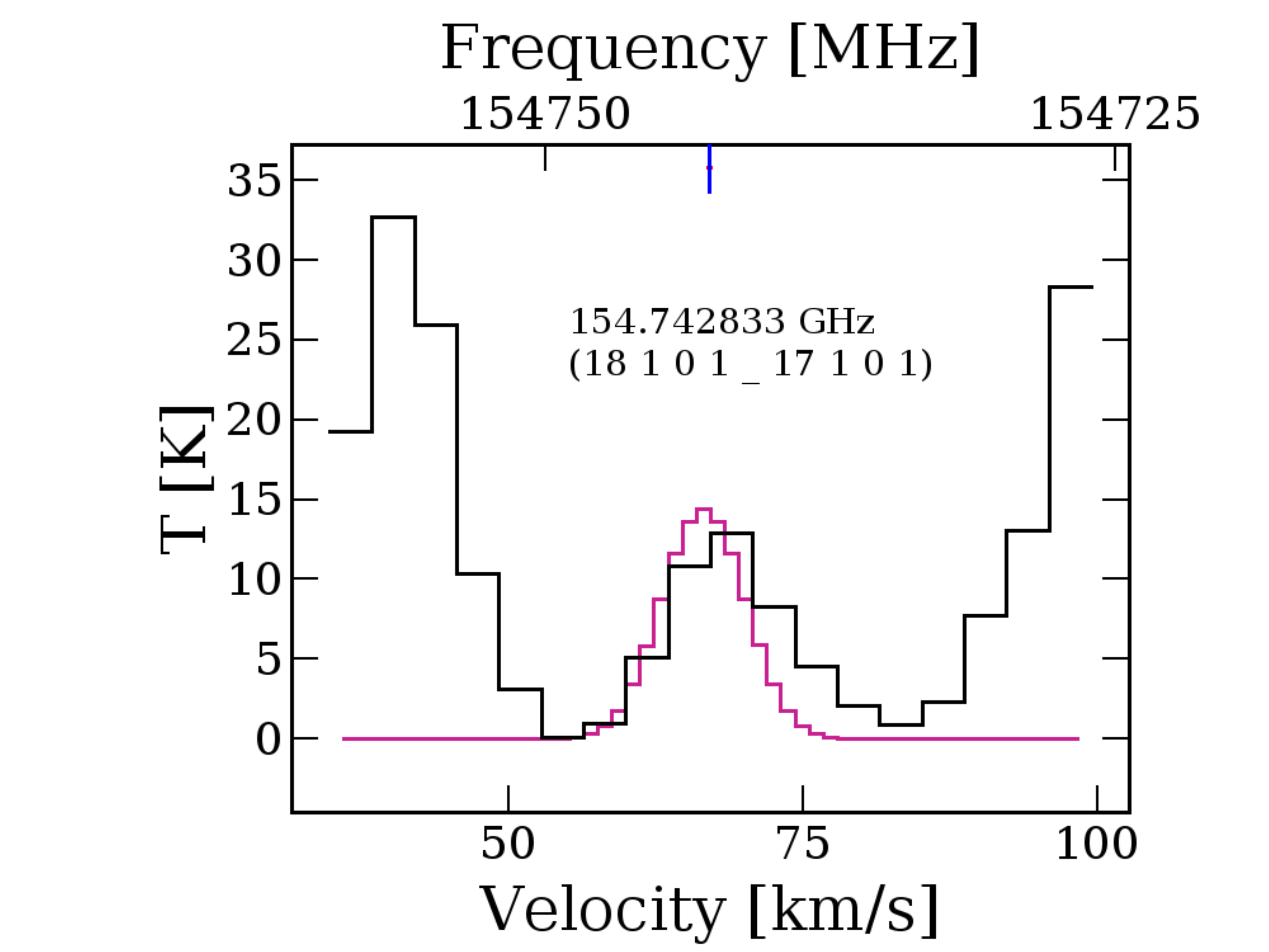}
\end{minipage}
\begin{minipage}{0.23\textwidth}
\includegraphics[width=\textwidth]{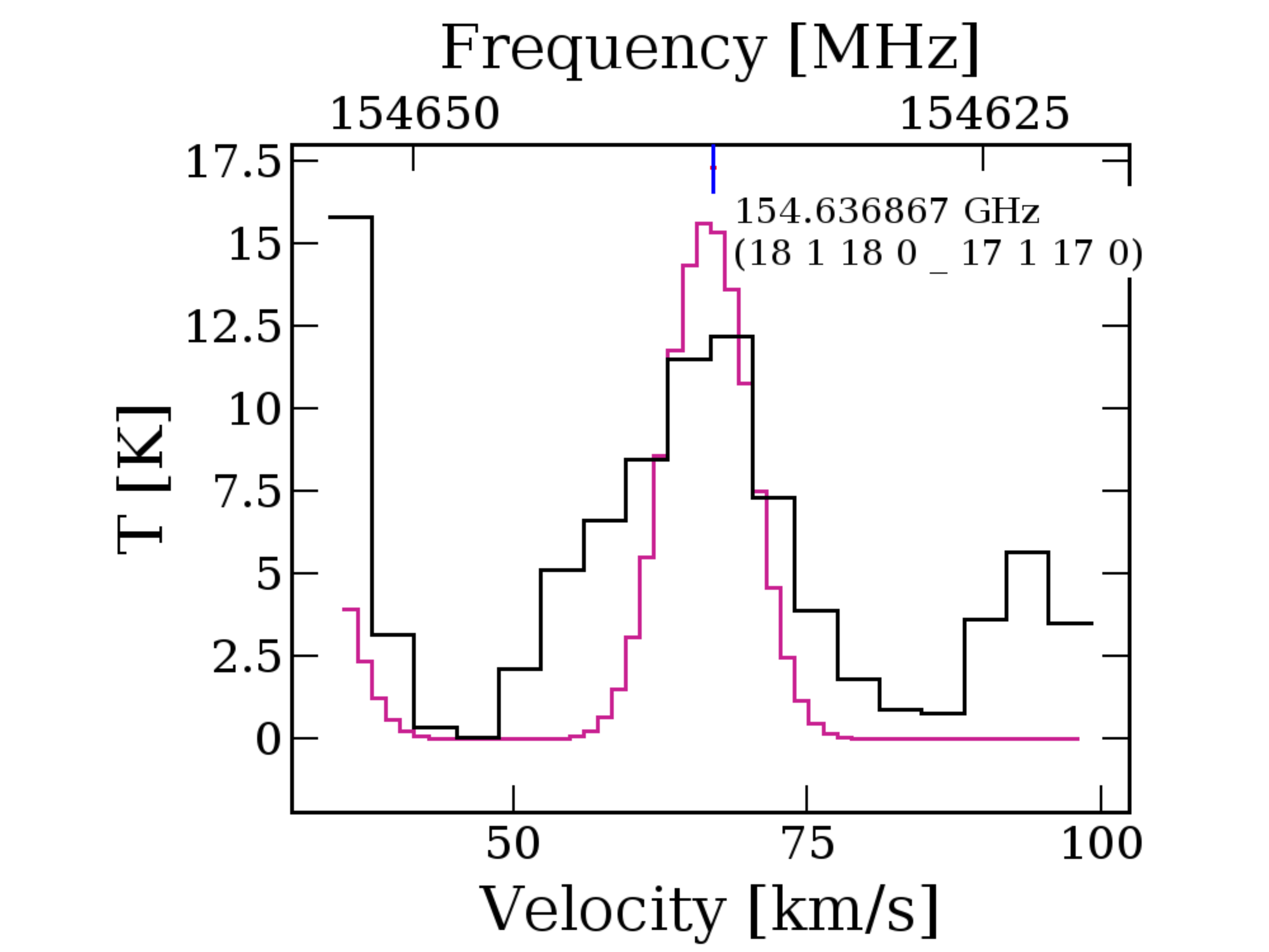}
\end{minipage}
\caption{LTE fitting of observed transitions of $\rm{CH_3NCO}$ towards G10. Black line represents the observed spectra and pink line 
is the fitted profile.}
\label{fig:CH3NCO-rot}
\end{figure*}

LTE model fitted line parameters of all the observed transitions are provided in Table \ref{table:fitted}. The observed 
spectra, along with the fitted one, are shown in Fig. \ref{fig:HNCO-rot},  \ref{fig:NH2CHO-rot}, and \ref{fig:CH3NCO-rot} for $\rm{HNCO}$, 
$\rm{NH_2CHO}$, and $\rm{CH_3NCO}$ respectively. We found that some transitions of $\rm{CH_3NCO}$ are blended and thus, we do not obtain better fits 
for those transitions. 
{ Some of the spectra shown in these figures contain multiple hyperfine transitions. For the LTE fitting, we have considered 
only that transition which has the highest value of Einstein coefficients among them. 
Since some of the transitions with the highest Einstein coefficient 
are slightly offset from the peak position, LTE fitting results show a slight offset from some observational spectra.}
Extracted physical parameters in Table \ref{table:fitted} shows that the optical 
depths ($\tau$) of all the lines are less than $1$. Our obtained best fitted column densities of these three species are shown in Table \ref{table:fitted}. For this MCMC 
fitting, we have used the different source sizes for different species as obtained from their two-dimensional Gaussian fitting. We have obtained higher 
excitation temperatures by the MCMC calculations (Table \ref{table:fitted}) which are consistent with the high rotational temperatures of these molecules obtained by the 
rotational diagram analysis which is described in the Section \ref{sec:rotdia} (Table \ref{table:abunobs}).

\begin{table*}
\centering
\tiny
\caption{Summary of the best fitted line parameters of observed molecules towards G10. \label{table:fitted}}
\begin{tabular}{|c|c|>{}c|p{1.0cm}|p{0.8cm}|c|c|c|c|p{0.7cm}|p{1.1cm}|p{0.8cm}|}
\hline
\hline
Species&Frequency&{Range used}
&Range used &Best fit FWHM &Best fit column&Optical depth&Range used &Best fitted
&Source size&Range used &Best fitted \\
&(GHz)&{Frequency (GHz)}&FWHM (Km s$^{-1}$) &(Km s$^{-1}$)&density (cm$^{-2}$)& ($\tau$) &T$_{ex}$ (K)&T$_{ex}$ (K)&($^{''}$)&V$_{lsr}$ (Km s$^{-1}$)&V$_{lsr}$ (Km s$^{-1}$)\\
\hline\hline
&131.394262&130.53092 - 131.46660&3-6&5.97&1.6$\times$10$^{17}$&2.85$\times$10$^{-1}$&200-350&201.19&1.12&67.2-67.9&67.59\\
HNCO&154.414770&154.03118 - 154.96636&5-7&6.97&1.6$\times$10$^{17}$&2.64$\times$10$^{-1}$&200-350&205.65&1.16&66.5-67.5&67.20\\
&153.865080&153.03117 - 153.96523&5-8&7.96&1.5$\times$10$^{17}$&2.94$\times$10$^{-1}$&200-330&211.98&1.35&66.5-67.5&67.32\\
&153.818870&153.03117 - 153.96523&5-8&7.96&1.5$\times$10$^{17}$&1.19$\times$10$^{-1}$&200-330&211.98&1.35&66.5-67.5&67.32\\
&153.291840&153.03117 - 153.96523&5-8&7.96&1.5$\times$10$^{17}$&2.34$\times$10$^{-1}$&200-330&211.98&1.35&66.5-67.5&67.32\\
\hline
&148.223354&148.46523 - 147.53117&3-9&8.98&1.3$\times$10$^{17}$&7.73$\times$10$^{-2}$&400-600&472.07&1.33&66.5-67.5&67.43\\
&148.556276&148.53115 - 149.46522&3-7&7.00&9.5$\times$10$^{16}$&1.78$\times$10$^{-2}$&400-550&450.09&1.37&66.5-67.5&67.09\\
&148.567249&148.53115 - 149.46522&3-7&7.00&9.5$\times$10$^{16}$&3.52$\times$10$^{-2}$&400-550&450.09&1.37&66.5-67.5&67.09\\
NH$_2$CHO&148.599727&148.53115 - 149.46522&3-7&7.00&9.5$\times$10$^{16}$&5.15$\times$10$^{-2}$&400-550&450.09&1.37&66.5-67.5&67.09\\
&148.667591&148.53115 - 149.46522&3-7&7.00&9.5$\times$10$^{16}$&6.53$\times$10$^{-2}$&400-550&450.09&1.37&66.5-67.5&67.09\\
&148.709316&148.53115 - 149.46522&3-7&7.00&9.5$\times$10$^{16}$&6.54$\times$10$^{-2}$&400-550&450.09&1.37&66.5-67.5&67.09\\
&153.432351&153.96523 - 153.03118&3-8&7.99&1.2$\times$10$^{17}$&6.70$\times$10$^{-2}$&400-600&474.33&1.18&66.5-67.5&67.43\\
\hline

&129.957471&129.53092 - 130.46674&4-8&7.99&6.9$\times$10$^{16}$&3.08$\times$10$^{-1}$&100-300&104.04&1.05&66.5-67.5&67.14\\
&129.669703&129.53092 - 130.46674&4-8&7.99&6.9$\times$10$^{16}$&1.77$\times$10$^{-1}$&100-300&104.04&1.05&66.5-67.5&67.14\\
&130.146799&129.53092 - 130.46674&4-8&7.99&6.9$\times$10$^{16}$&2.41$\times$10$^{-1}$&100-300&104.04&1.05&66.5-67.5&67.14\\
&130.300215&129.53092 - 130.46674&4-8&7.99&6.9$\times$10$^{16}$&1.57$\times$10$^{-1}$&100-300&104.04&1.05&66.5-67.5&67.14\\
&130.228419&129.53092 - 130.46674&4-8&7.99&6.9$\times$10$^{16}$&2.41$\times$10$^{-1}$&100-300&104.04&1.05&66.5-67.5&67.14\\
&130.541066&130.53092 - 131.46660&4-8&5.99&7.8$\times$10$^{16}$&1.91$\times$10$^{-1}$&100-300&115.48&1.15&66.5-67.5&66.89\\
&130.583038&130.53092 - 131.46660&4-8&7.99&7.8$\times$10$^{16}$&1.59$\times$10$^{-1}$&100-300&115.48&1.15&66.5-67.5&66.89\\
&130.653851&130.53092 - 131.46660&4-8&7.99&7.8$\times$10$^{16}$&1.68$\times$10$^{-1}$&100-300&115.48&1.15&66.5-67.5&66.89\\
&130.661691&130.53092 - 131.46660&4-8&7.99&7.8$\times$10$^{16}$&1.23$\times$10$^{-1}$&100-300&115.48&1.15&66.5-67.5&66.89\\
&130.88332&130.53092 - 131.46660&4-8&7.99&7.8$\times$10$^{16}$&1.58$\times$10$^{-1}$&100-300&115.48&1.15&66.5-67.5&66.89\\
CH$_3$NCO&148.833657&148.53115  - 149.46522&4-8&7.99&8.8$\times$10$^{16}$&7.25$\times$10$^{-2}$&100-300&122.21&1.23&66.5-67.5&66.68\\
&148.437799&148.46523  - 147.53117&4-8&7.92&7.7$\times$10$^{16}$&1.41$\times$10$^{-1}$&100-300&101.24&1.23&66.5-67.5&67.13\\
&148.376092&148.46523  - 147.53117&4-8&7.92&7.7$\times$10$^{16}$&2.32$\times$10$^{-1}$&100-300&101.24&1.23&66.5-67.5&67.13\\
&148.326896&148.46523  - 147.53117&4-8&7.92&7.7$\times$10$^{16}$&1.28$\times$10$^{-1}$&100-300&101.24&1.23&66.5-67.5&67.13\\
&148.280088&148.46523  - 147.53117&4-8&7.92&7.7$\times$10$^{16}$&1.27$\times$10$^{-1}$&100-300&101.24&1.23&66.5-67.5&67.13\\
&148.262442&148.46523  - 147.53117&4-8&7.92&7.7$\times$10$^{16}$&1.33$\times$10$^{-1}$&100-300&101.24&1.23&66.5-67.5&67.13\\
&148.075687&148.46523  - 147.53117&4-8&7.92&7.7$\times$10$^{16}$&1.88$\times$10$^{-1}$&100-300&101.24&1.23&66.5-67.5&67.13\\
&148.061901&148.46523  - 147.53117&4-8&7.92&7.7$\times$10$^{16}$&2.48$\times$10$^{-1}$&100-300&101.24&1.23&66.5-67.5&67.13\\
&147.673312&148.46523  - 147.53117&4-8&7.92&7.7$\times$10$^{16}$&2.12$\times$10$^{-1}$&100-300&101.24&1.23&66.5-67.5&67.13\\
&147.603962&148.46523  - 147.53117&4-8&7.92&7.7$\times$10$^{16}$&3.27$\times$10$^{-1}$&100-300&101.24&1.23&66.5-67.5&67.13\\
&154.742833&154.03118  - 154.96636&4-8&7.97&7.5$\times$10$^{16}$&3.53$\times$10$^{-1}$&100-300&101.70&1.21&66.5-67.5&66.54\\
&154.636867&154.03118  - 154.96636&4-8&7.97&7.5$\times$10$^{16}$&3.92$\times$10$^{-1}$&100-200&101.70&1.21&66.5-67.5&66.54\\
\hline
\end{tabular}
\end{table*}

\section{Chemical Modeling \label{sec:model-results}}
We carry out extensive modeling to study the abundance of three peptide bond related species in G10.  
To study the chemical evolution of these species, we used our previous chemical network \citep{das15a,das15b,das16,gora17a,gora17b,sil18}. 
Gas phase pathways were mainly adopted from the UMIST database \citep{mcel13}, whereas ice phase pathways and binding energies (BEs)
of the surface species were taken from the KIDA database \citep[unless otherwise stated,][]{ruau16}. We have considered
the diffusion energy of a species is $0.5$ times its adsorption energy and  
non-thermal desorption rate with a fiducial parameter $0.01$ \citep{gora17a}.
A cosmic ray rate of $1.3 \times 10^{-17}$ s$^{-1}$ is considered in all our models.
For the formation and destruction reactions of these species, we mainly have followed \cite{quen18}. In addition, 
following the recent study of \cite{haup19}, we have exclusively 
included a dual-cyclic hydrogen addition and abstraction reactions, which is connecting 
NH$_2$CHO, NH$_2$CO, and HNCO. We have shown the chemical linkages among HNCO, NH$_2$CHO, and CH$_3$NCO. Initial abundances of the 
model is provided in Table \ref{table:initial}. 


\begin{table}
\centering
{\tiny
\caption{Initial abundances with respect to total hydrogen nuclei.}
\begin{tabular}{|c|c|}
\hline
Species&Abundance\\
\hline\hline
$\mathrm{H_2}$ &    $5.00 \times 10^{-01}$\\
$\mathrm{He}$  &    $9.00 \times 10^{-02}$\\
$\mathrm{C}$ &    $7.30 \times 10^{-05}$\\
$\mathrm{O}$   &    $1.76 \times 10^{-04}$\\
$\mathrm{N}$   &    $2.14 \times 10^{-05}$\\
$\mathrm{Cl}$   &    $1.00 \times 10^{-09}$\\
$\mathrm{Fe}$&    $3.00 \times 10^{-09}$\\
$\mathrm{Mg}$&    $7.00 \times 10^{-09}$\\
$\mathrm{Na}$&    $2.00 \times 10^{-09}$\\
$\mathrm{S}$ &    $8.00 \times 10^{-08}$\\
$\mathrm{Si^+}$&    $8.00 \times 10^{-09}$\\
$\mathrm{e^-}$ &    $7.31 \times 10^{-05}$\\
\hline
\end{tabular}}
\label{table:initial}
\end{table}

\subsection{Physical condition of the adopted model}
We have considered a 3-phase model to study the chemical evolution of these species \citep{garr13}. This model is
best suited because G10 is a high mass star forming core. The detail considerations of each phase are
discussed as below.\\

\noindent {\it 1$^{st}$ phase:} In the first phase, we have considered that the cloud collapses from a low total hydrogen
density ($\rho_{min}=10^3$ cm$^{-3}$) to a high total hydrogen density ($\rho_{max}$). 
The initial gas temperature ($T_{gas}$) is assumed to be $40$ K, whereas the dust 
temperature is assumed to remain fixed at initial ice temperature ($T_{ice}$). {We have considered a time interval of $t_{coll}$ years to 
reach from $\rho_{min}$ to $\rho_{max}$}. Since at the highest density, the gas and dust are well coupled, 
we have considered $T_{gas}=T_{ice}$ at the highest density ($\rho_{max}$), i.e., at $t=t_{coll}$. 
From this stage onward, we have assumed that the temperature of the dust and the gas is the same. 
Thus, we have considered a negative slope for $T_{gas}$ for the collapsing phase. Throughout the first phase, the visual extinction 
parameter is considered constantly increasing from $\rm{A_V={A_V}_{min}=10}$ to finally at $\rm{{A_V}_{max}=200}$ in $t=t_{coll}$.\\

\noindent {\it 2$^{nd}$ phase:} The second phase of the simulation corresponds to a warm-up phase. Since G10 is a high mass star 
forming region, we consider a moderate warm-up time scale ($t_w$) $5 \times 10^4$ years \citep{garr13}. Therefore, during this short period, the 
temperature of the cloud from $T_{ice}$ can reach the highest hot core temperature $T_{max}$. 
Density, temperature, and visual extinction parameter remain constant at $\rho_{max}$, $T_{max}$, and ${A_V}_{max}$ respectively. \\

\noindent {\it 3$^{rd}$ phase:} This phase belongs to the post-warm-up time. Here, we have considered a post-warm-up
time scale ($t_{pw}$) of $10^5$ years. So, the total simulation time ($t_{tot}=t_{coll}+t_w+t_{pw}$). 
The parameters such as the density and visual extinction are assumed to be the same as they were 
in the warm-up phase. The temperature of the cloud is kept at $T_{max}$ throughout the last phase. \\

\subsection{Binding energies and reaction pathways}

\begin{table*}
\centering{\tiny
\caption{Obtained adsorption energies (BEs) and thermodynamic parameters.}
\begin{tabular}{|c|c|c|c|c|c|c|}
\hline
{\bf Species}& {\bf Optimized} & {\bf Calculated BE (K)} & {\bf Average BE (K)} & {\bf Scaled BE (K)} & {\bf Calculated BE (K)} & {\bf
Available BE (K)} \\
&{\bf Structure}& {\bf using water monomer} && {\bf ($\times 1.416$)} & {\bf using water hexamer} & {\bf in literatures} \\
\hline
\multicolumn{7}{|c|}{CHNO}\\
\hline
HNCO& \includegraphics[height=0.8cm, width=1.5cm]{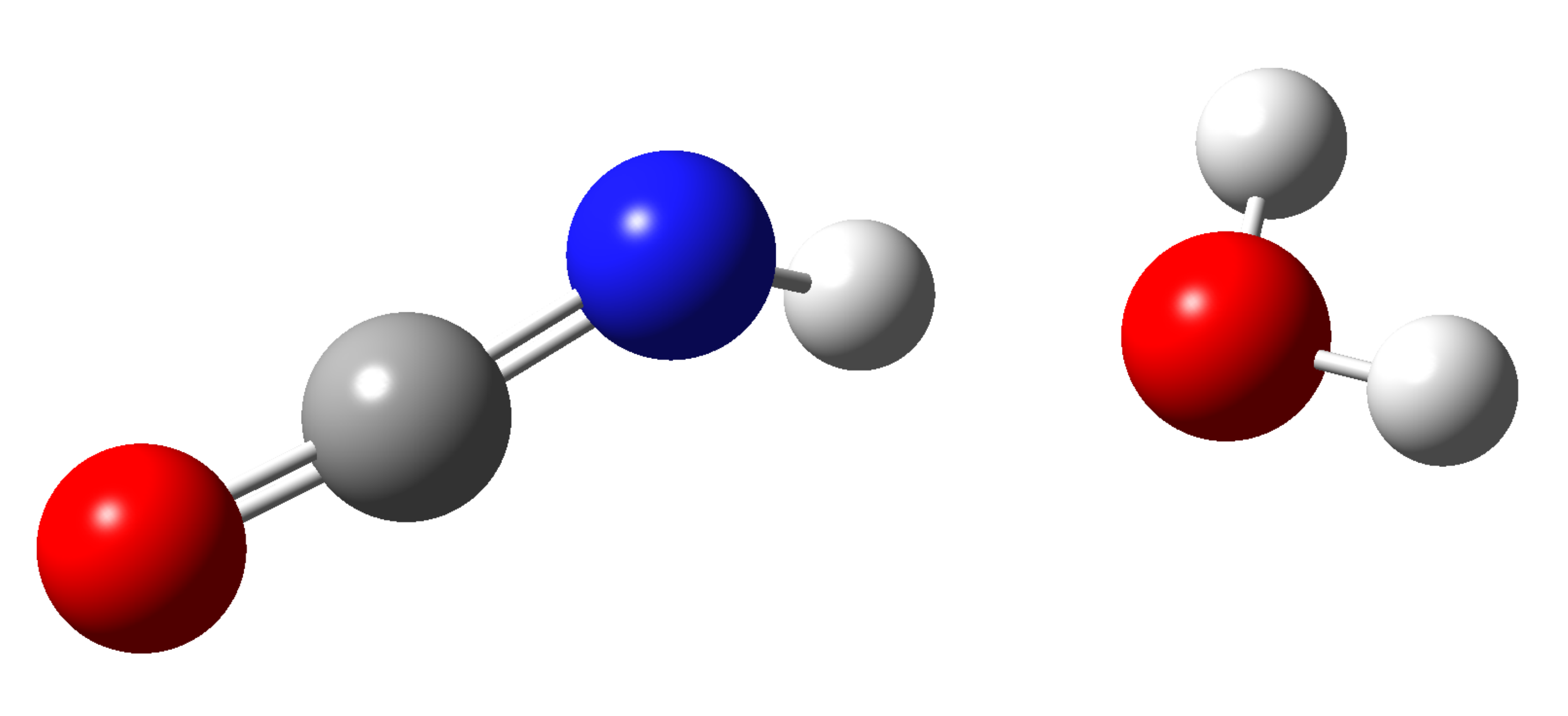}& 3308 & 3308 & 4684 & 6310, 5554 & $4400 \pm 1320$$^{a}$ \\
\hline
HCNO&\includegraphics[height=0.8cm, width=1.5cm]{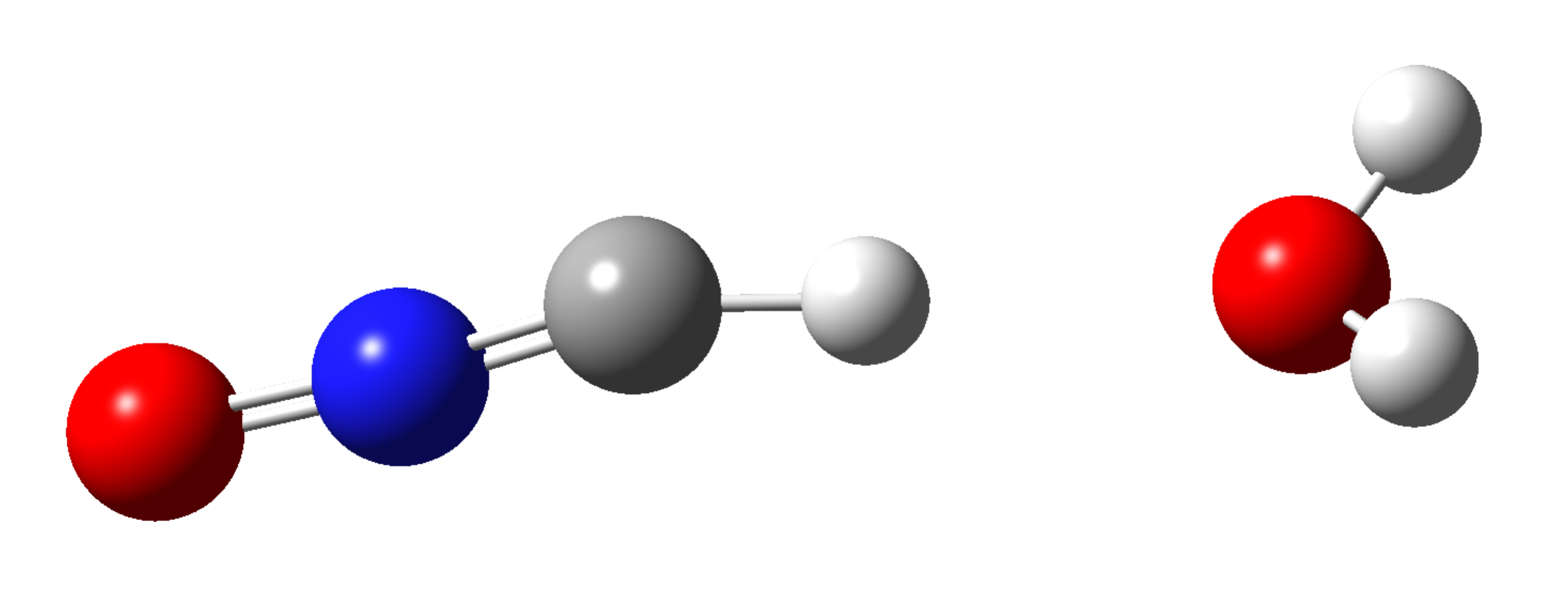}& 2640 & 2345 & 3320 & 6046 & 2800$^{b}$ \\
\cline{3-3}
&\includegraphics[height=1cm, width=1.5cm]{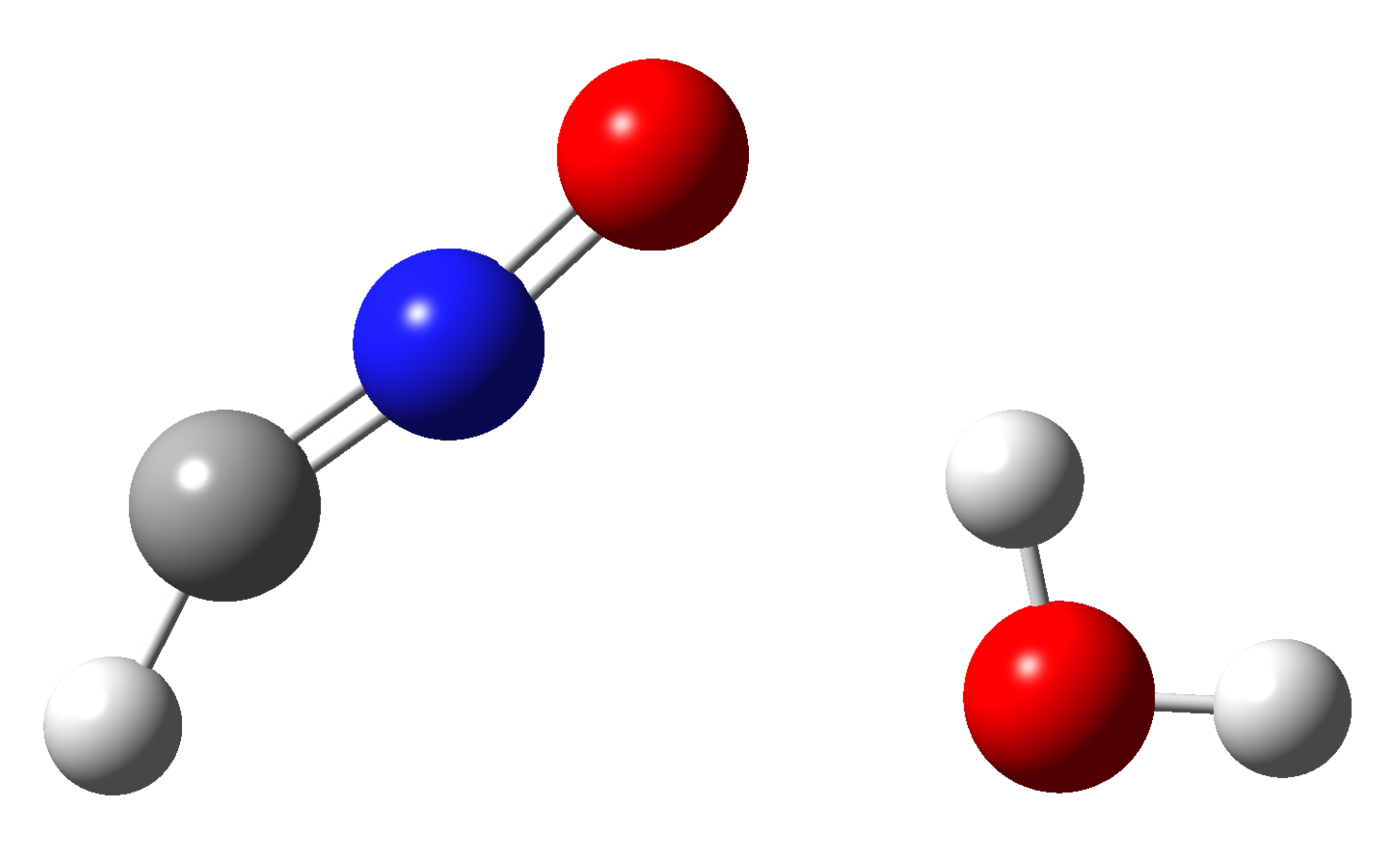}& 2050 &&&& \\
\hline
HOCN&\includegraphics[height=1cm, width=1.5cm]{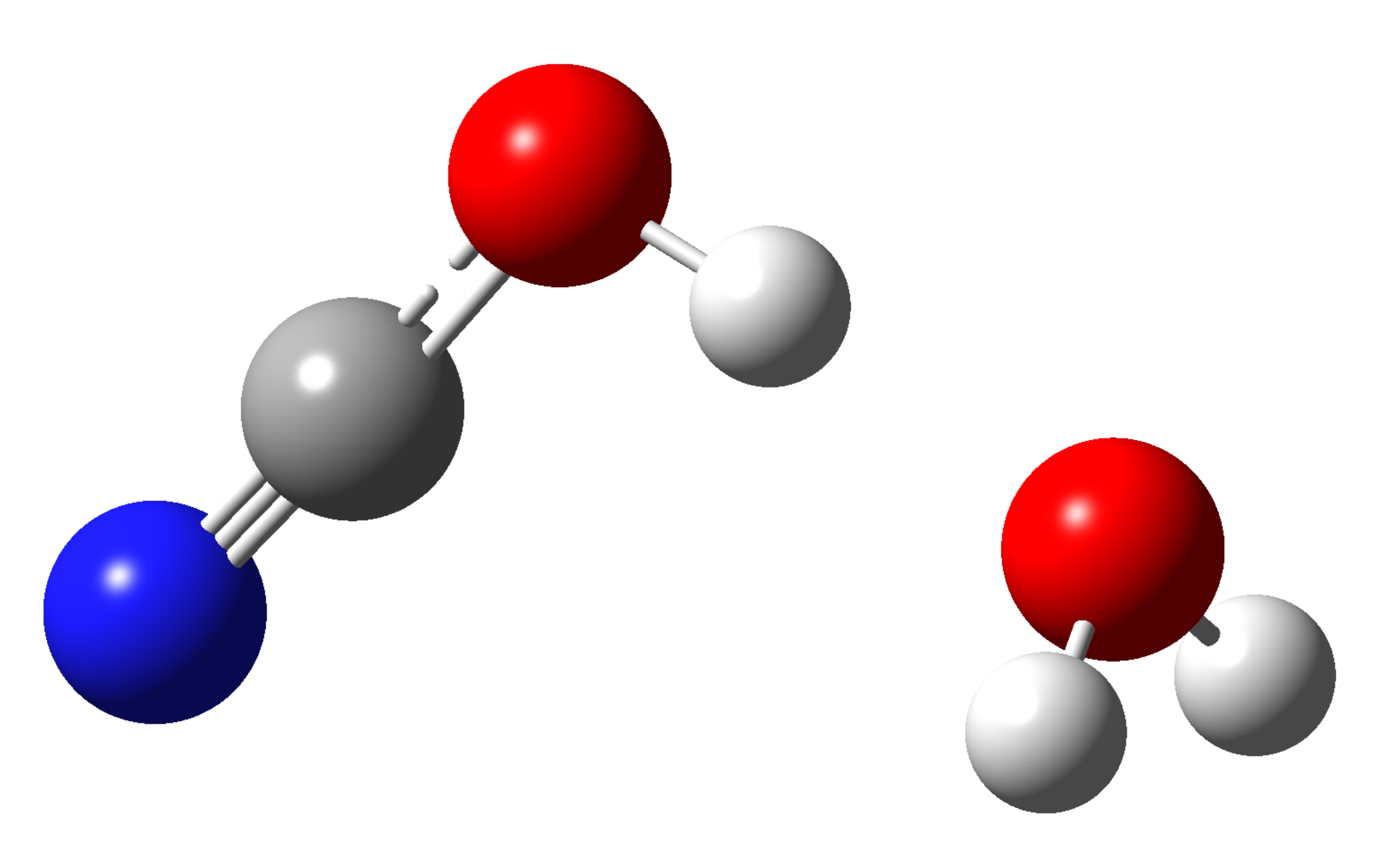}& 5936 & 4250 & 6018 & 2153, 8404 & 2800/$^{b}$ \\
\cline{3-3}
&\includegraphics[height=0.7cm, width=1.5cm]{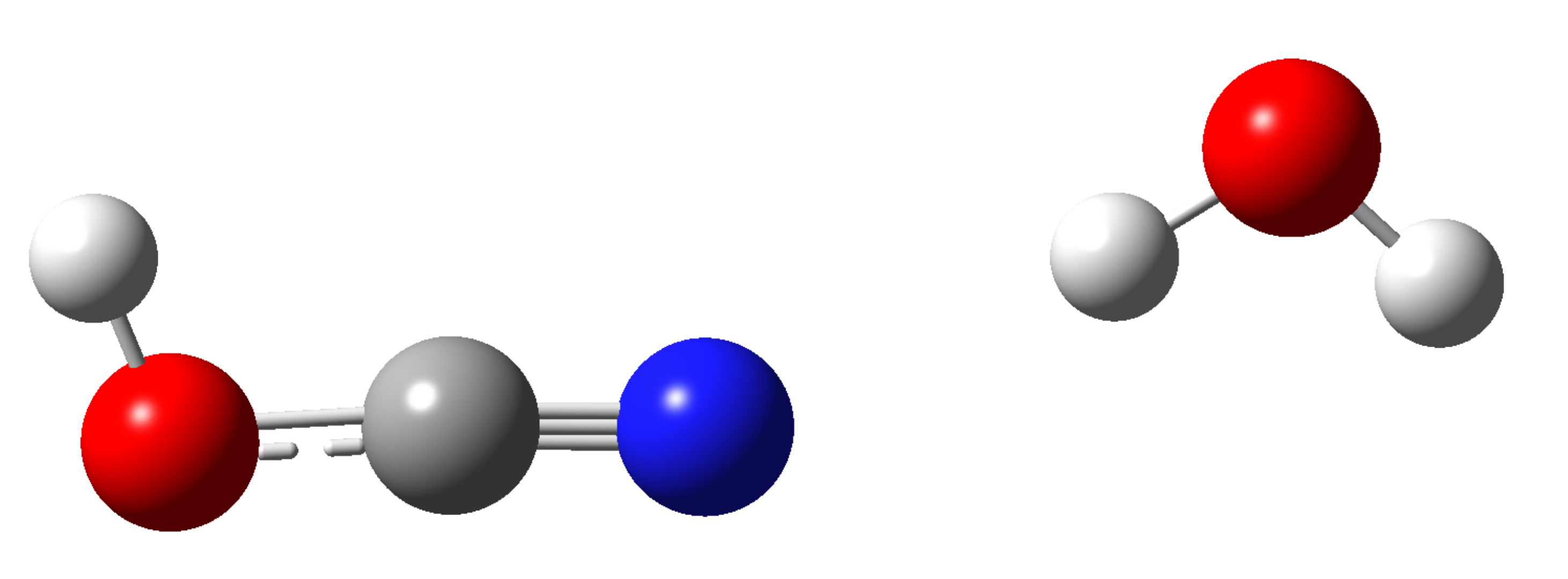}& 2563 &&&& \\
\hline
HONC&\includegraphics[height=1cm, width=1.5cm]{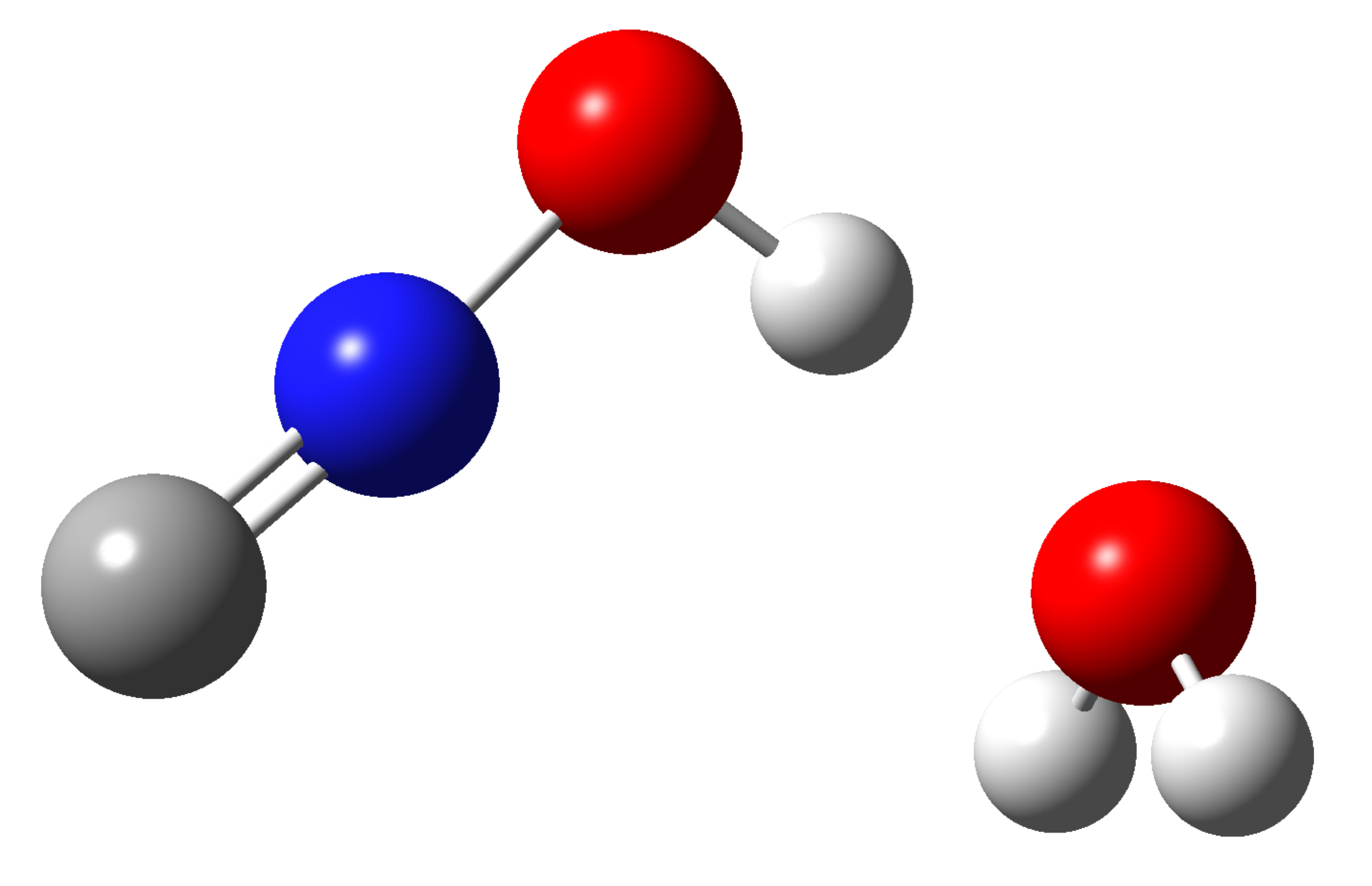}&5874& 4122 & 5837 & 3387, 8727 & 2800$^{b}$ \\
\cline{3-3}
&\includegraphics[height=0.7cm, width=1.5cm]{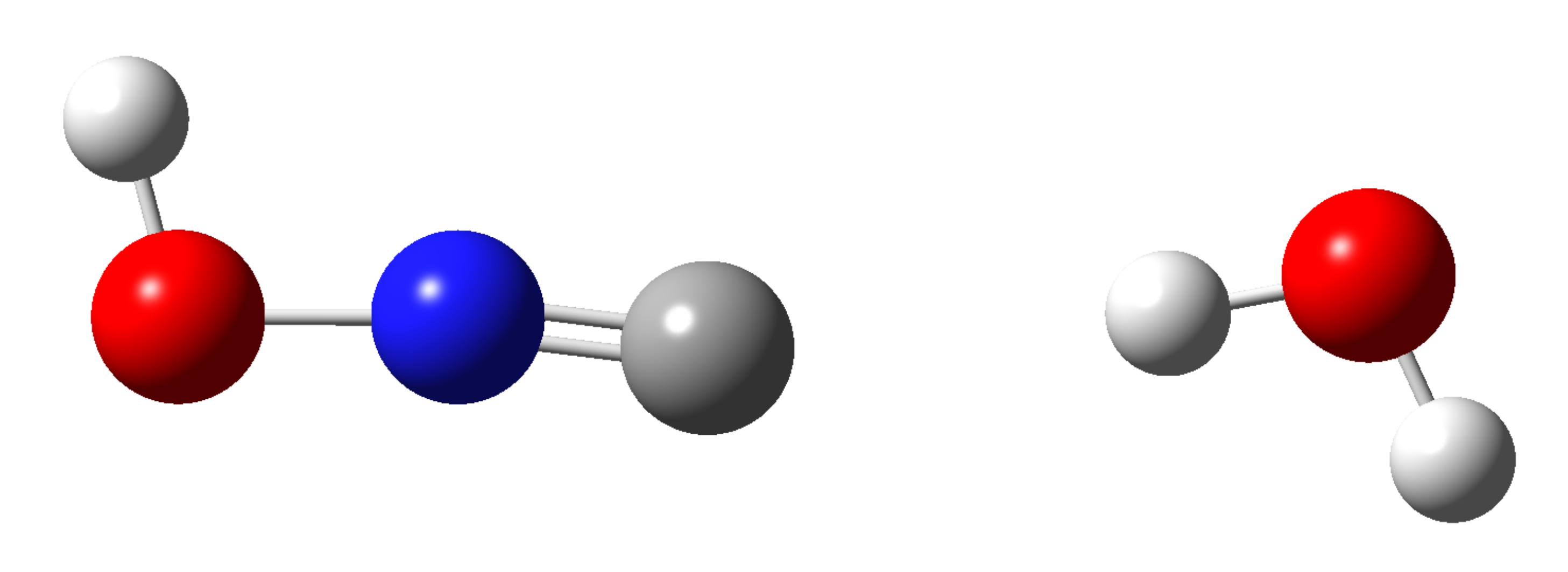}&2370&&&&\\
\hline
\multicolumn{7}{|c|}{$\rm{C_2NH_3O}$}\\
\hline
CH$_3$NCO&\includegraphics[height=1.2cm, width=1.5cm]{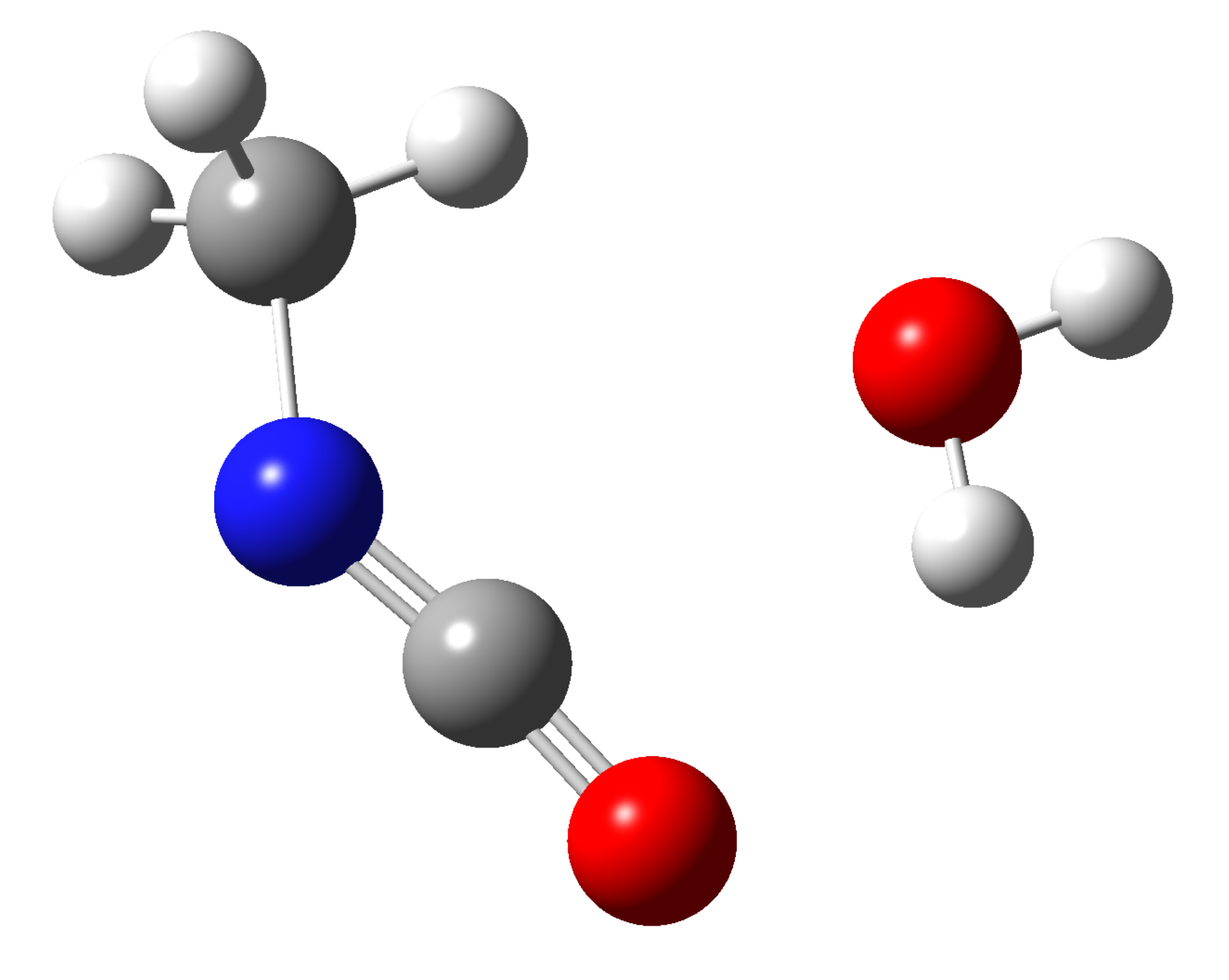}&3627& 3091 & 4377 & 4309 & $4700 \pm 1410$$^{a}$ \\
\cline{3-3}
&\includegraphics[height=1.2cm, width=1.2cm]{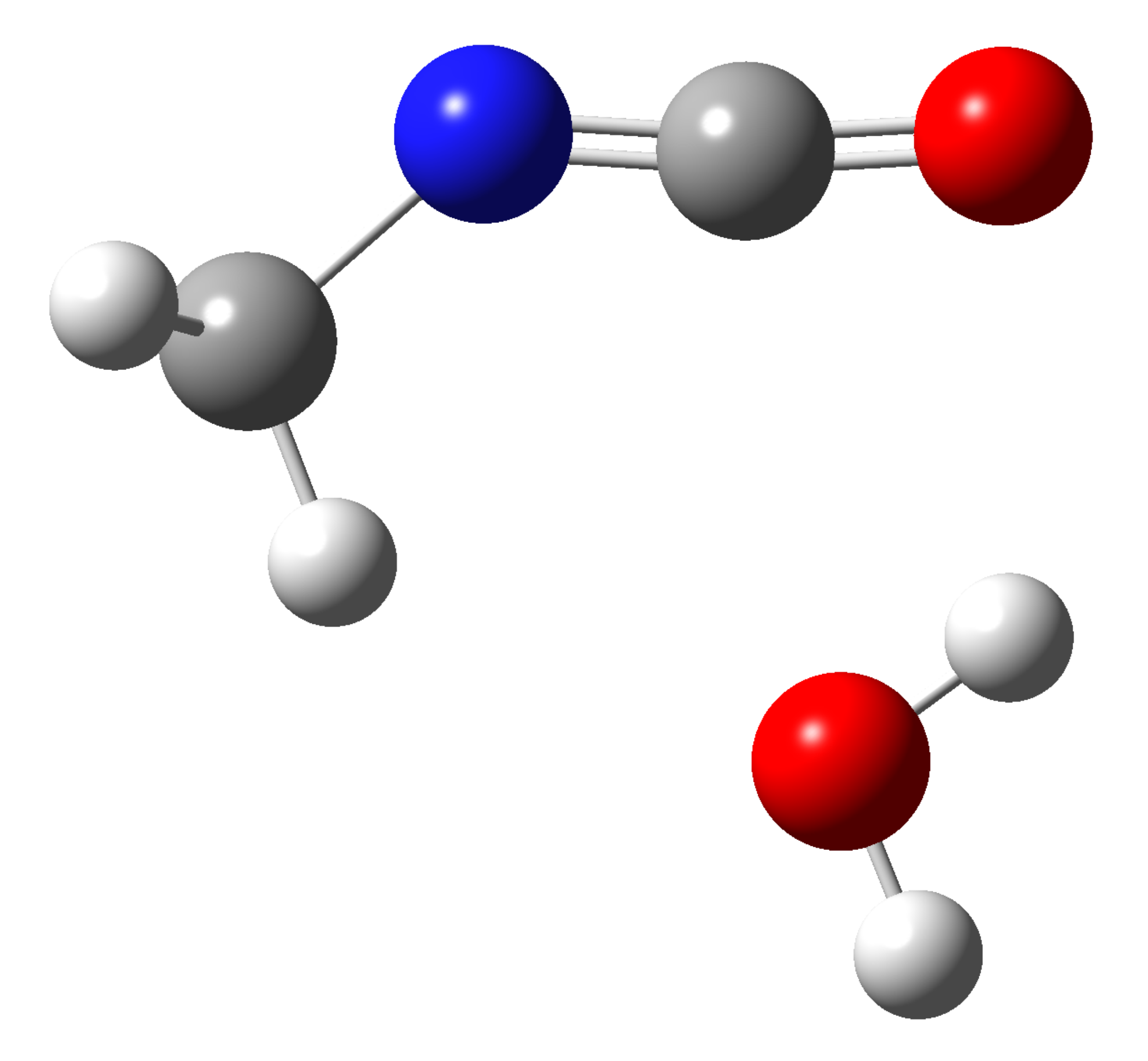}&2555&&&& \\
\hline
CH$_3$CNO&\includegraphics[height=1.2cm, width=1.5cm]{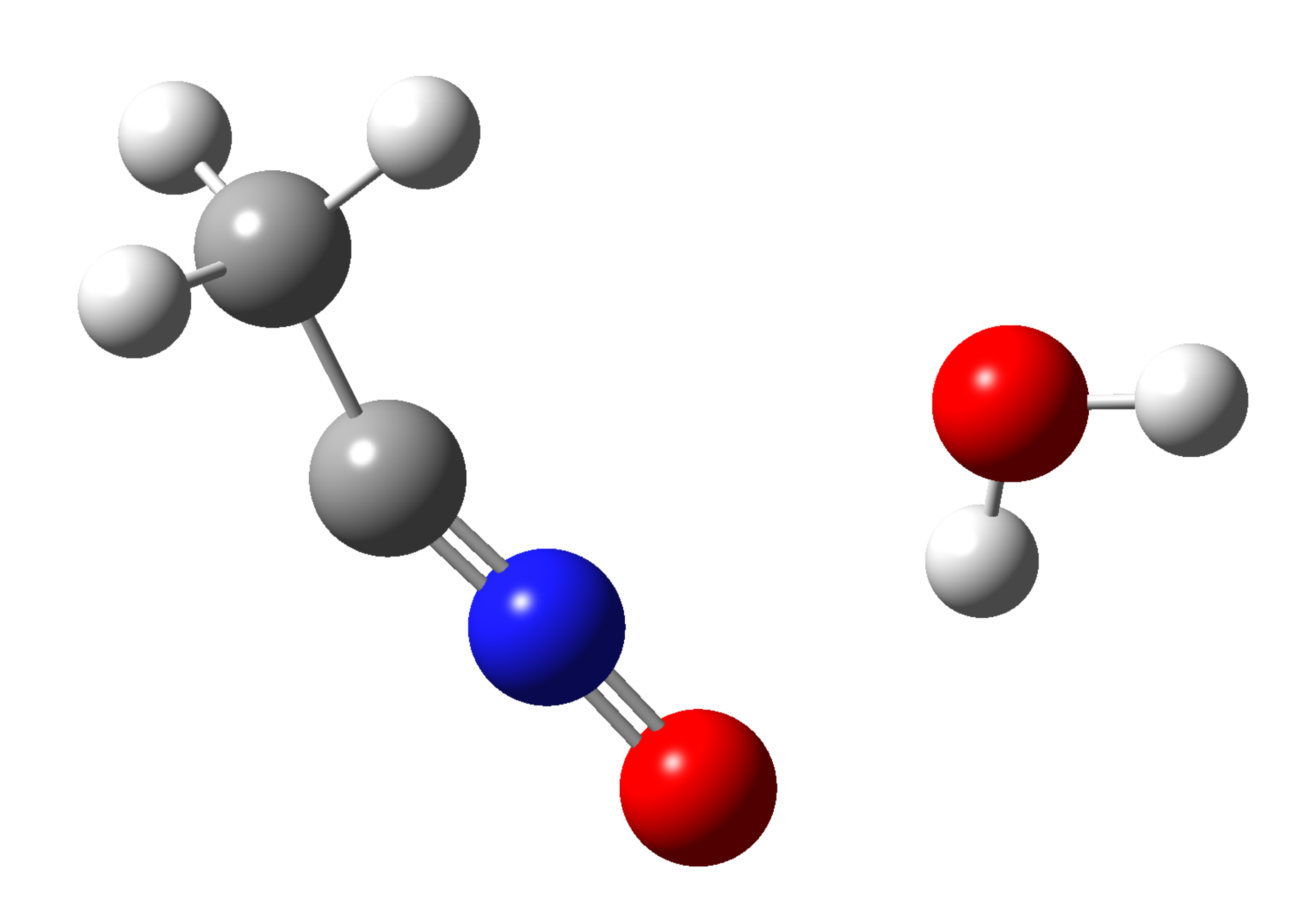}&2786& 2786 & 3945 & --- & --- \\
\hline
CH$_3$OCN&\includegraphics[height=1.2cm, width=1.5cm]{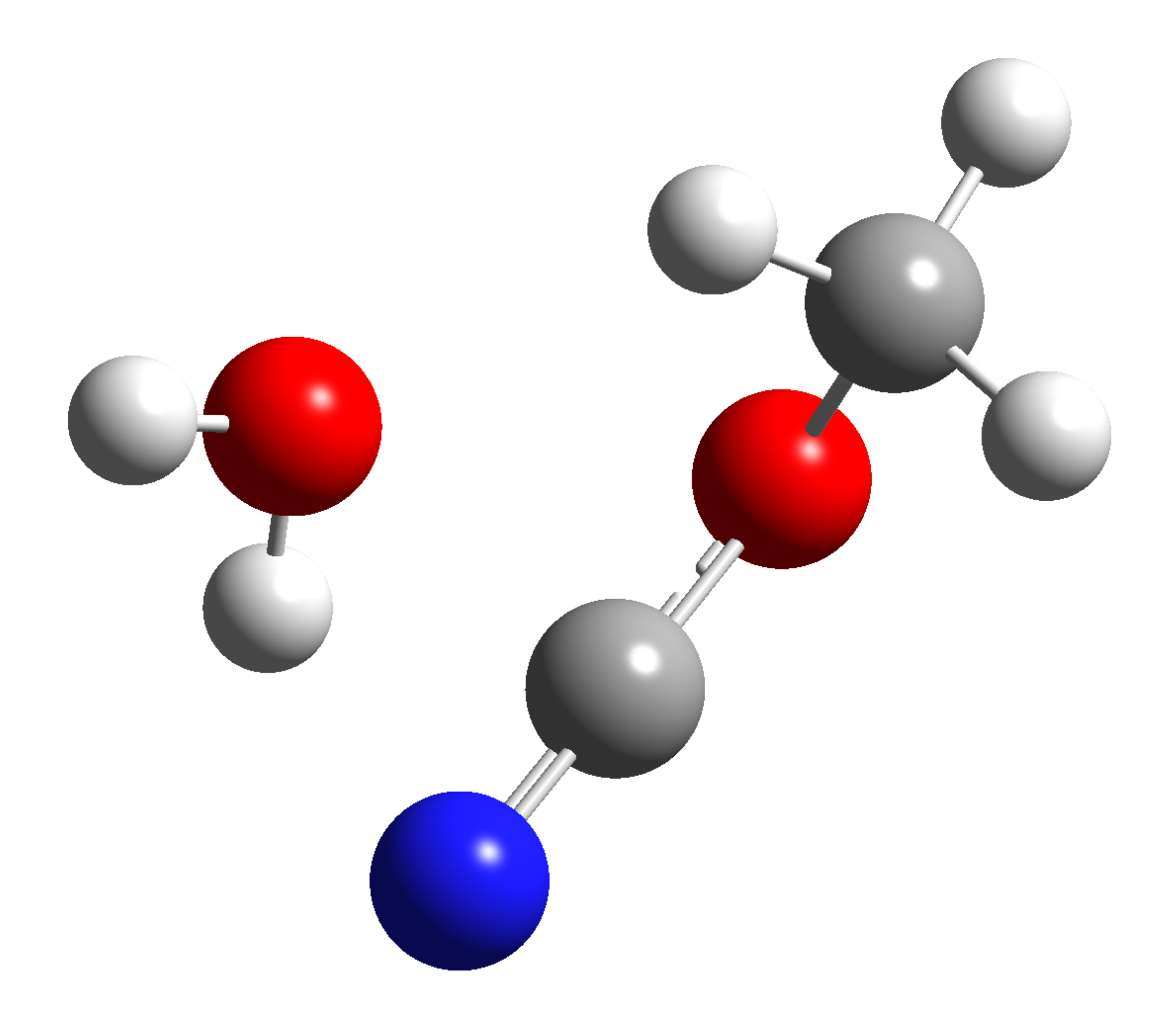}& 3534 & 3535 & 5006 & 6530 & --- \\
\cline{3-3}
&\includegraphics[height=1.2cm, width=1.5cm]{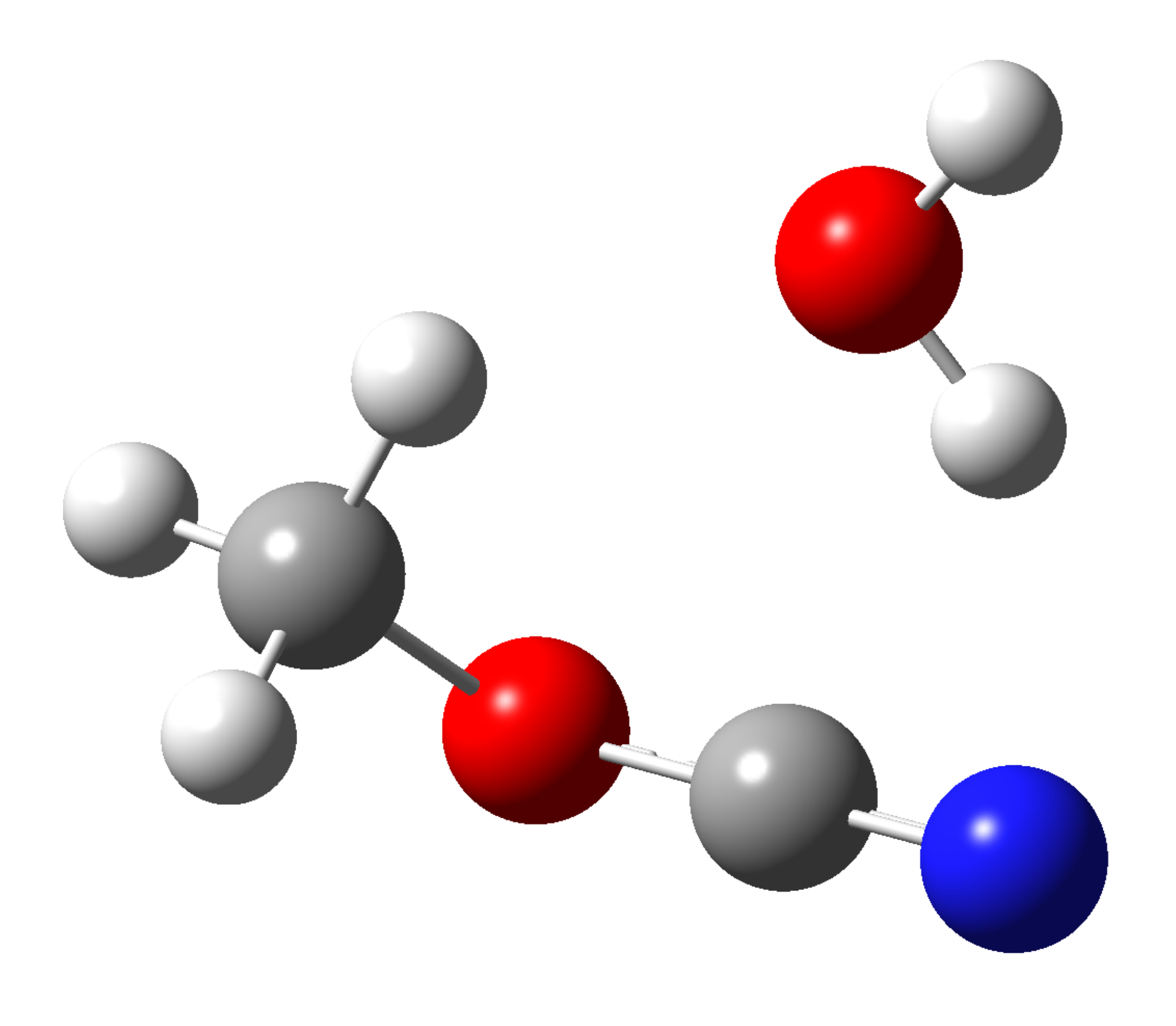}& 3536 &&&& \\
\hline
CH$_3$ONC&\includegraphics[height=1.2cm, width=1.5cm]{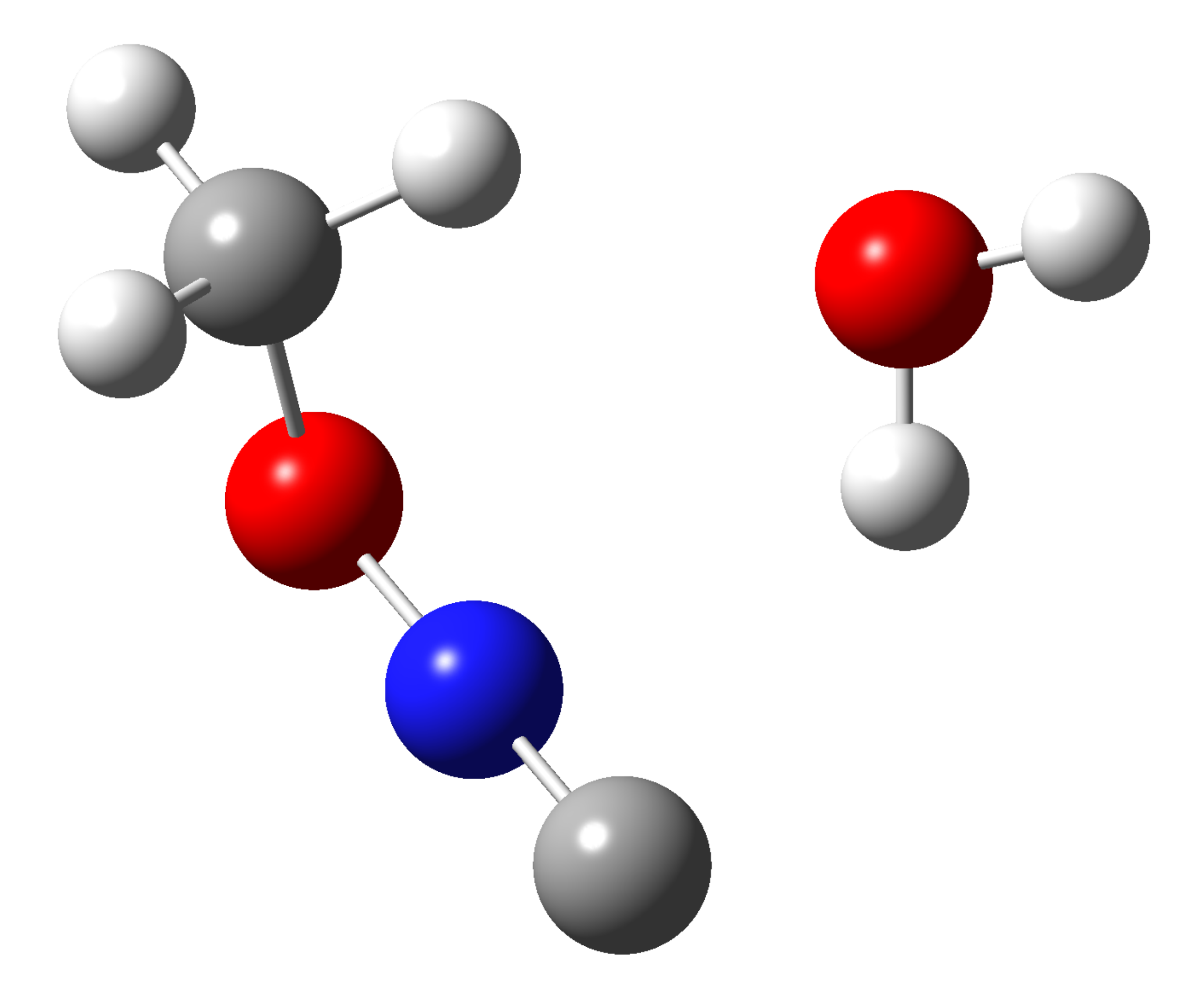}& 2939 & 2752 & 3897 & 4652 & --- \\
\cline{3-3}
&\includegraphics[height=1cm, width=2.2cm]{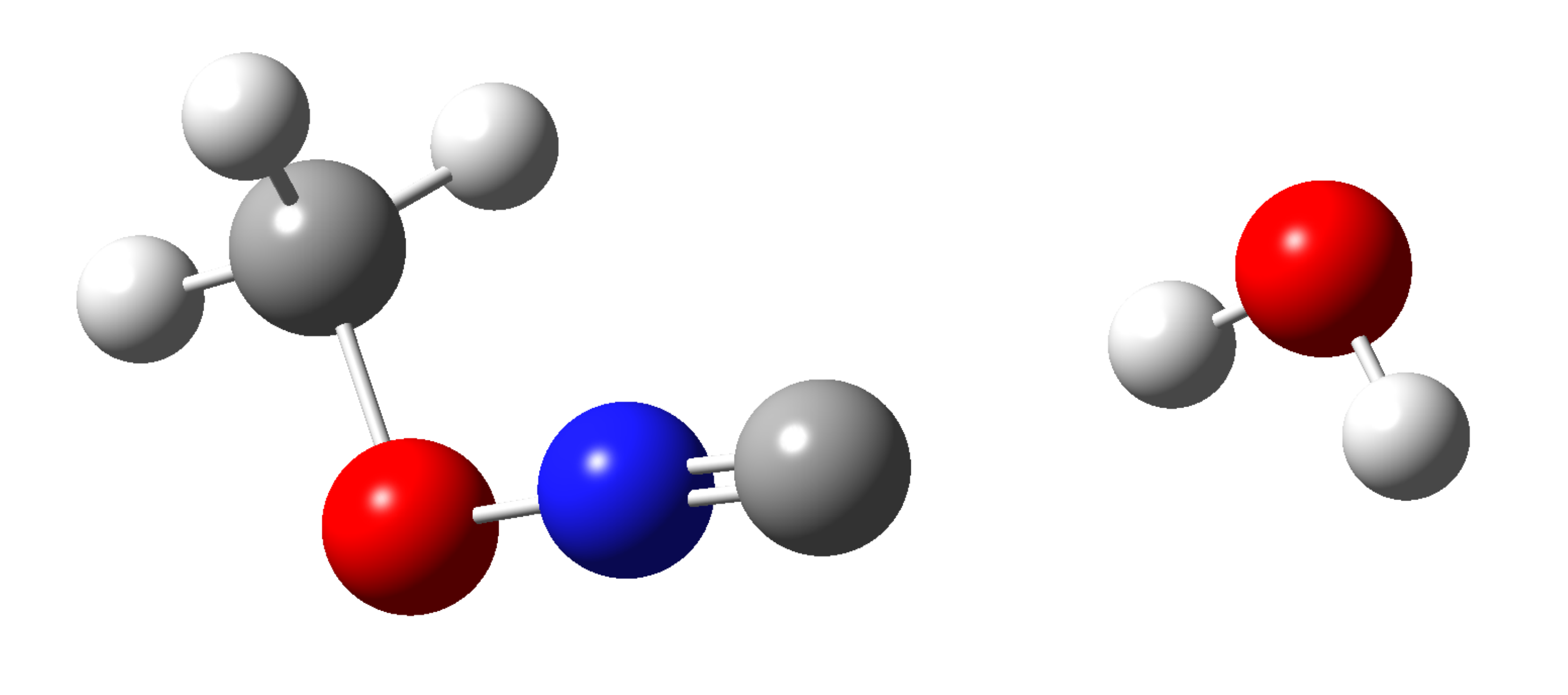}& 2565 &&&& \\
\hline
\multicolumn{7}{|c|}{$\rm{CNH_3O}$}\\
\hline
$\rm{NH_2CHO}$&\includegraphics[height=0.8cm, width=1.5cm]{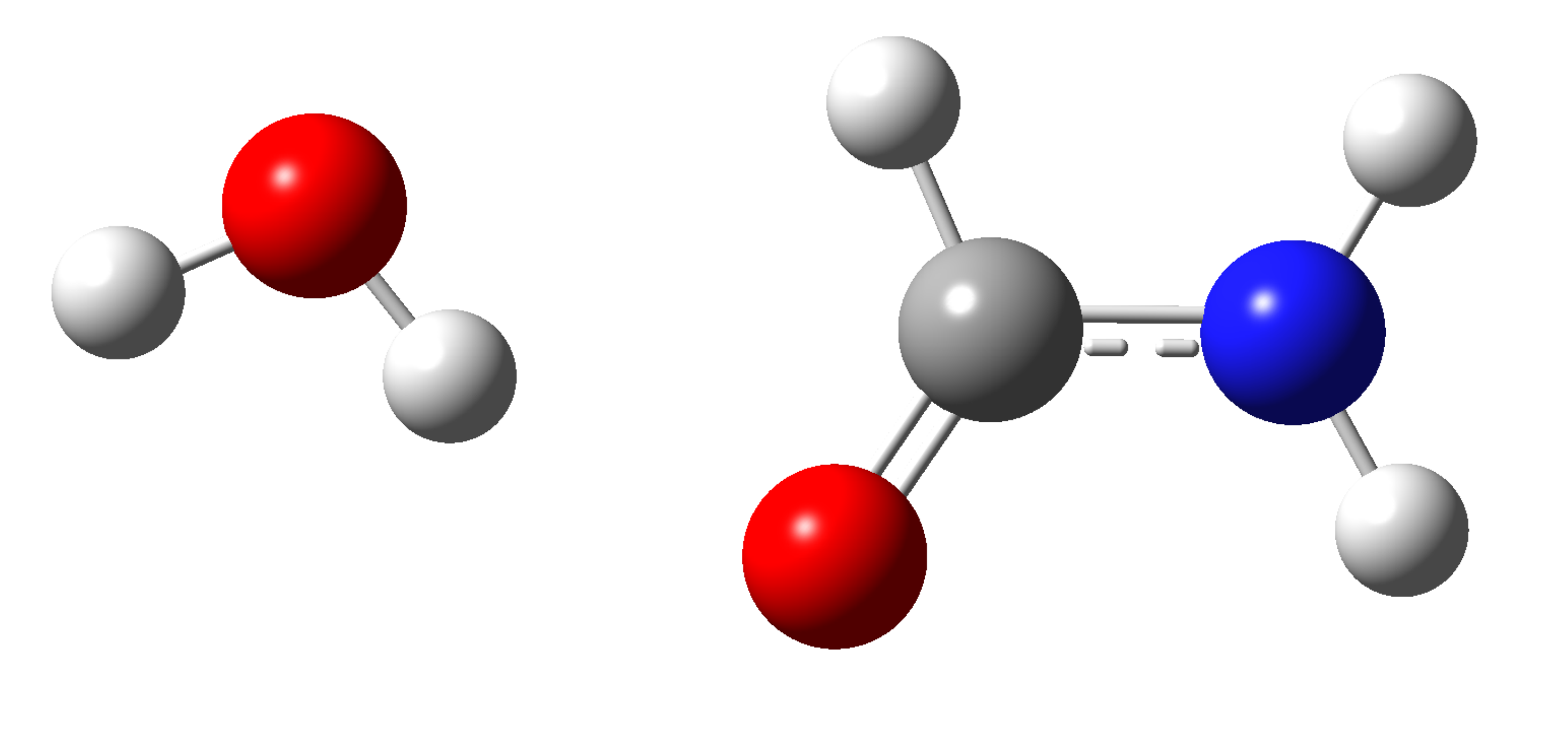}&3627&&&& \\
\cline{3-3}
&\includegraphics[height=1cm, width=1.5cm]{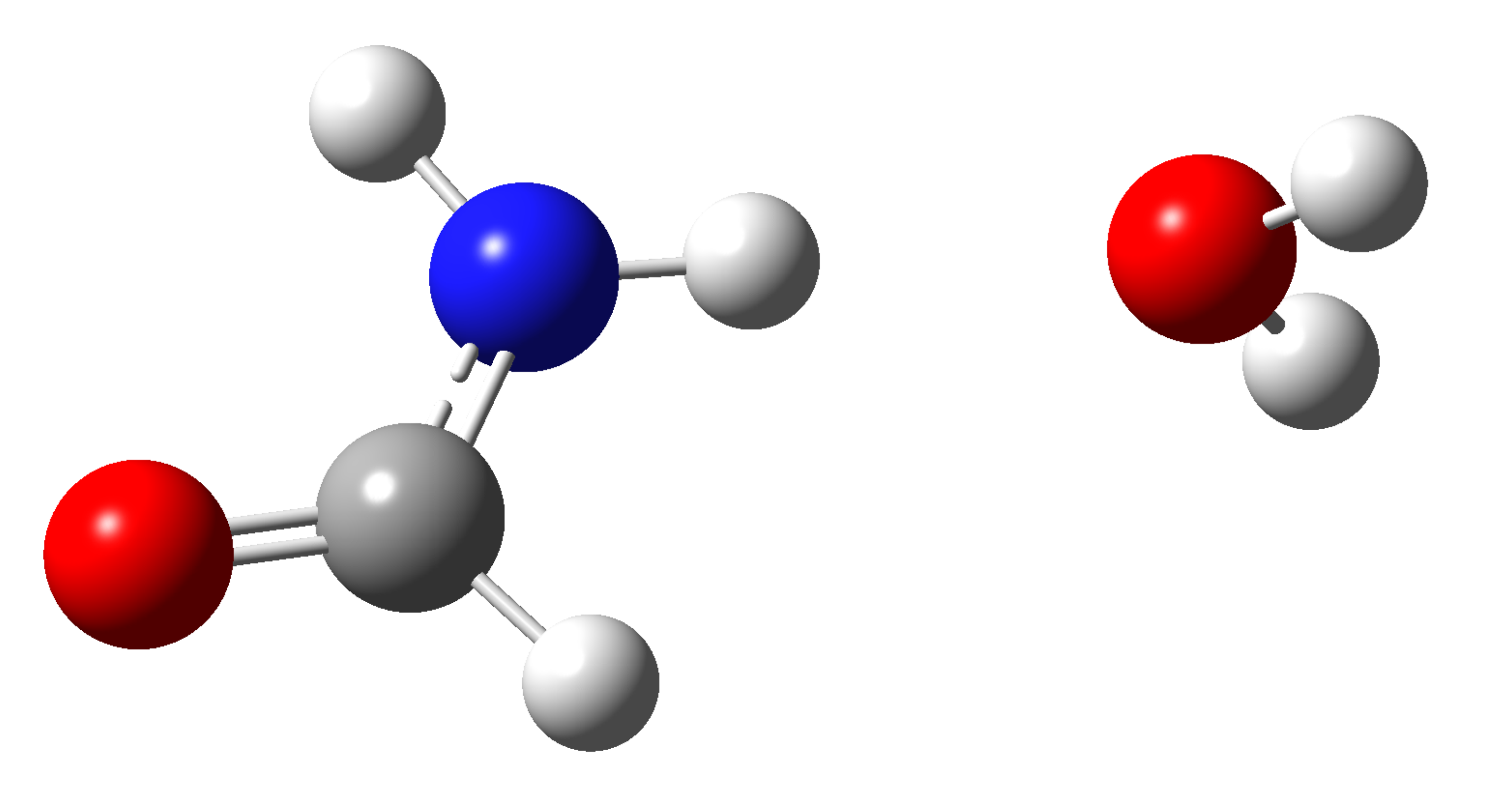}&2880& 3862 & 5468 & 6602 & $6300 \pm 1890$$^{a}$ \\
\cline{3-3}
&\includegraphics[height=1cm, width=1.5cm]{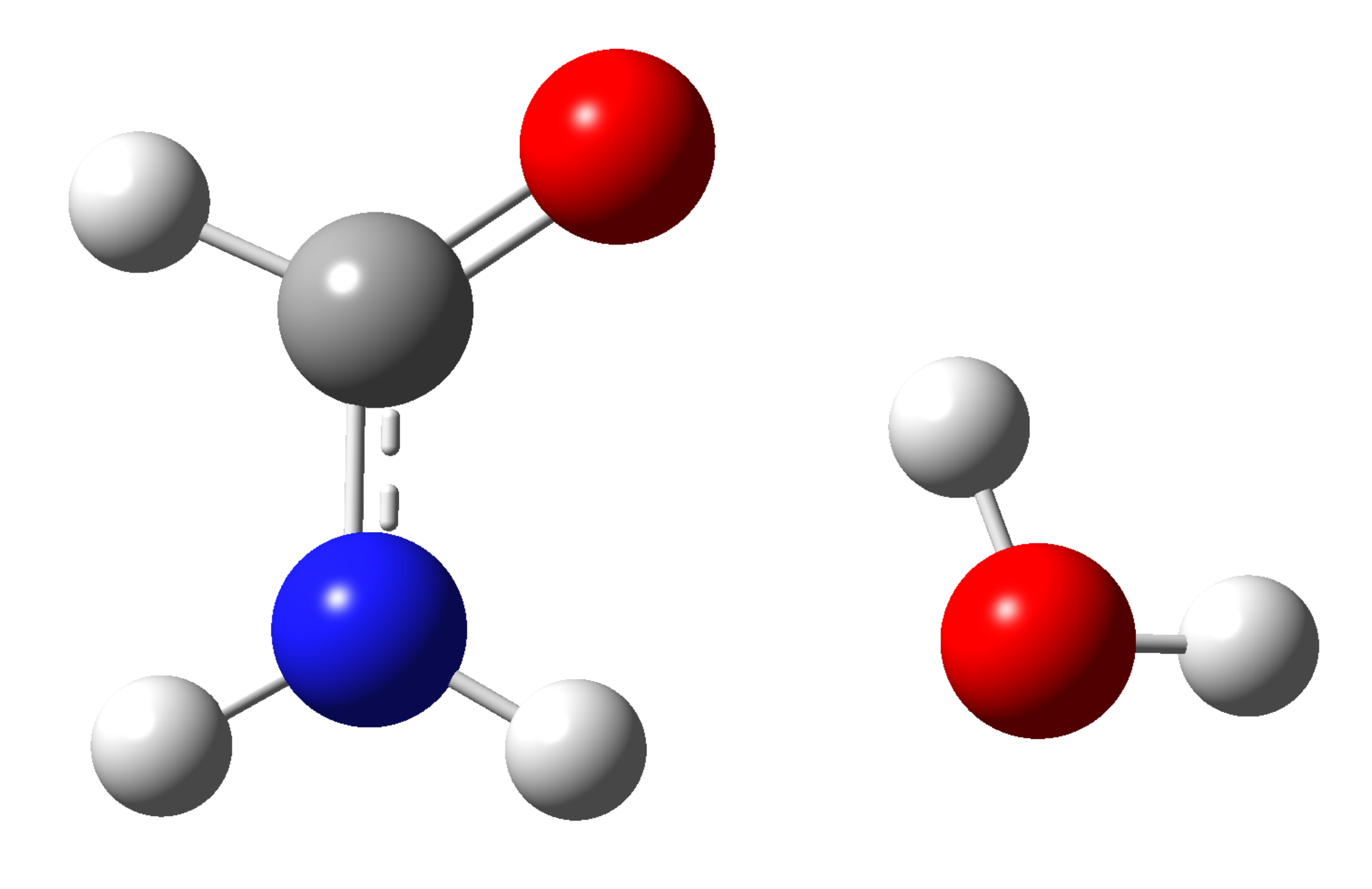}&5079&&&& \\
\hline
\end{tabular}}
\vskip 0.25cm
$^{a}$\cite{wake17}, $^{b}$\cite{quan10}\\
{\bf Notes.} We use our calculated scaled BE values of monomer for astrochemical modeling.
\label{table:BE}
\end{table*}

\begin{table}
\vbox{
\centering{\tiny
\caption{Calculated reaction enthalpies, type of reactions, and activation barriers of various reactions.}
\begin{tabular}{|c|c|c|c|c|}
\hline
{\bf Reactions} & {\bf Reaction} & {\bf Types of} & {\bf Activation} \\
&{\bf enthalpy}&& \\
& {\bf (kcal/mol)} & {\bf reactions} & {\bf barrier (K)} \\
\hline
\hline
\multicolumn{4}{|c|}{\bf Ice phase reactions} \\
\hline
\hline
$\rm{NH + CO \rightarrow HNCO}$ & -139.19 & Exothermic & 4200$^{a}$ \\
$\rm{CH + NO \rightarrow HCNO}$ & -124.62 & Exothermic & 4691 \\
$\rm{CN + OH \rightarrow HOCN}$ & -120.35 & Exothermic & 4857 \\
$\rm{CN + OH \rightarrow HONC}$ & -59.32 & Exothermic & 9855 \\
\hline
\hline
$\rm{HNCO + H \rightarrow H_2NCO}$ & -31.35 & Exothermic & 1962 \\
$\rm{H_2NCO + H \rightarrow NH_2CHO}$ & -92.76 & Exothermic & 0 \\
$\rm{H_2NCO + H \rightarrow HNCO + H_2}$ & -73.15 & Exothermic & 0 \\
\hline
$\rm{HCNO + H \rightarrow H_2CNO}$ & -57.29 & Exothermic & 1073 \\
$\rm{H_2CNO + H \rightarrow HCNO + H_2}$ & -47.20 & Exothermic & 0 \\
\hline
\hline
\multicolumn{4}{|c|}{\bf Gas phase reactions} \\
\hline
\hline
$\rm{HNCO + O \rightarrow CO + HNO}$ & -23.70 & Exothermic & - \\
$\rm{HNCO + O \rightarrow OH + OCN}$ & 5.70 & Endothermic & - \\
\hline
$\rm{HCNO + O \rightarrow CO + HNO}$ & -92.52 & Exothermic & - \\
$\rm{HCNO + O \rightarrow OH + CNO}$ & 0.064 & Endothermic & - \\
\hline
$\rm{HOCN + O \rightarrow OH + OCN}$ & -22.90 & Exothermic & - \\
\hline
$\rm{HONC + O \rightarrow OH + CNO}$ & -18.80 & Exothermic & - \\
\hline
\end{tabular}}
\vskip 0.25cm
$^{a}$\cite{himm02}}
\label{table:Enthalpy}
\end{table}

To study the desorption energy (BE) and reaction pathways of three peptide bond-like species HNCO, NH$_2$CHO, and CH$_3$NCO 
and their isomers/precursors, we use Gaussian 09 suite of programs \citep{fris13}.
Recently, \cite{das18} made an extensive effort to estimate the BE
of $100$ interstellar species on water ice surface by applying quantum chemical approach and compared 
their values with the available experimental results. They found that on an average, the computed BE shows larger deviation from experiments when they considered a single water molecule as a substrate. 
The deviation is minimum when they used pentamer or hexamer configuration of water cluster.
They provided a scaling factor for the extrapolation as the computation was performed with smaller water structures. We carry out quantum 
chemical study to find out the BEs of the three peptide bond related species considered in this work along with their potentially observable 
isomers. To estimate the BEs of these species, we have used similar method and basis set (MP2/aug-cc-pVDZ) as mentioned in \cite{sil17} and \cite{das18}.
Our calculated BE values are given in Table \ref{table:BE}. For some cases, we have found multiple 
probable sites for the adsorption and thus obtained multiple BE values. 
In that case, we take the average of the multiple BEs. Calculated BEs with the single water molecule are 
then scaled up by a factor of $1.416$ \citep{das18} to have the realistic estimation. 
Additionally, in Table \ref{table:BE} we present the BE values for some of these species with the
hexamer configuration of water cluster. Since the BE with the pentamer/hexamer configuration show minimum deviation \citep{das18}, one can use 
these BEs values in the model without scaling. We are unable to provide the BE values of all the
species with the hexamer configuration and with all the probable sites of adsorption. Thus for the modeling, we use BE values obtained
with single water molecule with appropriate scaling.

For the formation of ice phase HNCO, \cite{quen18} considered the reaction between NH and CO. They considered
an activation barrier of $4200$ K for this reaction \citep{himm02}. For the formation of its other isomers no such 
reactions were available. Due to this reason, for the sake of completeness, here, we run quantum chemical
calculation to check the reaction enthalpy of the following reactions.
\begin{equation}
\rm{NH+CO\rightarrow HNCO},\\ 
\end{equation}
\begin{equation}
\rm{CH+NO\rightarrow HCNO},\\ 
\end{equation}
\begin{equation}
\rm{CN+OH\rightarrow HOCN},\\
\end{equation}
\begin{equation}
\rm{CN+OH\rightarrow HONC}.\\ 
\end{equation}

{ We have found that the above four reactions are exothermic in nature. 
Exothermicity values are given in Table \ref{table:Enthalpy}.
The activation barrier for the reaction between NH and CO was known to be $4200$ K but for the others it was unknown.
The above reactions are mostly between radicals and finding a true transition state is a difficult task. 
Instead, we have calculated the reaction enthalpy of these four reactions. Based on the reaction enthalpies, 
we have prepared the most probable reaction sequence in between these four reactions. Since the activation barrier 
of the first reaction was known to be $4200$ K, we scaled the activation barriers of the rest of the reactions. 
Though the reaction enthalpy (exothermicity values) is not directly related to the activation barrier of the 
reaction but it is eventually a better-educated approximation rather than using any other crude approximation. 
Scaled activation barriers are provided in Table \ref{table:Enthalpy}.} 

\cite{quen18} studied the peptide bond related molecules in protostar (IRAS 16293-2422) and pre-stellar 
core (L1554) by using chemical model. Earlier, it was claimed that HNCO and $\rm{NH_2CHO}$ are 
chemically linked because $\rm{NH_2CHO}$ could be formed by the successive hydrogenation
reactions of HNCO ($\rm{HNCO\rightarrow H_2NCO\rightarrow NH_2CHO}$) \citep{mend14,lope15,song16,lope19}. 
The first step of this hydrogenation sequence have the activation barrier of $1962$ K and the second step is a radical-radical 
reaction and thus could be barrier-less. Recent experimental study by \cite{nobl15} and \cite{fedo15} questions this fact. { They opposed
the formation of $\rm{NH_2CHO}$ by the reaction between $\rm{H_2NCO}$ and hydrogen, rather they proposed that eventually
it would return back to HNCO again ($\rm{H_2NCO+H\rightarrow HNCO+H_2}$).}  
Here, we have considered only the formation of $\rm{NH_2CHO}$ in our ice phase network. 
In order to continue a comparative study between the various isomers of HNCO, we are interested to check the hydrogenation 
reactions with the various isomeric forms of HNCO. Thus, we have studied the reaction enthalpies of the following reactions:

\begin{equation}
\rm{HNCO+H\rightarrow H_2NCO},\\
\end{equation}
\begin{equation}
\rm{HCNO+H\rightarrow H_2CNO},\\
\end{equation}
\begin{equation}
\rm{HOCN+H\rightarrow H_2OCN},\\
\end{equation}
\begin{equation}
\rm{HONC+H\rightarrow H_2ONC}.\\
\end{equation}

However, no valid neutral structure for $\rm{H_2OCN}$ and $\rm{H_2ONC}$ were obtained and thus we did not consider the
last two hydrogenation reactions of this sequence. In Table \ref{table:Enthalpy}, we summarize the obtained reaction
enthalpies of the reactions 12 and 13. Based on the obtained reaction enthalpy for the second reaction with respect to the 
first reaction, we scale the activation barrier of the second reaction to $1073$ K.

Recently, \cite{haup19} proposed the successive hydrogen abstraction reactions to $\rm{NH_2CHO}$ for the formation of HNCO:
\begin{equation}
\rm{NH_2CHO+H\rightarrow H_2+H_2NCO},\\ 
\end{equation}
\begin{equation}
\rm{H_2NCO+H\rightarrow H_2+HNCO}.\\ 
\end{equation}
They pointed out that reaction 16 has an activation barrier of $240-3130$ K depending on the level of
theory used for the quantum chemical calculation. They found that the reaction 17 is barrier-less. 
This reaction is very interesting as it might supports the earlier 
claim of chemical linkage between HNCO and ${\rm NH_2CHO}$. 
They also performed quantum chemical calculations for the hydrogen addition reactions to H$_2$NCO and HNCO:
\begin{equation}
\rm{H + H_2NCO\rightarrow NH_2CHO},
\end{equation}
\begin{equation}
\rm{H+HNCO\rightarrow H_2NCO}.
\end{equation}
They found that reaction 18 is barrier-less whereas the reaction 19 is having an activation barrier $2530-5050$ K depending on the level of theory used for the computation.

For the computation of the gas phase reaction rate of these
four reactions, we have used,
\begin{equation}
{\rm  rate=\alpha \Big(\frac{T}{300}\Big)^{\beta}exp{({-\gamma/T})},}
\end{equation}
where $\alpha$, $\beta$, and $\gamma$ are the three constants of the reaction. We have considered $\alpha=10^{-10}$, $\beta=0$ and 
$\gamma=240-3130$ for reaction 16. For the reaction 17 and 18, we have considered $\alpha=10^{-10}$, $\beta=0$, and $\gamma=0$ and for reaction 19, we have considered $\alpha=10^{-10}$, $\beta=0$, and $\gamma=2530-5050$. 
Since a valid structure for H$_2$CNO was obtained, we have considered the reaction $\rm{H +H_2CNO \rightarrow HCNO + H_2}$ in both gas and ice phases. 

\cite{quen18} used the gas phase destruction of HOCN, HCNO, HONC by the oxygen atom. For all the three destruction
reactions they considered an activation barrier of $195$ K. However, \cite{quan10} considered the activation barrier
of $2470$, $195$ and $3570$ K respectively for these three destruction reactions by oxygen atom and these are the default
in the UMIST 2012 network. Here, we consider the default destruction reactions as it was used in UMIST 2012. 
For the destruction of HNCO by the oxygen atom, no reaction was considered. In this effect, we calculate 
the reaction enthalpies for the reactions $\rm{HNCO+O\rightarrow CO+ HNO}$ 
and $\rm{HNCO+O\rightarrow OH+ OCN}$. We have found that the second reaction in this sequence is endothermic whereas the first one is
exothermic and thus we are not considering the second one. $\rm{HNCO+O\rightarrow CO+ HNO}$ is very similar to  $\rm{HCNO+O \rightarrow CO +HNO}$ 
for which a $195$ K activation barrier was considered in UMIST 2012. Based on the exothermicity values between $\rm{HCNO + O}$ and $\rm{HNCO + O}$, 
we have used a scaling factor and obtained an activation barrier of $765$ K for $\rm{HNCO + O}$. 
Calculated reaction enthalpies and the activation barriers are noted in Table \ref{table:Enthalpy}.

\begin{figure*}
\vbox{
\includegraphics[height=10cm, width=10cm, angle=-90]{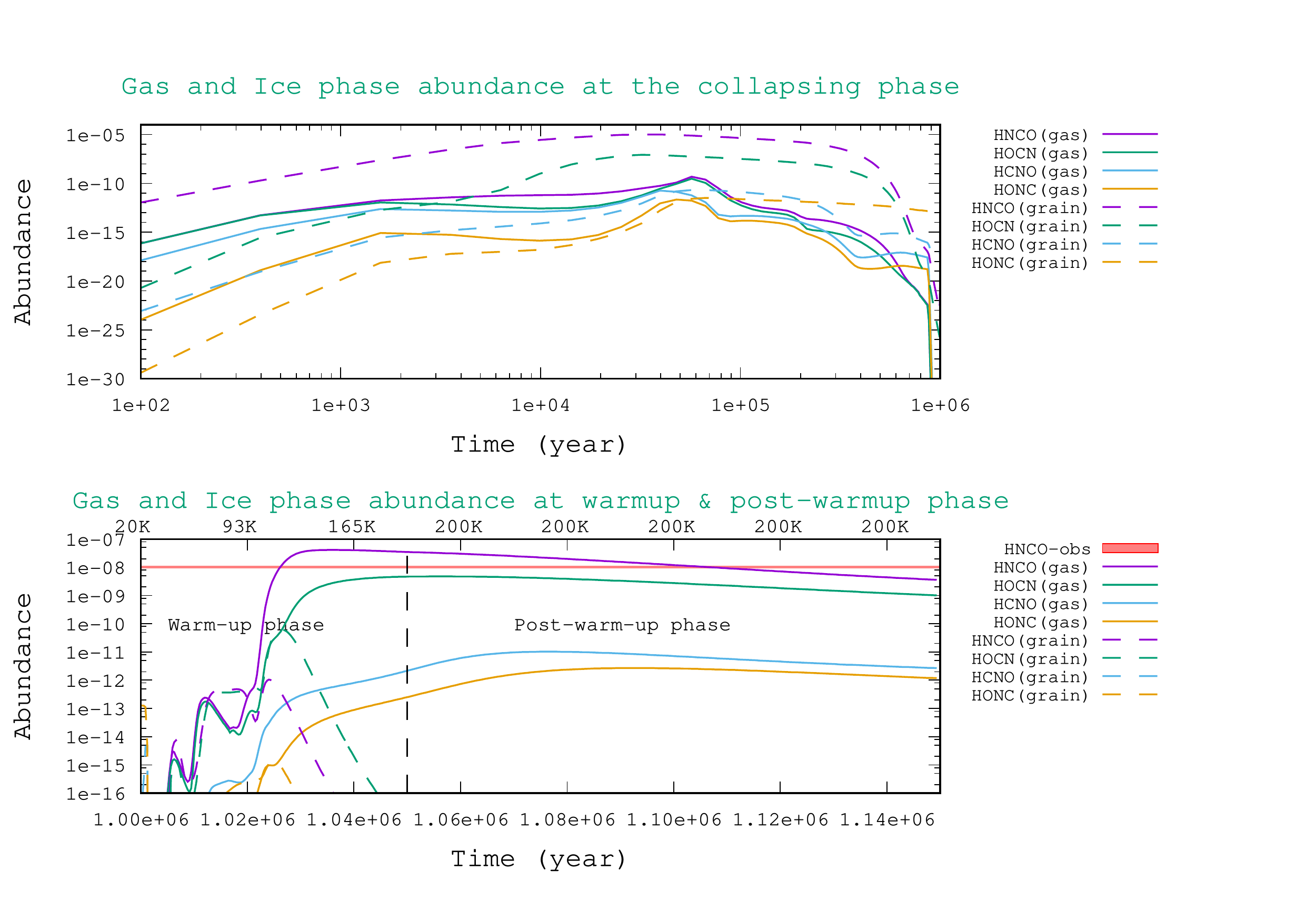}
\includegraphics[height=10cm, width=10cm, angle=-90]{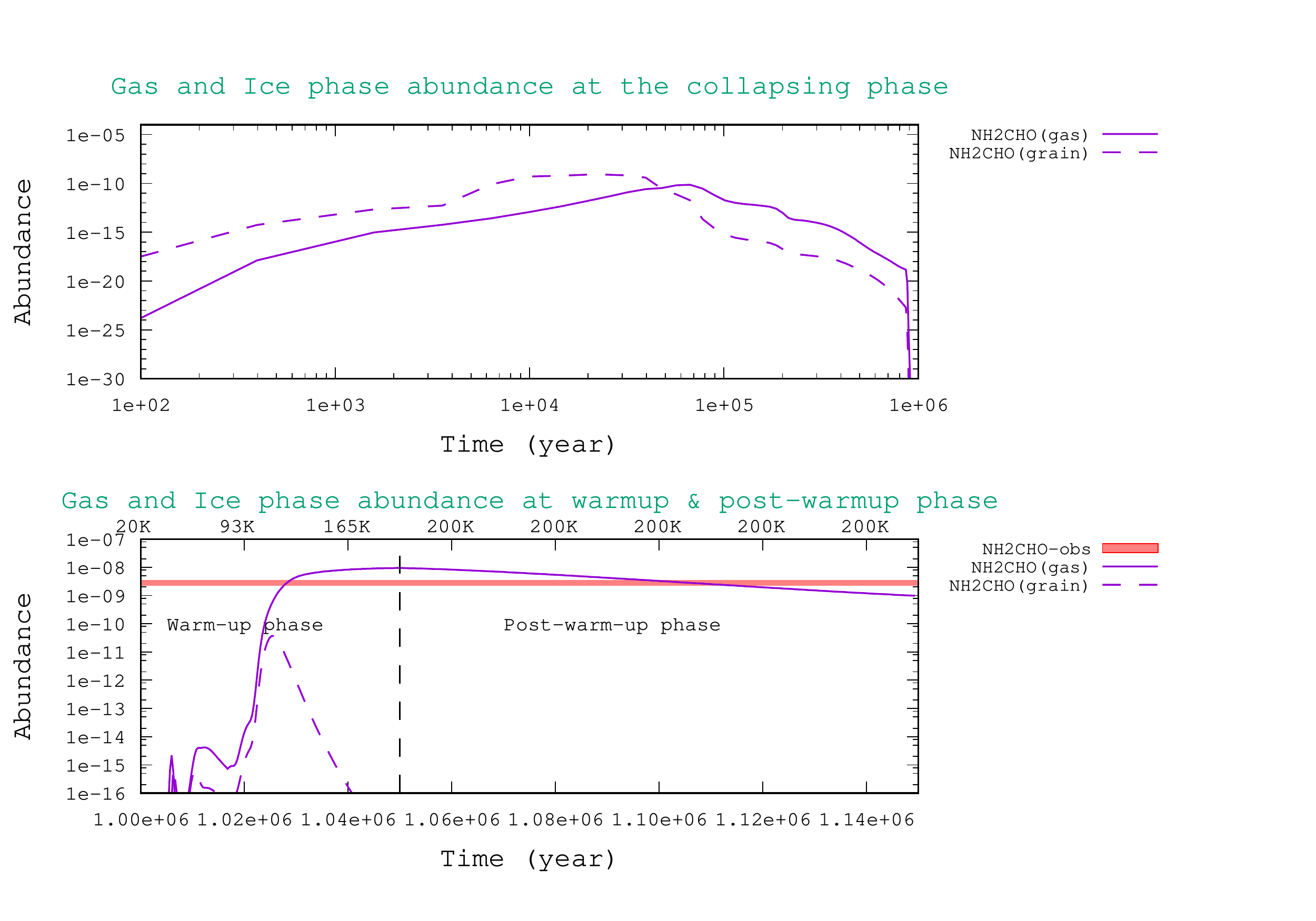}
\includegraphics[height=10cm, width=10cm, angle=-90]{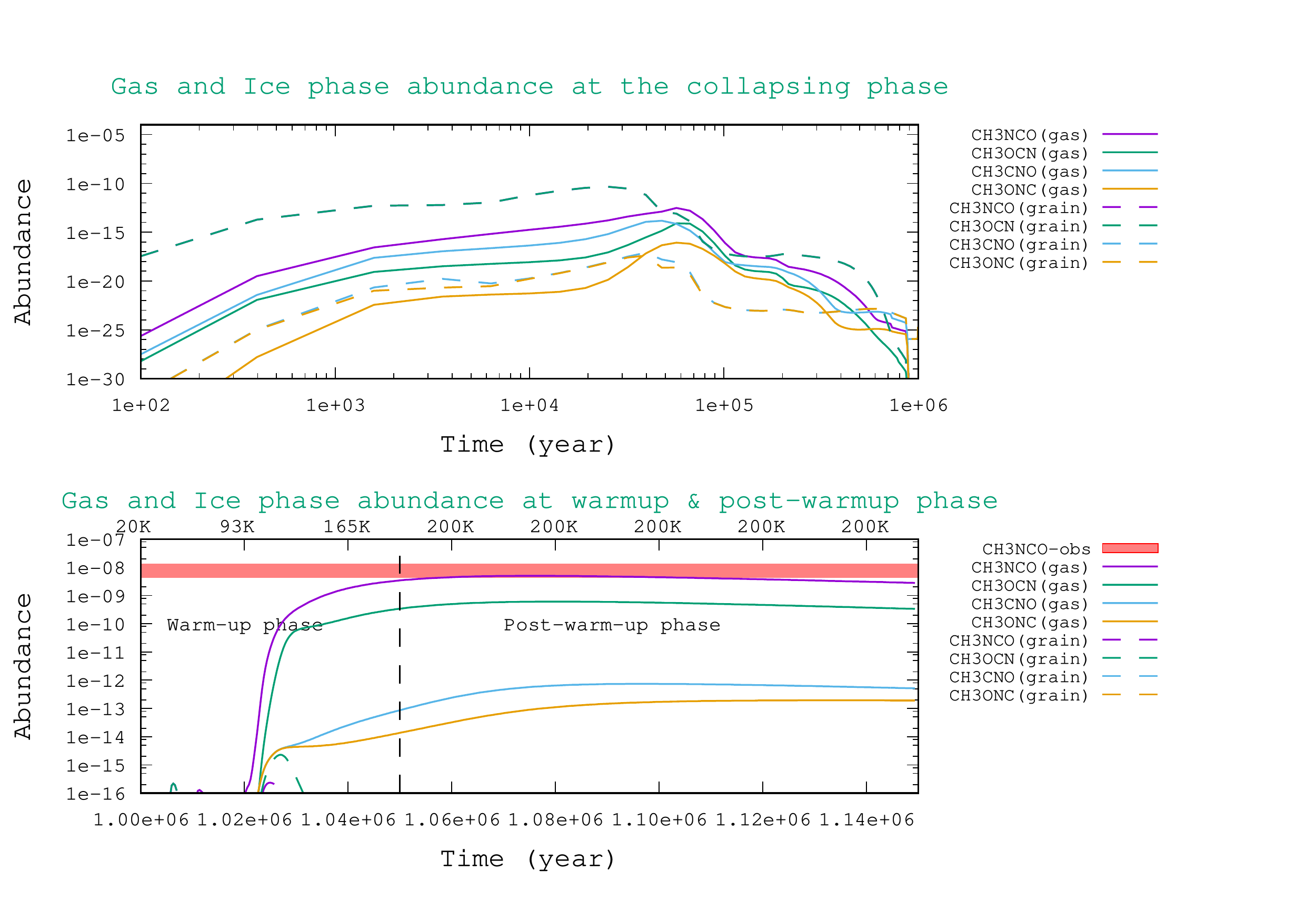}}
\caption{Chemical evolution of the three peptide bond related molecules (HNCO, NH$_2$CHO, CH$_3$NCO) and 
their isomers (if any). This is shown for $\rho_{max}=1.0 \times 10^7$ cm$^{-3}$ and $T_{ice}=20$ K by considering the best-fitted parameters with Model A. Red shaded lines represent the observed abundances obtained in G10.}
\label{fig:ModelA-isomer}
\end{figure*}

\begin{figure}
\vbox{
\includegraphics[height=10cm, width=9cm, angle=-90]{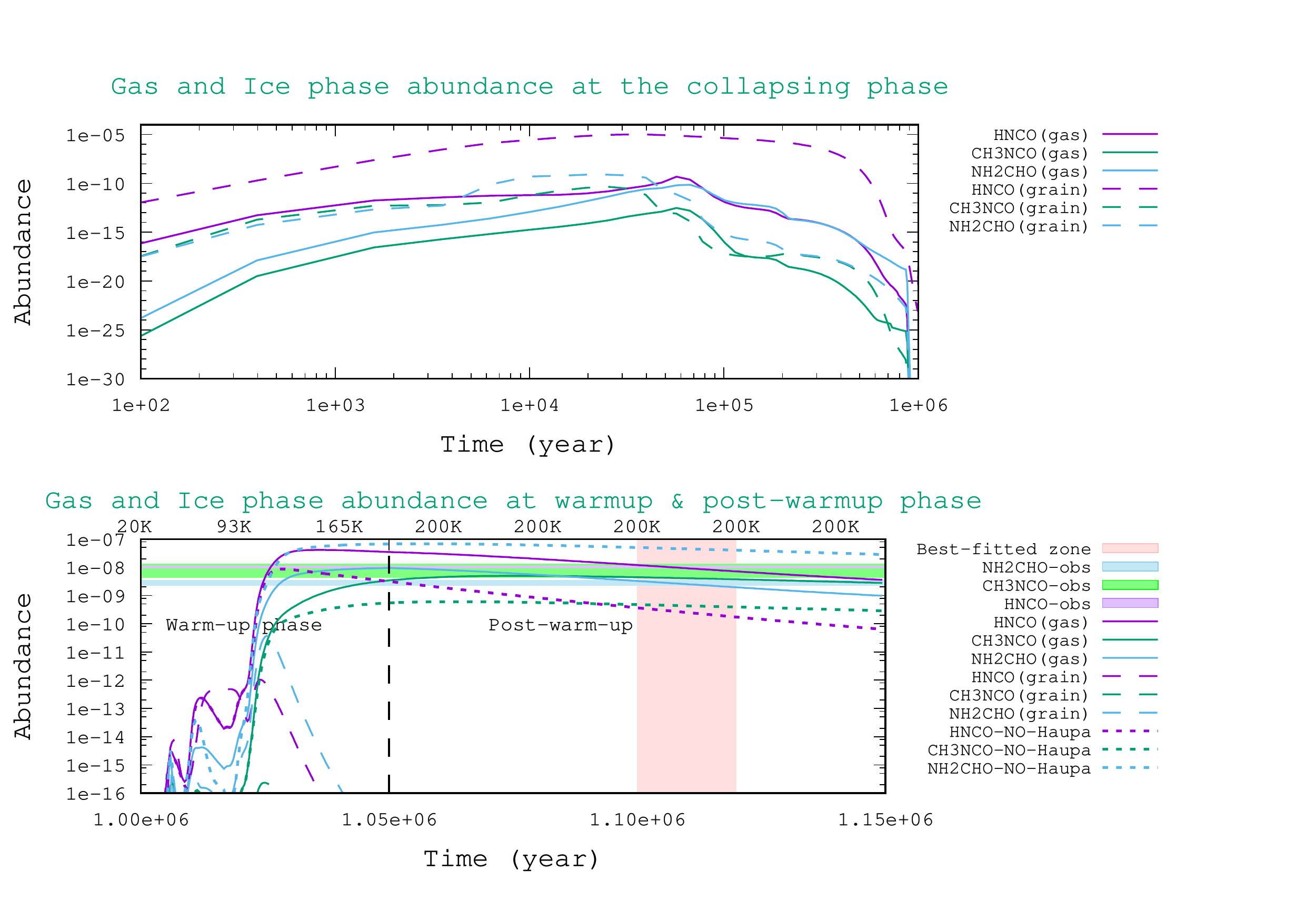}
\caption{ 
Chemical evolution of HNCO, CH$_3$NCO, and NH$_2$CHO during the three phases by considering the best-fit parameters of Model A. Best-fitted time zone is also highlighted.
Abundance variation by avoiding \cite{haup19} pathways in gas phase is also shown.
}
\label{fig:best}}
\end{figure}

\begin{figure*}
\centering
\vbox{
\includegraphics[height=18cm,angle=-90]{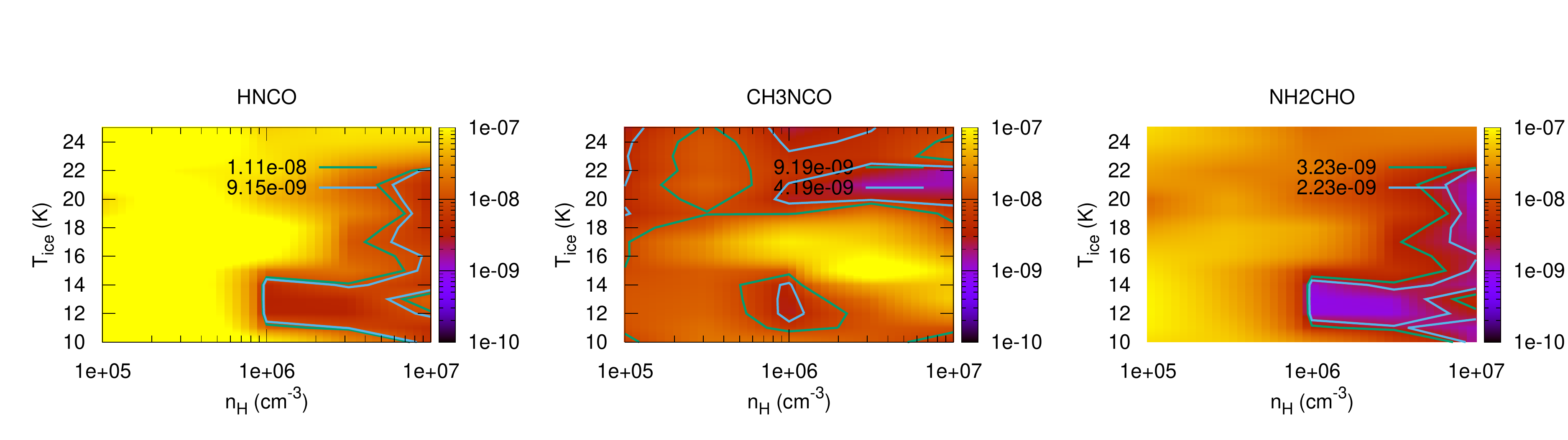}
\caption{ Parameter space obtained with Model A by considering the best-fitted parameters noted in Table \ref{tab:Model} at an age
position $1.12 \times 10^6$ years. Color code in the right side of the each panel represents the abundance with respect to 
H$_2$.\label{fig:param}}}
\end{figure*}

\begin{figure*}
\centering
\vbox{
\includegraphics[width=5cm,height=18cm,angle=-90]{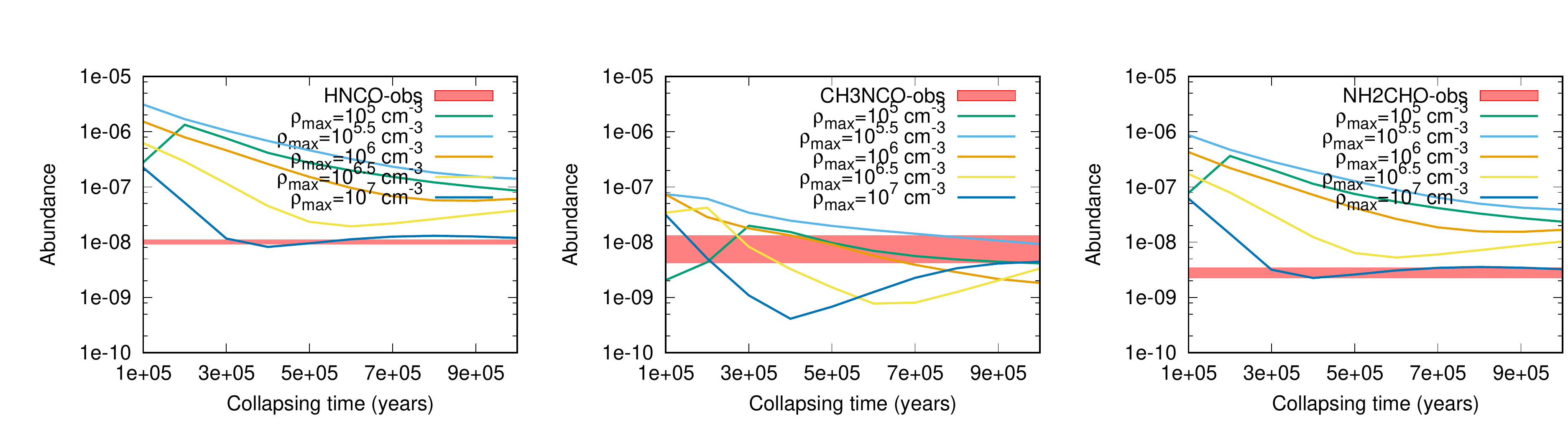}
\caption{ Abundances of HNCO, CH$_3$NCO, and NH$_2$CHO by considering different $T_{coll}$, $\rho_{max}$ with Model B.\label{fig:ModelB-den-time}}}
\end{figure*}

\begin{figure*}
\centering
\vbox{
\includegraphics[height=18cm,angle=-90]{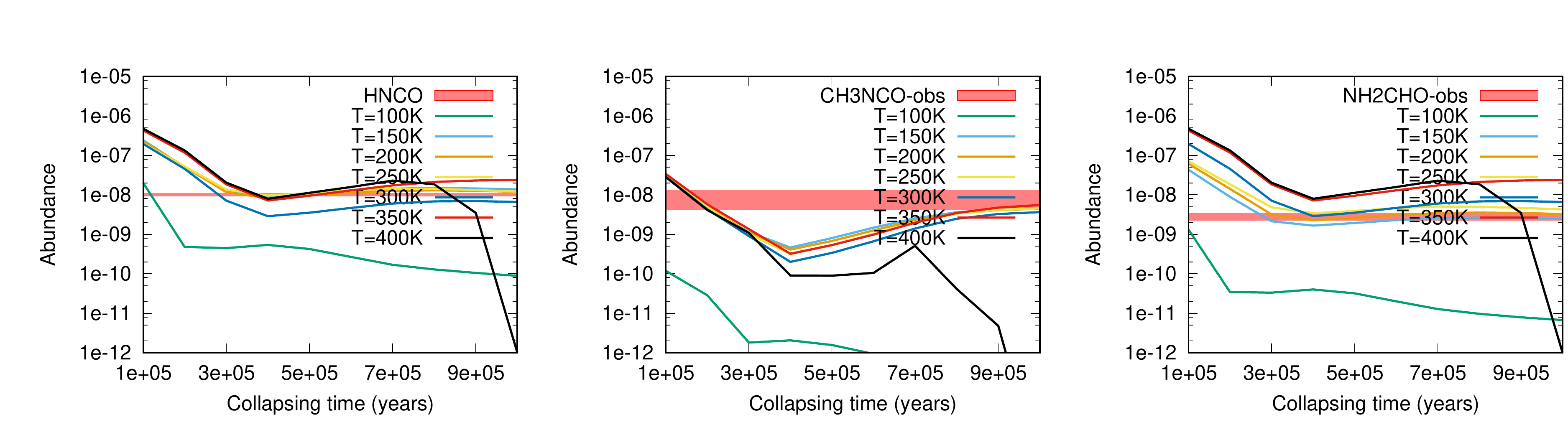}
\caption{ Abundances of HNCO, CH$_3$NCO, and NH$_2$CHO by considering $\rho_{max}=10^7$ cm$^{-3}$, different $T_{max}$ and $T_{coll}$ with Model B.\label{fig:ModelB-temp}}}
\end{figure*}

\subsection{Modeling results}

The observed abundances of HNCO, $\rm{NH_2CHO}$, and $\rm{CH_3NCO}$ are provided in Table \ref{table:abunobs}. From the
chemical modeling, we have seen that
our obtained abundance is very much sensitive to the { physical parameters ($T_{ice}$, $\rm{\rho_{max}}$, $T_{max}$, $t_{coll}$, and $t_{pw}$) and adopted rate constants. 
Here, we have put an extensive effort to find out the simultaneous appearance of these 
three nitrogen-bearing species 
by varying the sensitive physical parameters and rate constants of some of the key reactions.
More precisely, we have prepared two models: Model A and Model B. The difference between the two models 
is highlighted in Table \ref{tab:Model}.}

\begin{deluxetable*}{ccc}
{ 
\tabletypesize{\footnotesize}
\tablewidth{0pt}
\scriptsize
\tablecaption{Key differences between the Model A and Model B. \label{tab:Model}}
\tablehead{\colhead{Physical parameters}&\colhead{Model A}&\colhead{Model B}}
\startdata
${\rm{\rho_{max}}}$ ($cm^{-3}$)&$10^7$&$10^{5-7}$\\
${\rm{T_{max}}}$ (K)&200&100-400\\
$t_{coll}$ (years)&$10^6$&$10^{5-6}$\\
$t_{w}$ (years)&$5 \times 10^4$&$5 \times 10^4$\\
$t_{pw}$ (years)&$6.2-10 \times 10^4$&$10^5$\\
$T_{ice}$ (K)&10-25&20\\
\hline
&Gas phase reactions parameterized&\\
&Gas phase rate constants used in Model A and Model B& Rate constant used in literature\\
\hline
$\rm{NH_2+H_2CO \rightarrow NH_2CHO +H}$&$\alpha=5.00 \times 10^{-12}, \ \beta=-2.56, \ \gamma=4.88$&$\alpha=7.79 \times 10^{-15}, \ \beta=-2.56, \ \gamma=4.88$ \citep{skou17}\\
$\rm{CH_3+HNCO \rightarrow CH_3NCO +H}$&$\alpha= 1.0 \times 10^{-12}, \ \beta=0, \ \gamma=0$&$\alpha= 5 \times 10^{-11}, \ \beta=0, \ \gamma=0$ \citep{quen18}\\
$\rm{H + NH_2CHO \rightarrow H_2NCO+ H}$&$\alpha= 1 \times 10^{-10}, \ \beta=0, \ \gamma=240$&$\alpha=-, \ \beta=-, \ \gamma=240-3130$ \citep{haup19}\\
$\rm{H + H_2NCO \rightarrow HNCO+ H_2}$&$\alpha= 1 \times 10^{-10}, \ \beta=0, \ \gamma=0$&$\alpha=-, \ \beta=-, \ \gamma=0$ \citep{haup19}\\
$\rm{H + H_2NCO \rightarrow NH_2CHO}$&$\alpha= 1 \times 10^{-10}, \ \beta=0, \ \gamma=0$&$\alpha=-, \ \beta=-, \ \gamma=0$ \citep{haup19}\\
$\rm{H + HNCO \rightarrow H_2NCO}$&$\alpha= 1 \times 10^{-10}, \ \beta=0, \ \gamma=5050$&$\alpha=-, \ \beta=-, \ \gamma=2530-5050$ \citep{haup19}\\
\enddata}
\end{deluxetable*}

{ 
\subsubsection{Results obtained with Model A}
To constrain the best possible model, we have explored the parameter space around which the modeling results are in well
agreement with the observational results. In this effect, we run several cases by varying the 
initial dust temperature ($T_{ice}$) 
in between $10$ to $25$ K and $\rho_{max}$ in between $10^5$ to $10^7$ cm$^{-3}$ for Model A.
Since, G10 is a hot core, higher density ($10^6-10^7$ cm$^{-3}$) is preferable. 
It has been earlier pointed out that the G10 region is extended roughly by $0.1$ parsec and has of $\sim10^3$ 
solar mass of matter \citep{cesa94}. From that estimation, the average density of the source is around $10^7$ cm$^{-3}$.
It is also interesting to note that our observational analysis suggests that continuum temperatures vary within $19$ to $27$ K.
Based on the observational results as a preliminary guess, we have used $\rho_{max}=10^7$ cm$^{-3}$ 
and $T_{ice}=20$ K for Model A. Initially, we have started with Model A with the rate constants of the 
gas phase reactions available in the literature \citep{skou17,quen18,haup19}. 
Based on some preliminary iterations of our simulation, we have varied the rate constants of the some key gas phase reactions which are controlling the abundances of the three targeted species.  
We have obtained a nice correlation 
between these three species when the rate constants listed in Table \ref{tab:Model} is used. 
\cite{quen18} considered the reaction between HNCO and CH$_3$ in the ice phase for the formation of CH$_3$NCO and CH$_3$OCN
both the isomers with the same rate. However, for the gas phase formation of other 
isomers of CH$_3$NCO (CH$_3$CNO, CH$_3$OCN, CH$_3$ONC), \cite{quen18} considered some rate coefficients of $\sim 10^{-20}$ and $5 \times 10^{-11}$ cm$^3$ s$^{-1}$. Here, instead of the rate constant $5 \times 10^{-11}$ cm$^3$ s$^{-1}$, we have considered a rate constant $10^{-12}$ cm$^3$ s$^{-1}$ for some gas phase reactions and have used a rate $10^{-20}$ cm$^3$ s$^{-1}$ for those reactions as it was used in \cite{quen18}. For the ice phase formation reactions, we kept it as it was considered
by \cite{quen18}.
To study the abundances of various isomers considered in the network, we thus choose our best fitted parameters listed in 
Table \ref{tab:Model} for Model A. Time evolution of the abundances of HNCO isomers, $\rm{CH_3NCO}$ isomers, and 
$\rm{NH_2CHO}$ is shown in Fig. \ref{fig:ModelA-isomer}. 
Results obtained with the best fitted rate constants are shown separately in Fig. \ref{fig:best} for HNCO, CH$_3$NCO, and NH$_2$CHO. 
It clearly shows that around the age of $\sim 1.12 \times 10^6$ years, we are having a good correlation 
between these three species.
Parameter space obtained with the best-fitted rate constants after a suitable age position ($1.12 \times 10^6$ years) 
is shown in Fig. \ref{fig:param}. 
For the better understanding, abundances closer to the observed 
values are shown with 
the contours. 

Among the other isomers of HNCO, HOCN is found to be significantly abundant (during the warm-up and post-warm-up phase, it
has attained a peak value $4.8 \times 10^{-9}$ for the best fitted parameters of Model 
A). This is of $\sim 10$ times lower than the lowest energy isomer, HNCO (peak abundance $4.13 \times 10^{-8}$). Similarly, in between all the other isomers of 
$\rm{CH_3NCO}$, abundance of $\rm{CH_3OCN}$ is found to be higher. 
This is due to the gas phase formation of CH$_3$OCN 
by the reaction between CH$_3$ and HOCN.
With the best-fitted parameters, we have obtained the peak abundance of $\rm{CH_3OCN}$ as $6.1 \times 10^{-10}$ which is
 of $\sim 8$ times lower than that of the $\rm{CH_3NCO}$ (peak abundance $5.0 \times 10^{-9}$). Here, we have reported the
identification of HNCO and CH$_3$NCO in G10. However, looking at the abundances of HOCN and CH$_3$OCN, it should also be
proposed as potential candidates for the future astronomical detection in G10. 
\cite{cern16} predicted an upper limit of $6 \times 10^{13}$ cm$^{-2}$ 
for the another isomer, CH$_3$CNO in Orion. 
Here, we have found its peak abundance  $7.4 \times 10^{-13}$. Converting this peak abundance in terms of the column density,
we have of $\sim 10^{13}$ cm$^{-2}$ (by using a hydrogen column density $1.35 \times 10^{25}$ cm$^{-2}$) which is in line with the observed upper limit.

\subsubsection{Results obtained with Model B}
For Model B, we did not vary any rate constants. We kept it as it was obtained with the best fitted Model A which is noted in Table \ref{tab:Model}. To find out the best-fit physical parameters for Model B, we have started 
with $T_{ice}=20$ K and have varied $\rho_{max}$. Figure \ref{fig:ModelB-den-time} shows the variation of HNCO, CH$_3$NCO and NH$_2$CHO 
abundance by considering a post warm-up time ($t_{pw}$) of $10^5$ years. Observed abundances are also marked in each panel. 
We found that the abundance of these three species is highly sensitive on the chosen collapsing time scale ($t_{coll}$) and the 
maximum density ($\rho_{max}$) achieved during the collapsing phase. 
As we have increased $\rho_{max}$, abundance significantly decreased. 
Similarly, as we have increased $t_{coll}$, 
the abundances gradually decreased. 
Based on Figure \ref{fig:ModelB-den-time}, we found that $\rho_{max}=10^7$ cm$^{-3}$ and 
$T_{coll} \sim 2-3 \times 10^5$ years are most suitable to explain the abundance of these three species 
simultaneously. We further have varied $T_{max}$ in between $100-400$ K 
by considering $\rho_{max} =10^7$ cm$^{-3}$. Fig. \ref{fig:ModelB-temp} shows that an increase in 
$T_{max}$ from $100$ K to $150$ K shows a strong increasing trend in abundance profile. 
We have a reasonable match when we have used $T_{max}=200$ K. In between $T_{max}=150-350$ K abundance profile
shows moderate changes. Beyond $350$ K the abundances drastically decreased while we have considered a comparatively
longer collapsing time scale. 
Thus, by considering all types of variation with Model B, we have obtained a good fit between the three targeted
nitrogen-bearing species when we have used the parameters listed in Table \ref{tab:Model}. 
We found that our model B with $T_{coll}=(2-3) \times 10^5$ years, $T_w=5 \times 10^4$ years and $T_{pw}=10^5$ years can be able to 
explain the observation of these three species simultaneously when we have considered $\rho_{max}=10^7$ cm$^{-3}$, $T_{max}=200$ K and $T_{ice}=20$ K. 
Obtained lower time scale with Model B is very interesting because G10 is a high-mass star forming region.
Gas phase pathways required to establish the linkage between these three species 
are summarized in Fig. \ref{fig:path}}.

\begin{figure}
\centering
\vbox{
\includegraphics[height=8cm,angle=0]{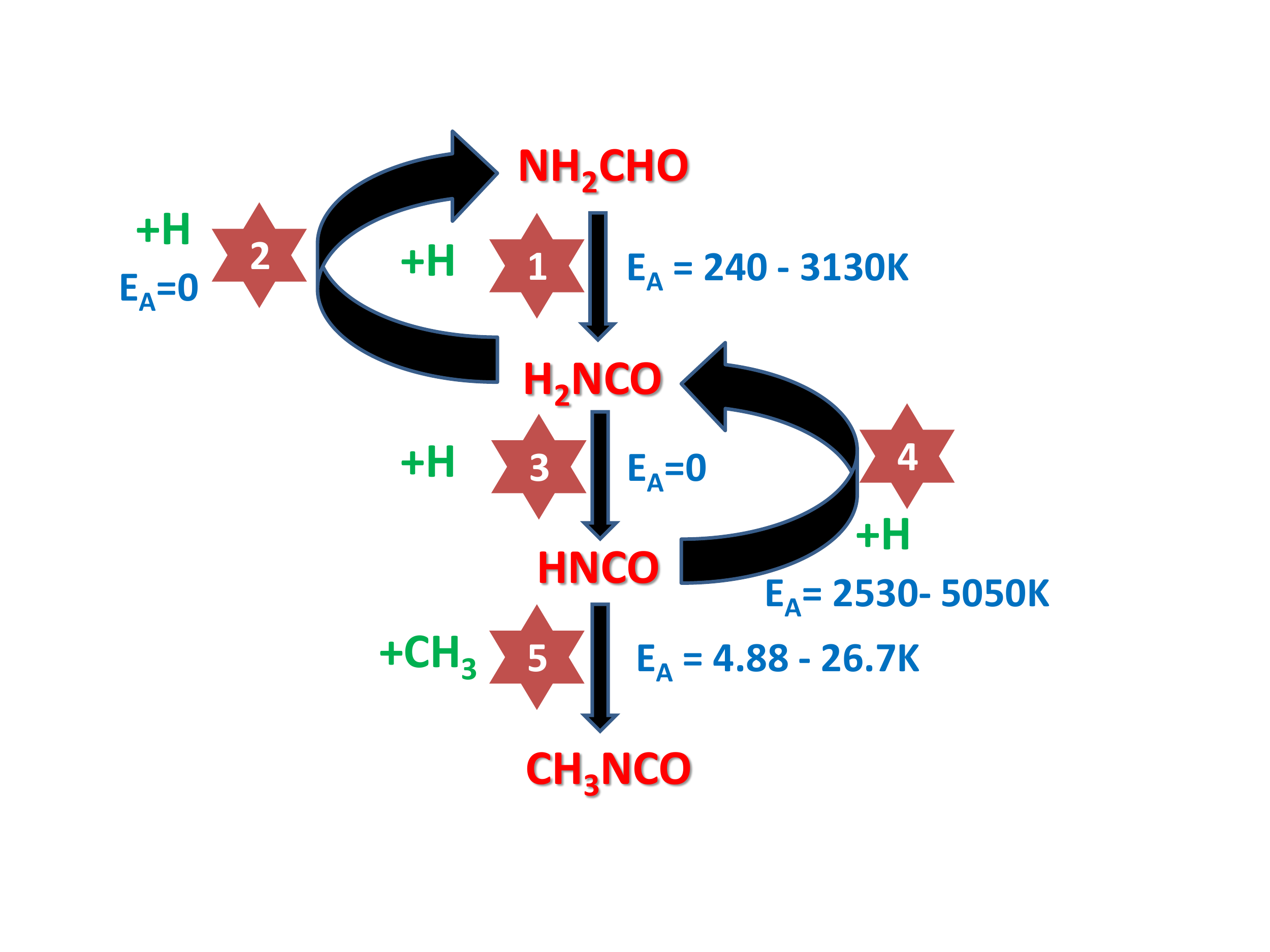}
\caption{ Chemical linkage between three nitrogen bearing molecules.\label{fig:path}}}
\end{figure}

\subsubsection{Chemical linkage between $\rm{HNCO}$, $\rm{NH_2CHO}$, and $\rm{CH_3NCO}$}
Earlier it was proposed that HNCO and NH$_2$CHO are chemically linked. The successive hydrogenation reactions of HNCO 
was proposed for the 
formation of $\rm{NH_2CHO}$. However, the validity of the second hydrogenation reaction is ruled out by the experimental study \citep{nobl15}. 
Recently a theoretical work by \cite{haup19} proposed dual-cyclic hydrogen addition and abstraction reactions to support the
chemical linkage between HNCO and NH$_2$CHO.
The chemical evolution of HNCO, $\rm{NH_2CHO}$, and CH$_3$NCO with Model A are shown in Fig. \ref{fig:best}.
{ Gradual enhancement in the abundance of ice phase HNCO and its isomers arises because radicals become 
mobile enough with the increase in temperature. Beyond $80-90$ K, the diffusion time scale of the radicals become comparable 
to their desorption time scale and thus desorbed back to the gas phase very quickly. Also, 
HNCO starts to sublimate beyond $90$ K 
and resulting in a sharp decrease in the ice phase.}
{ The gas phase production of CH$_3$NCO mainly occurs by the reaction between 
CH$_3$ and HNCO. The formation rate of CH$_3$NCO enhances during the later phases of the simulation.} 
In the case of $\rm{NH_2CHO}$, ice phase production is sufficient in the collapsing phase,
but gas phase production is not adequate. In the warm-up period, a smooth transfer of $\rm{NH_2CHO}$ from the
ice phase to the gas phase can occur. The location of this transfer depends on the adopted BE of $\rm{NH_2CHO}$.
In the warm-up and post-warm-up phase,  major portion of $\rm{NH_2CHO}$ is formed by the gas phase reaction
between $\rm{NH_2}$ and $\rm{H_2CO}$. 
Due to the increased temperature, activation barrier for the hydrogen abstraction reaction of $\rm{NH_2CHO}$ 
(by reaction 16) become probable and thus produce HNCO by reaction 17 by the barrier-less reaction. 
To check the
effects of the addition of the \cite{haup19} pathways, we have checked with $\alpha=0$ for the gas phase
reactions 16-19. Figure \ref{fig:best}  shows the abundances of these three species by considering $\alpha=0$ (marked
as "NO-Haupa").
We have noticed that the abundance of gas phase
HNCO is significantly affected with the inclusion of the gas phase pathways of \cite{haup19}. 
Consideration of reactions $16-19$ shows 
more $\rm{HNCO}$ at the end and absence of these pathways (i.e., with $\alpha=0$) reflects  comparatively
lower $\rm{HNCO}$.
In brief, we have found that the pathways proposed by \cite{haup19} are relevant for the gas phase production of HNCO around the post-warm-up period.
In the first phase, $\rm{CH_3NCO}$ is mainly formed in the grain surface by the reaction between CH$_3$ and OCN.
$\rm{CH_3NCO}$ also have formed by the reaction between $\rm{CH_3}$ and HNCO \citep{quen18} in the ice phase. 
{ However, it is clear from the
warm-up and post-warm-up phase that the major contribution of the gas phase $\rm{CH_3NCO}$ is not coming from the ice phase; instead it is producing inside the
gas phase itself. The gas phase formation is efficient by the HNCO channel at the warm-up and post-warm-up phase.}

\section{Conclusions \label{sec:conclusions}}

\begin{itemize} 
\item We identified three molecules HNCO, NH$_2$CHO, and CH$_3$NCO in G10 which contain peptide-like bond. 
Earlier HNCO and $\rm{NH_2CHO}$ had been
identified in G10, but this is the first identification of $\rm{CH_3NCO}$ in this source.

\item We estimated the hydrogen column density of this source to be { $\rm{N_{H_2} = 1.35\times10^{25} \ cm^{-2}}$}. Our
estimated optical depth is { $0.136$} which suggest that the dust is optically thin. Kinetic temperatures of the gas 
is found to vary between { $248$ to $439$} K. We estimated the
column densities and fractional abundances of three observed peptide like bond containing molecules.

\item From the obtained spatial distribution of these three species we speculated that they are chemically linked. 
Since all the transitions were marginally resolved, one need to have high angular and spatial resolution data
to make a rigorous comment on their spatial distribution in G10.

\item From our chemical modeling results, we also noticed that these three species are chemically linked. 
We found that HNCO and NH$_2$CHO are 
chemically linked by a dual-cyclic hydrogen addition and abstraction reactions proposed by \cite{haup19} during the warm-up and post-warm-up phase. HNCO and 
CH$_3$NCO are also chemically related because HNCO reacts with CH$_3$ to form $\rm{CH_3NCO}$ (Fig. \ref{fig:path}).

\item Our modeling results suggest that the abundance of HOCN and $\rm{CH_3OCN}$ are significantly higher and could be observed in G10.

\end{itemize}

\acknowledgments
 This paper makes use of the following ALMA data: ADS/JAO.ALMA\#2016.1.00929.S. ALMA is a partnership of ESO (representing its member states), NSF (USA) and 
 NINS (Japan), together with NRC (Canada), MOST and ASIAA (Taiwan), and KASI (Republic of Korea), in cooperation with the Republic of Chile. The Joint ALMA 
 Observatory is operated by ESO, AUI/NRAO and NAOJ. P.G. acknowledges CSIR extended SRF fellowship (Grant No. 09/904 (0013) 2018 EMR-I). B.B. acknowledges 
 DST-INSPIRE Fellowship [IF170046] for providing partial financial assistance. M.S. acknowledges DST, the Government of India, for providing financial 
 assistance through the DST-INSPIRE Fellowship [IF160109] scheme. A.D. and R.G. acknowledge ISRO respond project (Grant No. ISRO/RES/2/402/16-17) for 
 partial financial support. S.K.M. acknowledges CSIR fellowship (Ref no. 18/06/2017(i) EU-V). This research was possible in part due to a Grant-In-Aid 
 from the Higher Education Department of the Government of West Bengal. { We would like to thanks the reviewer whose extensive comments help in improving the quality of this manuscript}.

%

\newpage
\appendix

\begin{figure}
\begin{minipage}{0.35\textwidth}
\includegraphics[width=\textwidth]{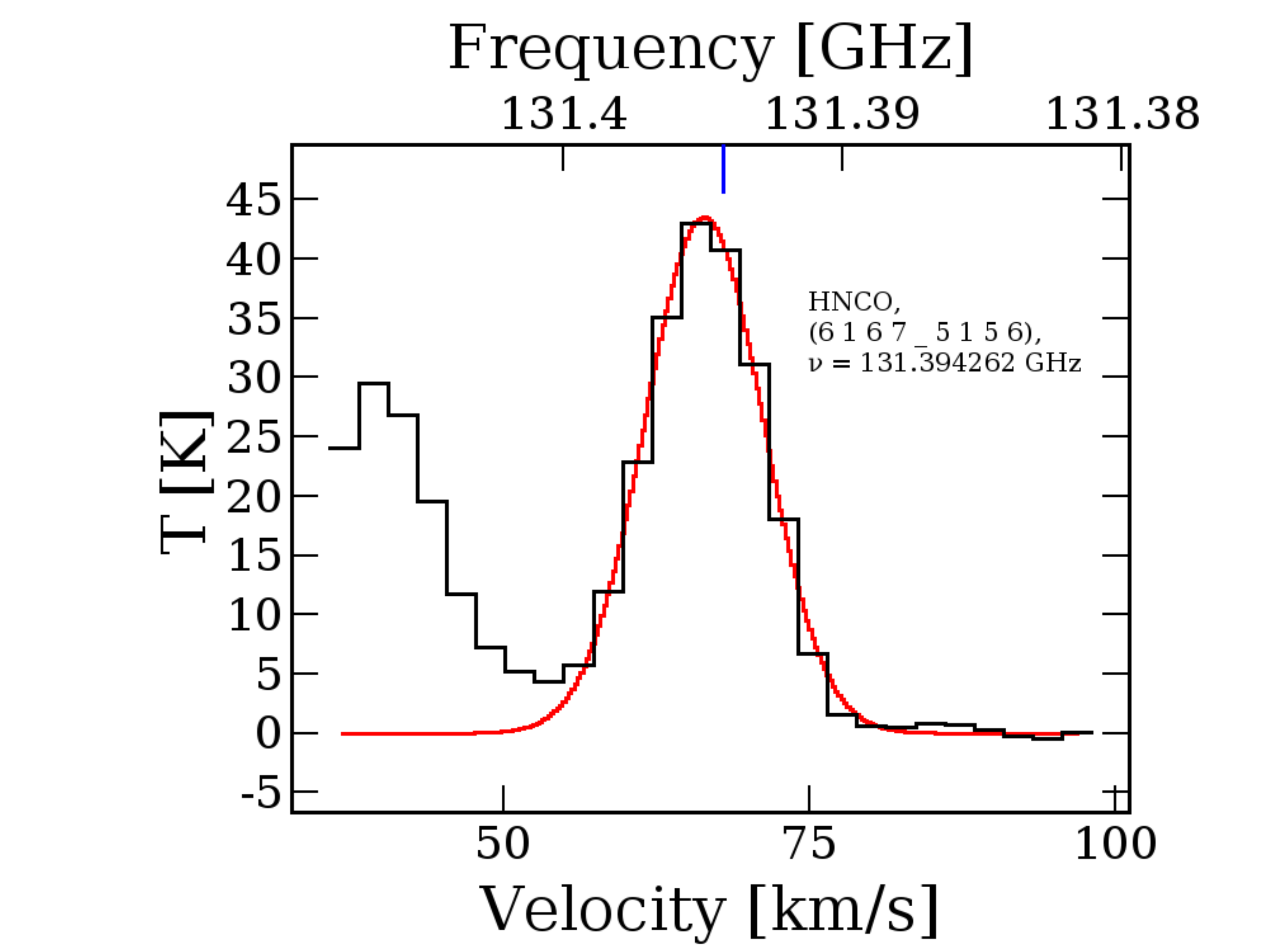}
\end{minipage}
\begin{minipage}{0.35\textwidth}
\includegraphics[width=\textwidth]{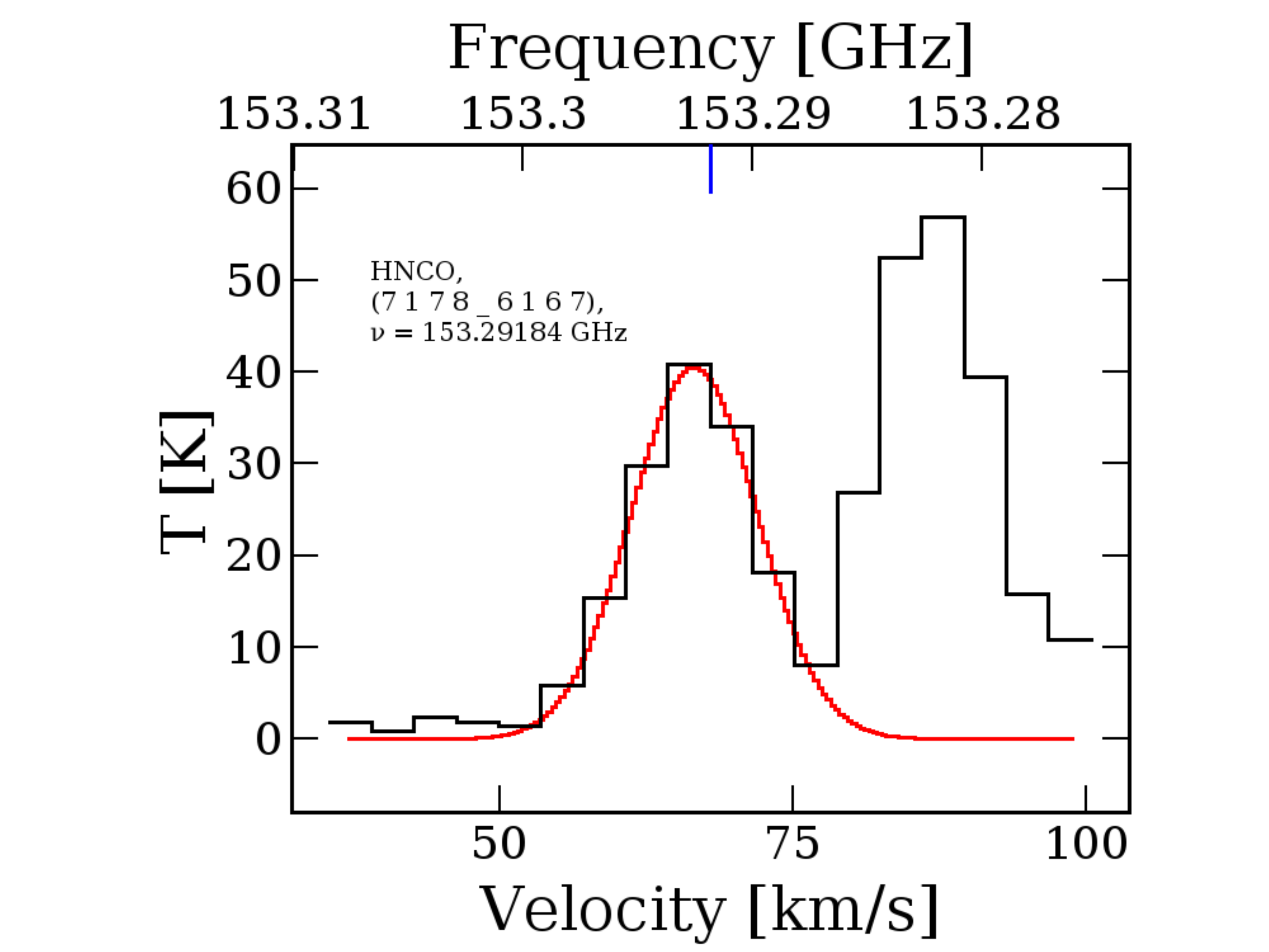}
\end{minipage}
\begin{minipage}{0.35\textwidth}
\includegraphics[width=\textwidth]{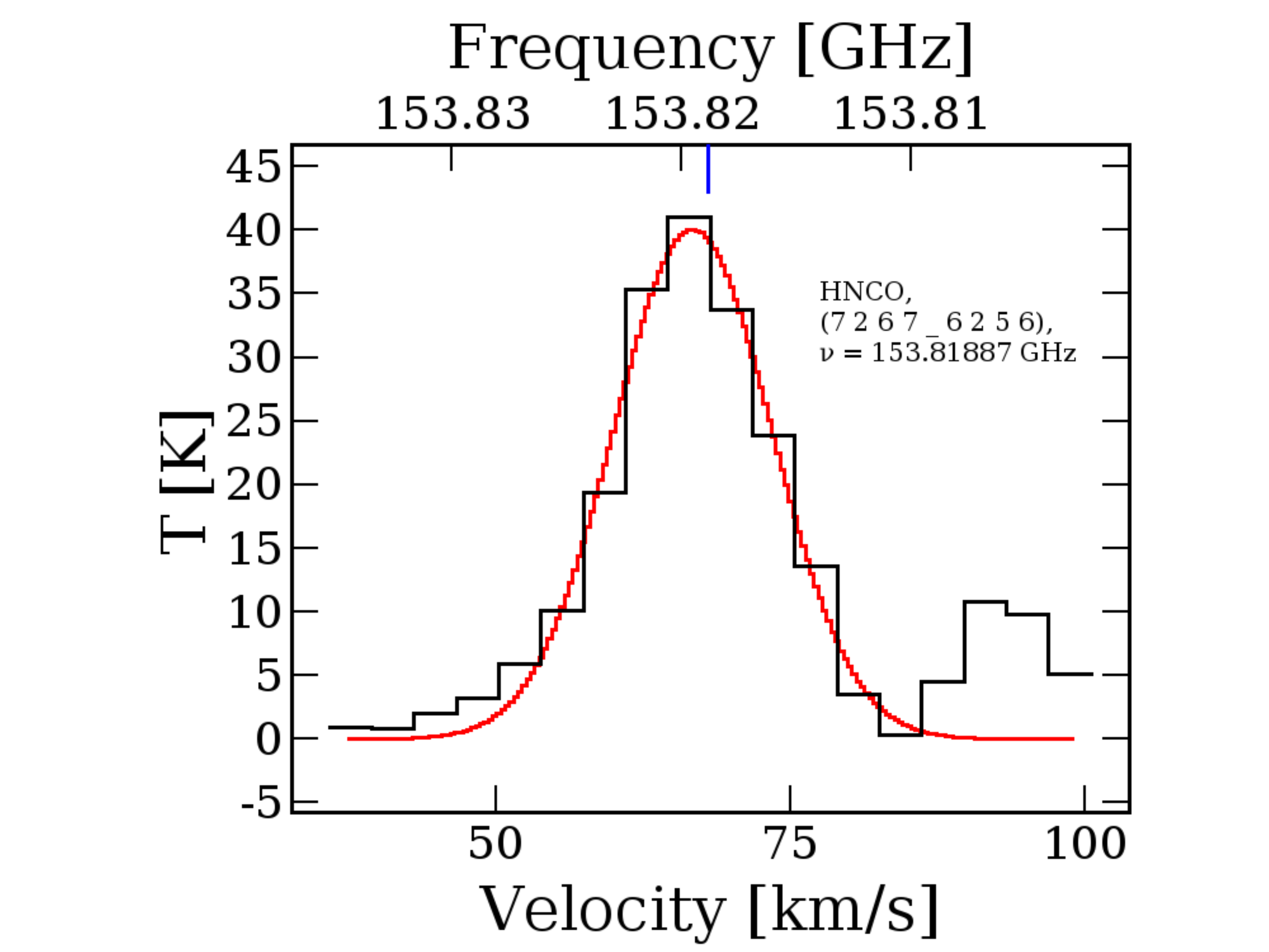}
\end{minipage}
\begin{minipage}{0.35\textwidth}
\includegraphics[width=\textwidth]{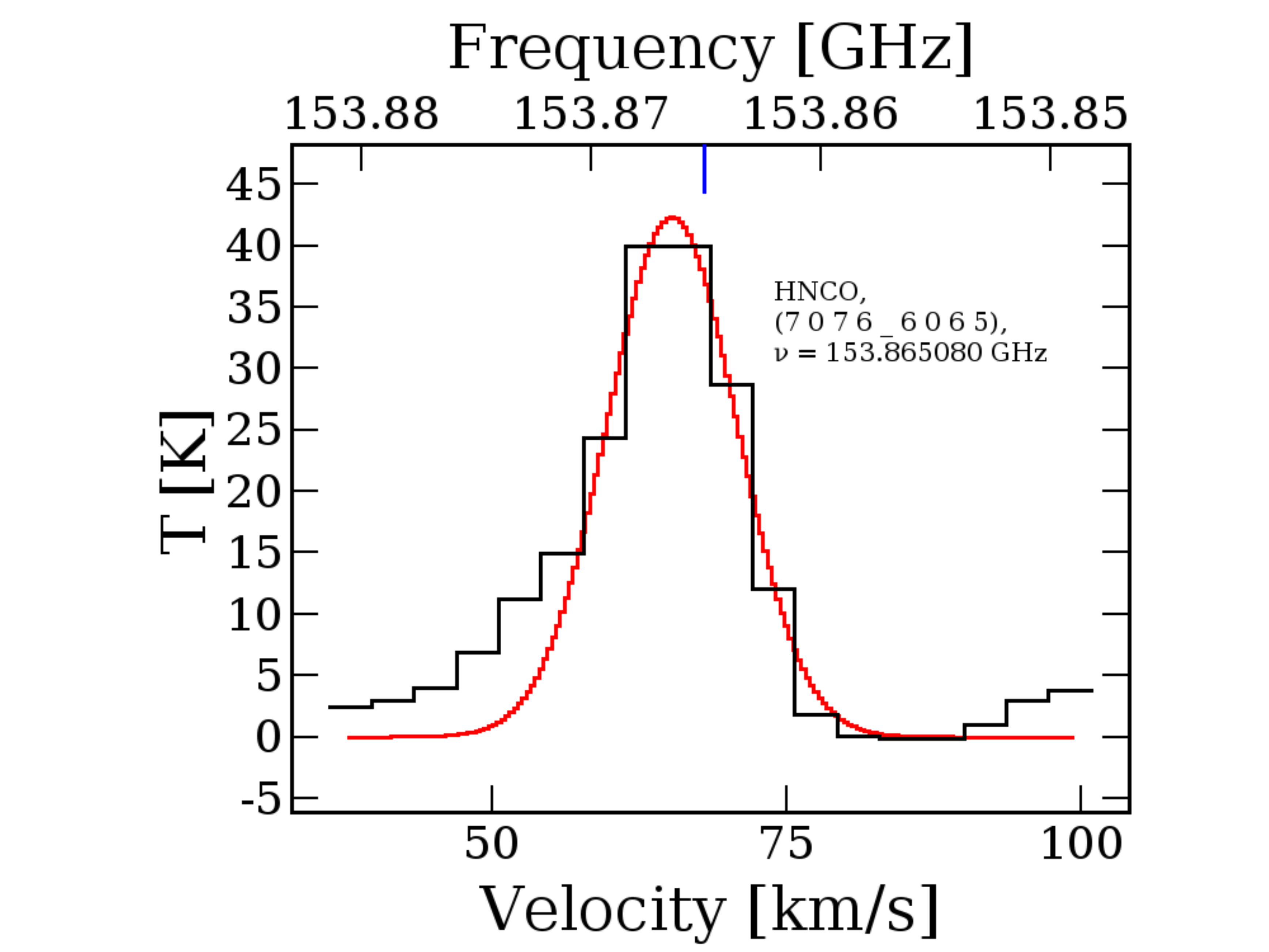}
\end{minipage}
\begin{minipage}{0.35\textwidth}
\includegraphics[width=\textwidth]{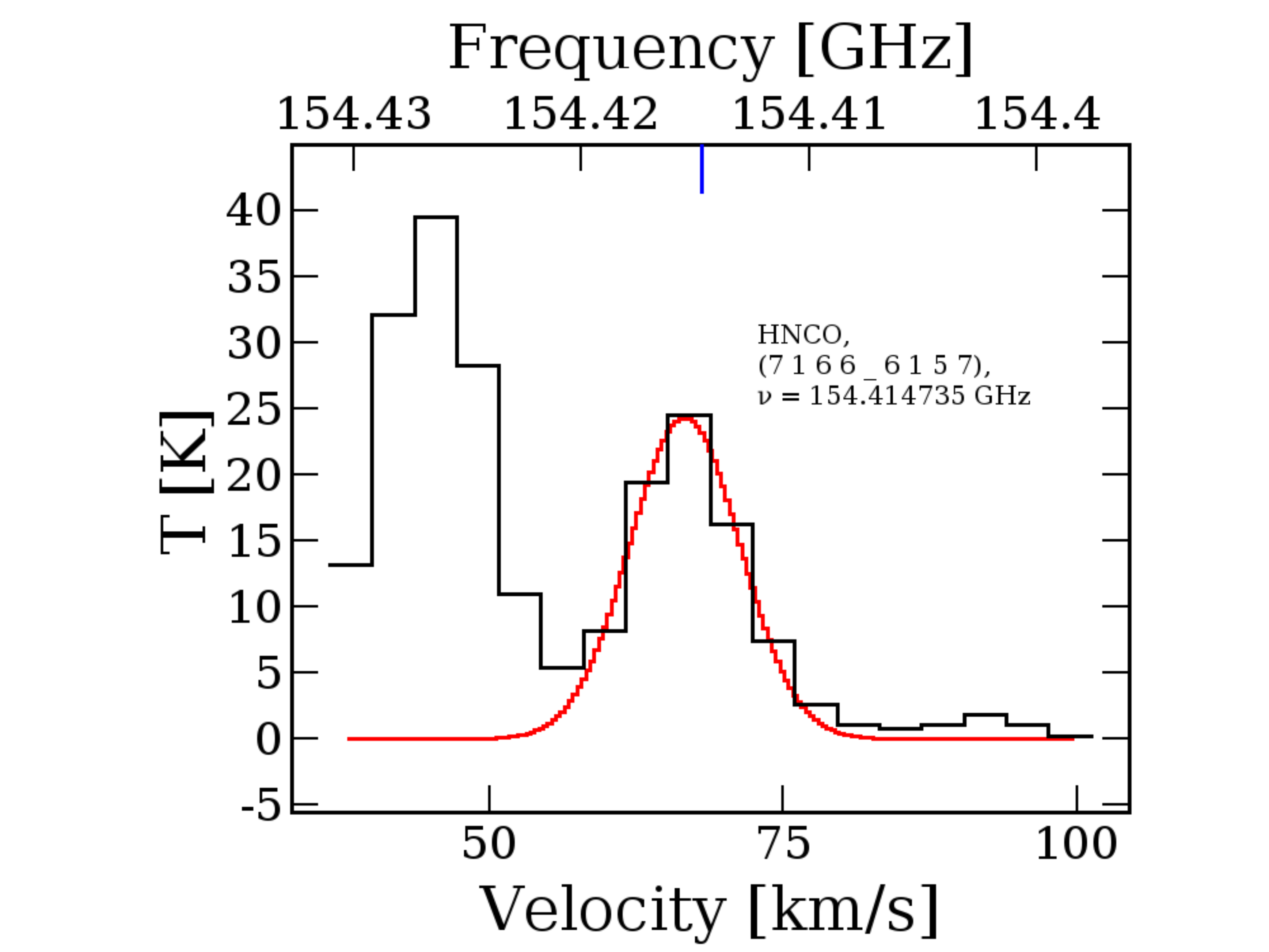}
\end{minipage}
\caption{{ Gaussian fitting of the observed emission spectra of HNCO towards G10. Black line represents observed emission spectra and red line represents a Gaussian profile fitted to the observed 
spectra.}}
\label{Gfit-hnco}
\end{figure}

\begin{figure}
\begin{minipage}{0.35\textwidth}
\includegraphics[width=\textwidth]{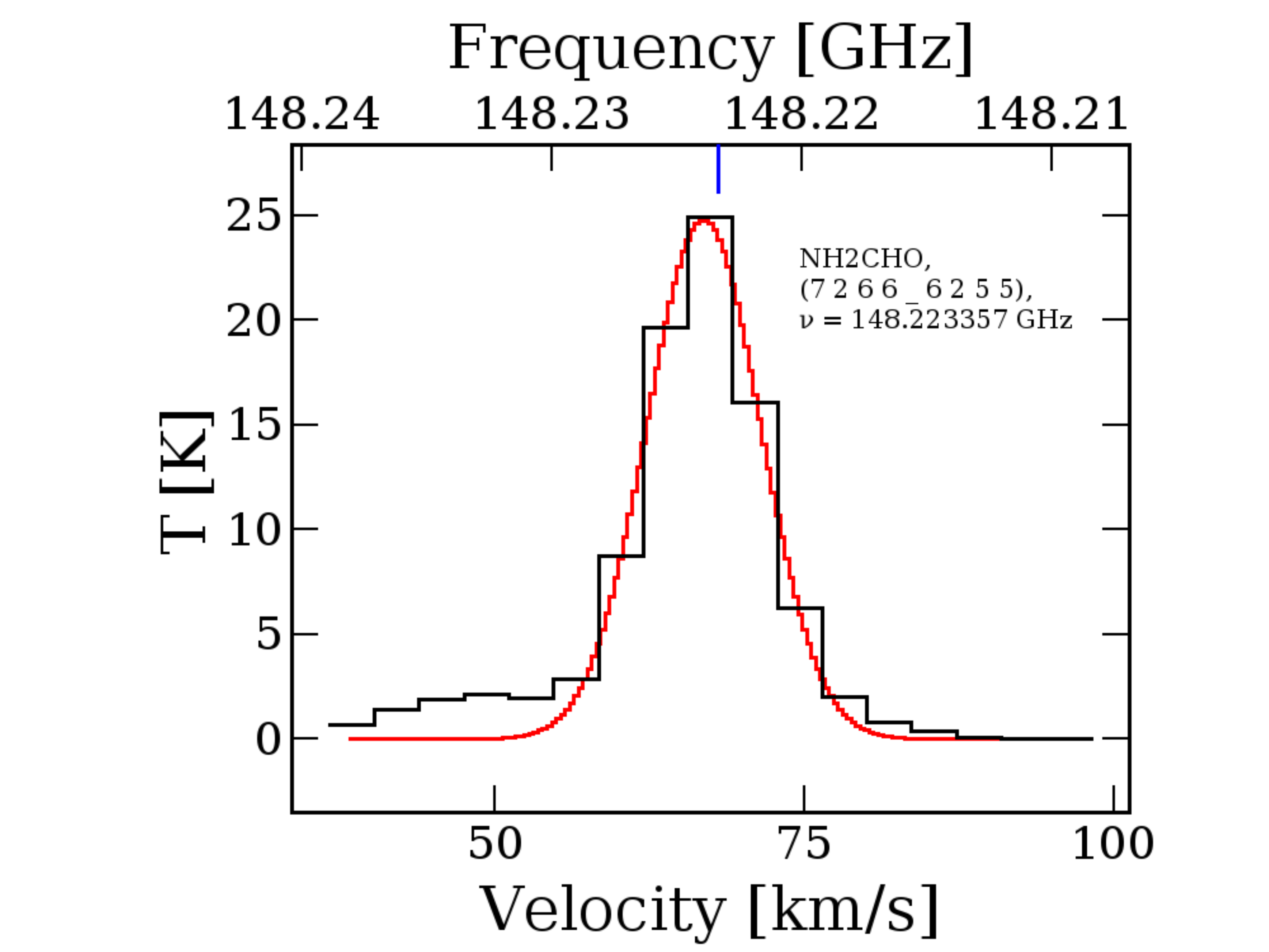}
\end{minipage}
\begin{minipage}{0.35\textwidth}
\includegraphics[width=\textwidth]{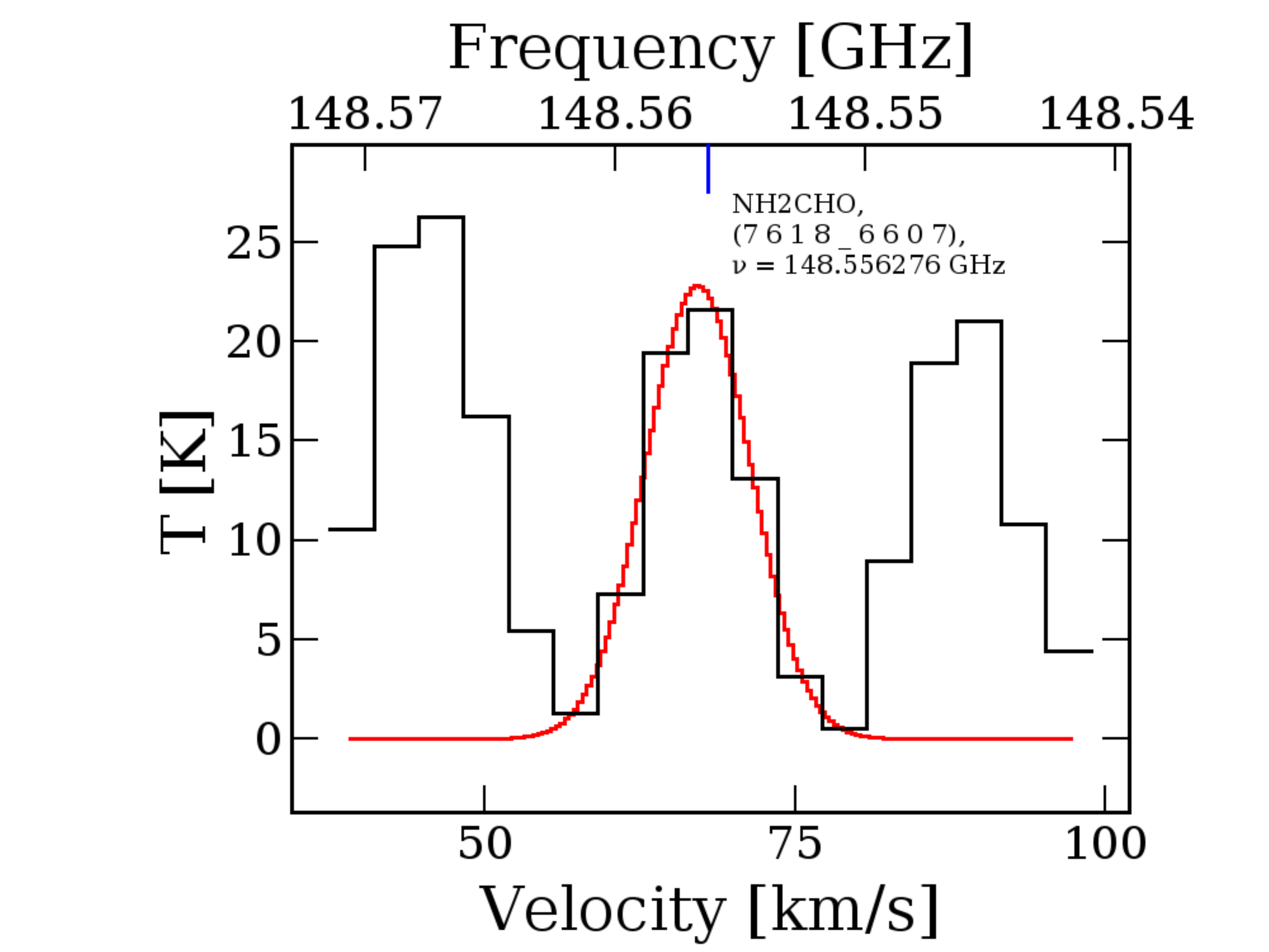}
\end{minipage}
\begin{minipage}{0.35\textwidth}
\includegraphics[width=\textwidth]{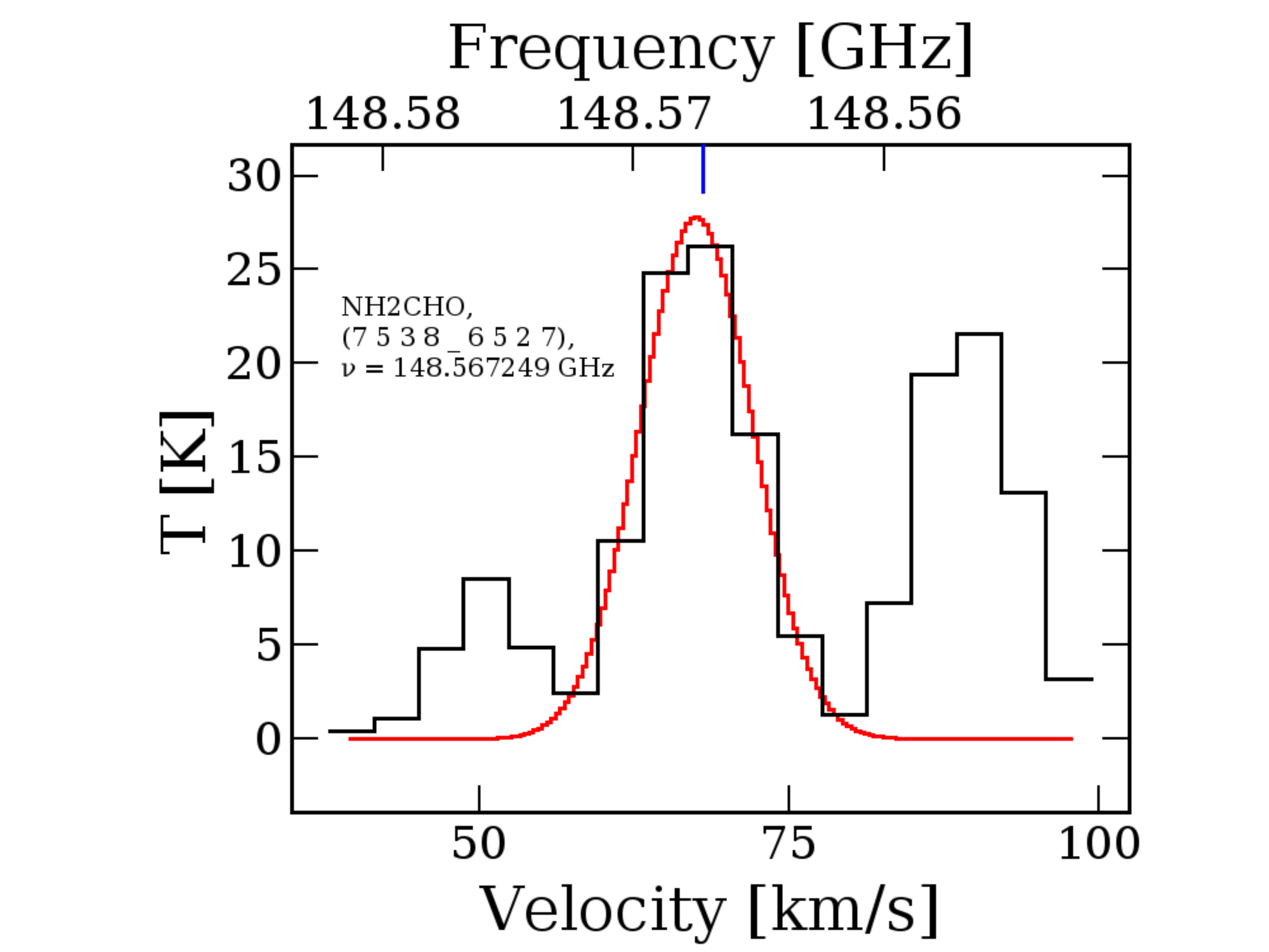}
\end{minipage}
\begin{minipage}{0.35\textwidth}
\includegraphics[width=\textwidth]{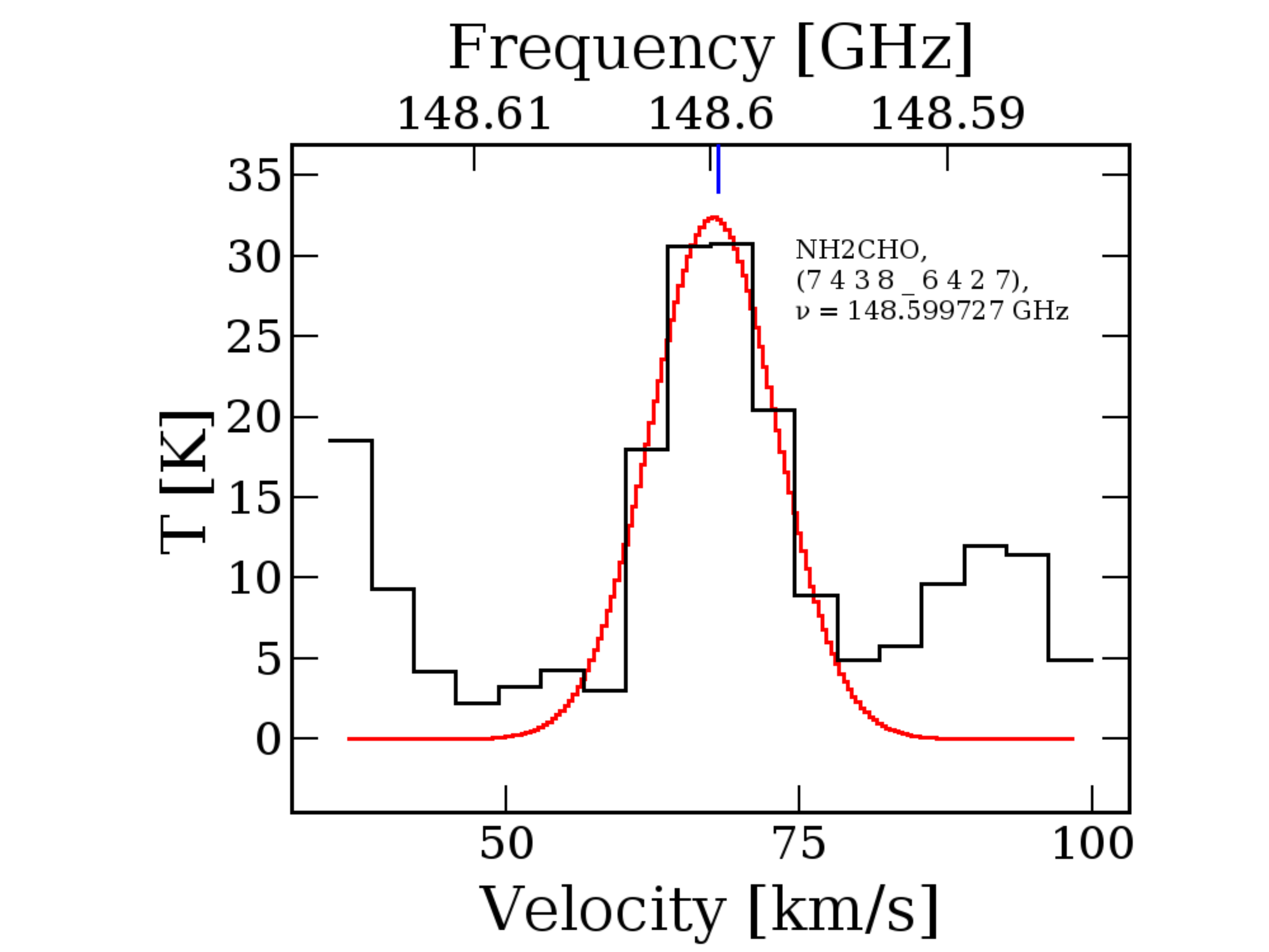}
\end{minipage}
\begin{minipage}{0.35\textwidth}
\includegraphics[width=\textwidth]{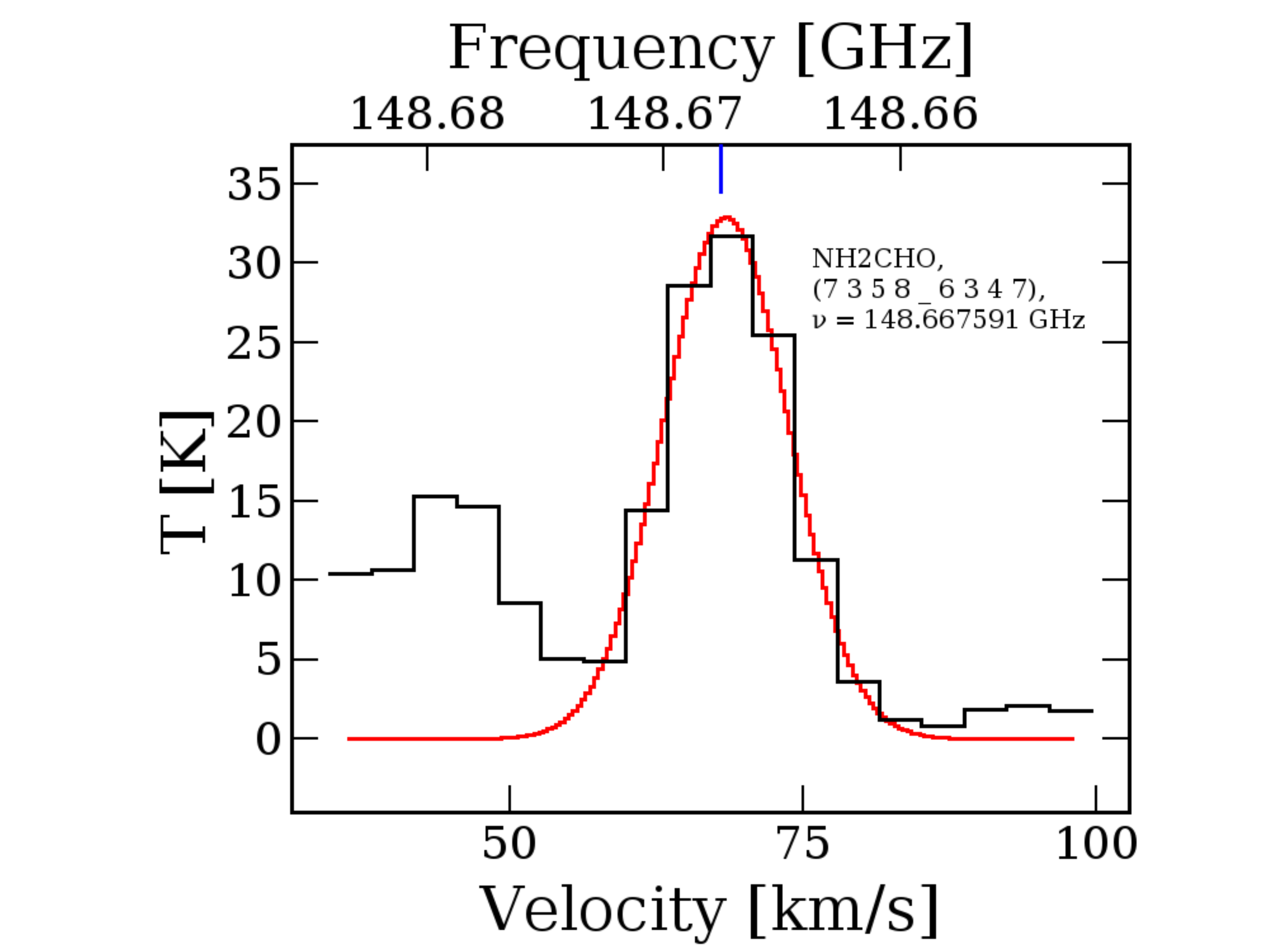}
\end{minipage}
\begin{minipage}{0.35\textwidth}
\includegraphics[width=\textwidth]{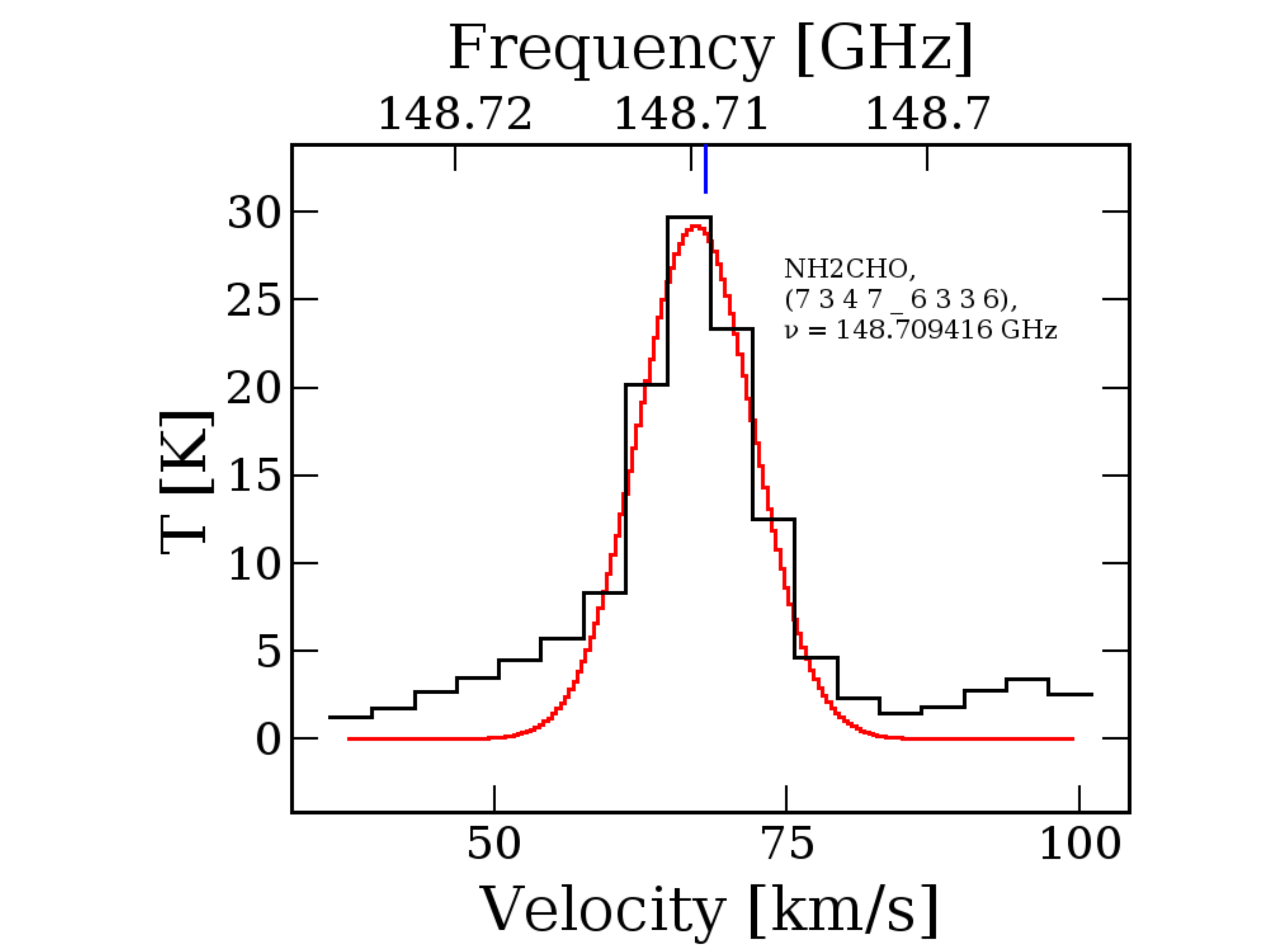}
\end{minipage}
\begin{minipage}{0.35\textwidth}
\includegraphics[width=\textwidth]{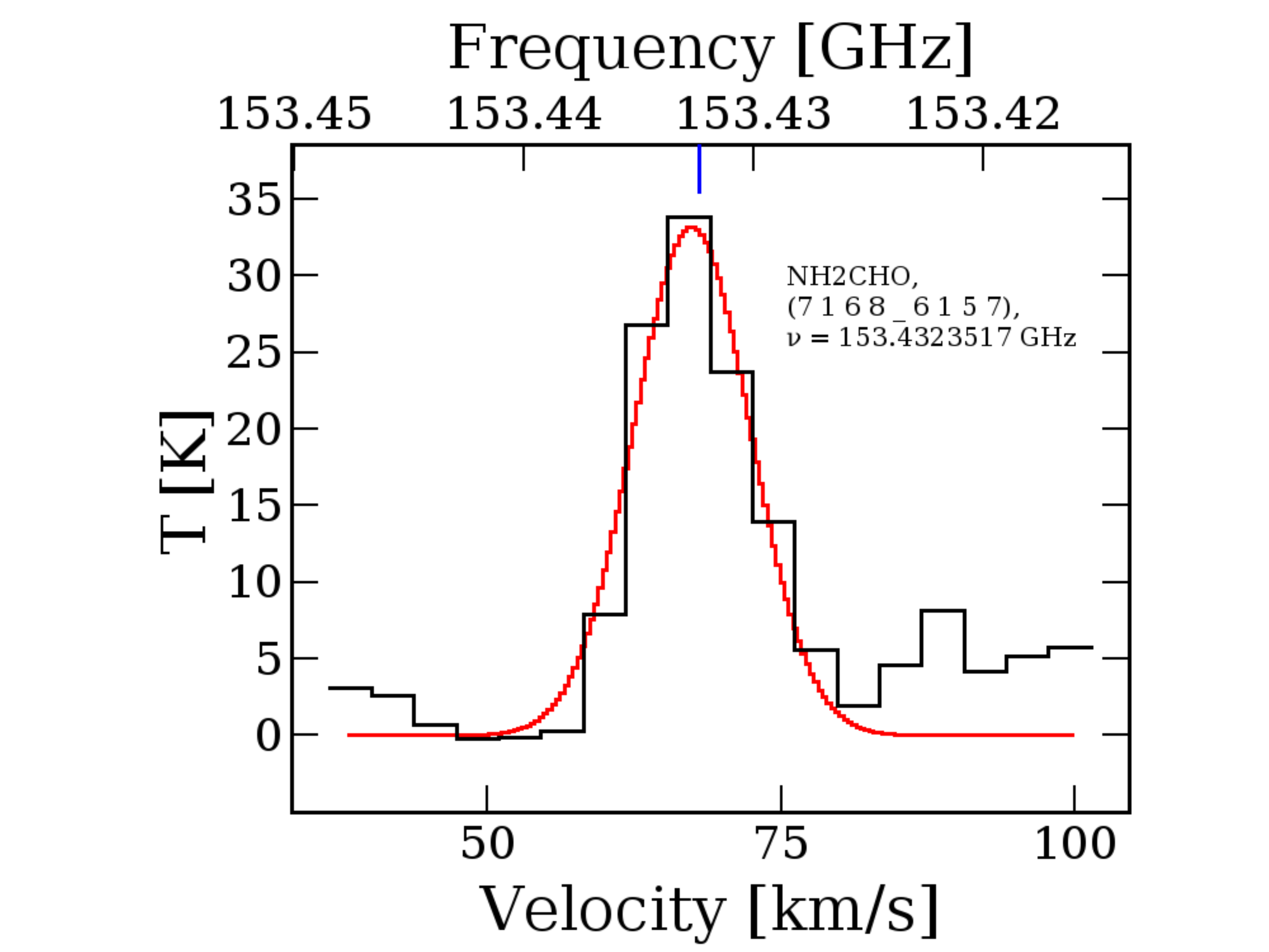}
\end{minipage}
\caption{{ Gaussian fitting of the observed emission spectra of NH$_2$CHO towards G10. Black line represents observed emission spectra and red line represents a Gaussian profile fitted to the observed 
spectra.}}
\label{Gfit-nh2cho}
\end{figure}

\begin{figure}
\begin{minipage}{0.35\textwidth}
\includegraphics[width=\textwidth]{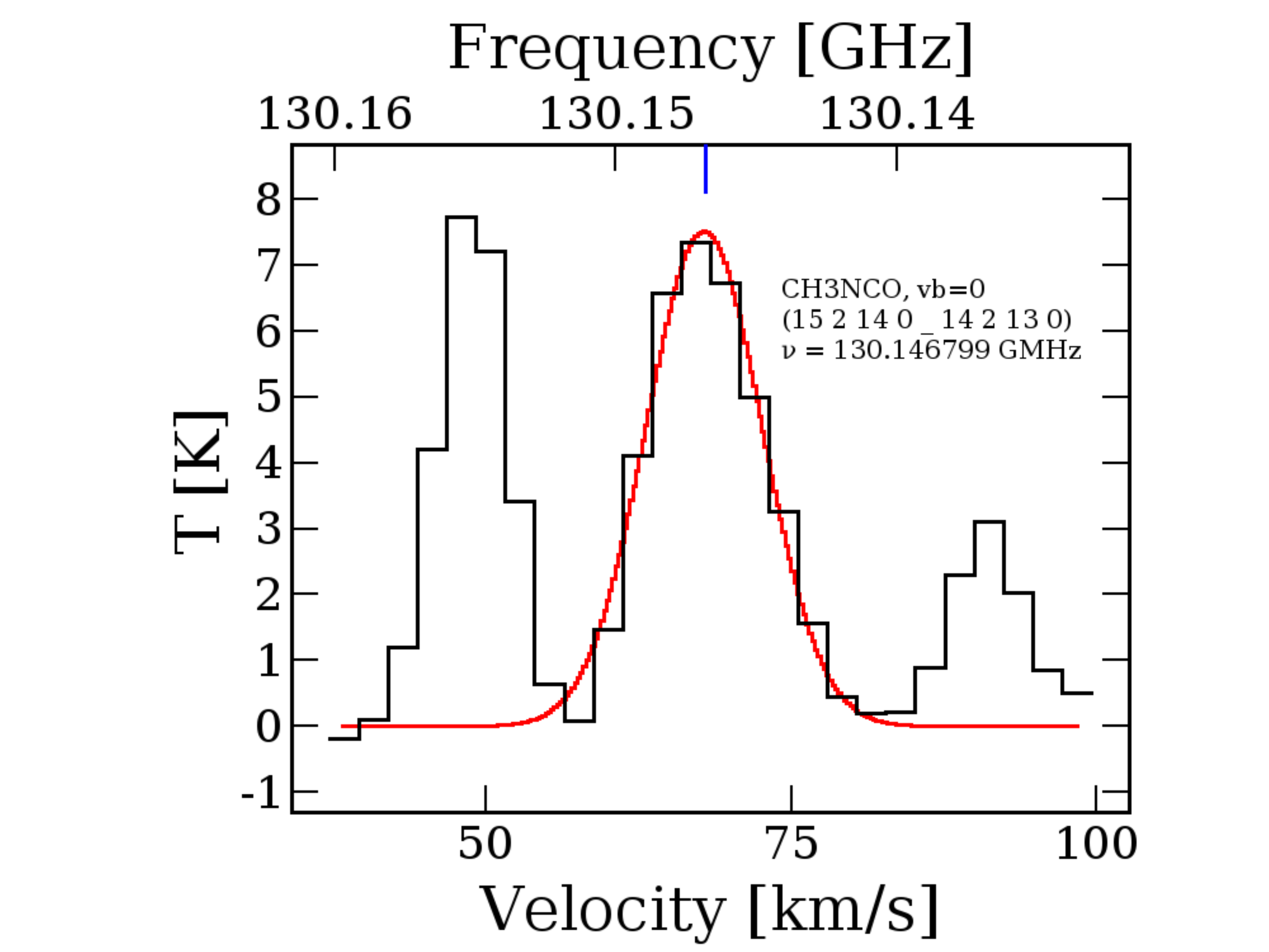}
\end{minipage}
\begin{minipage}{0.35\textwidth}
\includegraphics[width=\textwidth]{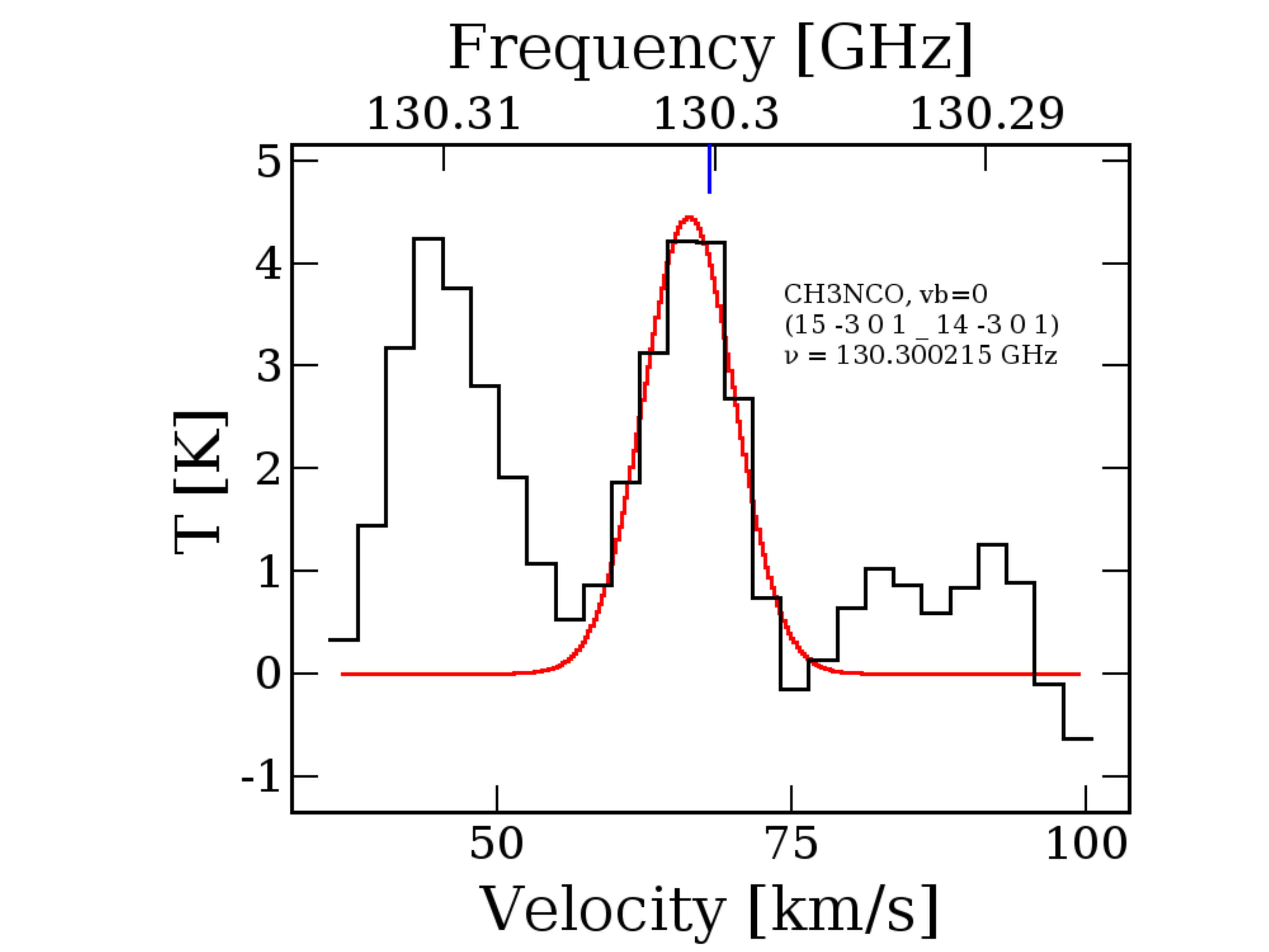}
\end{minipage}
\begin{minipage}{0.35\textwidth}
\includegraphics[width=\textwidth]{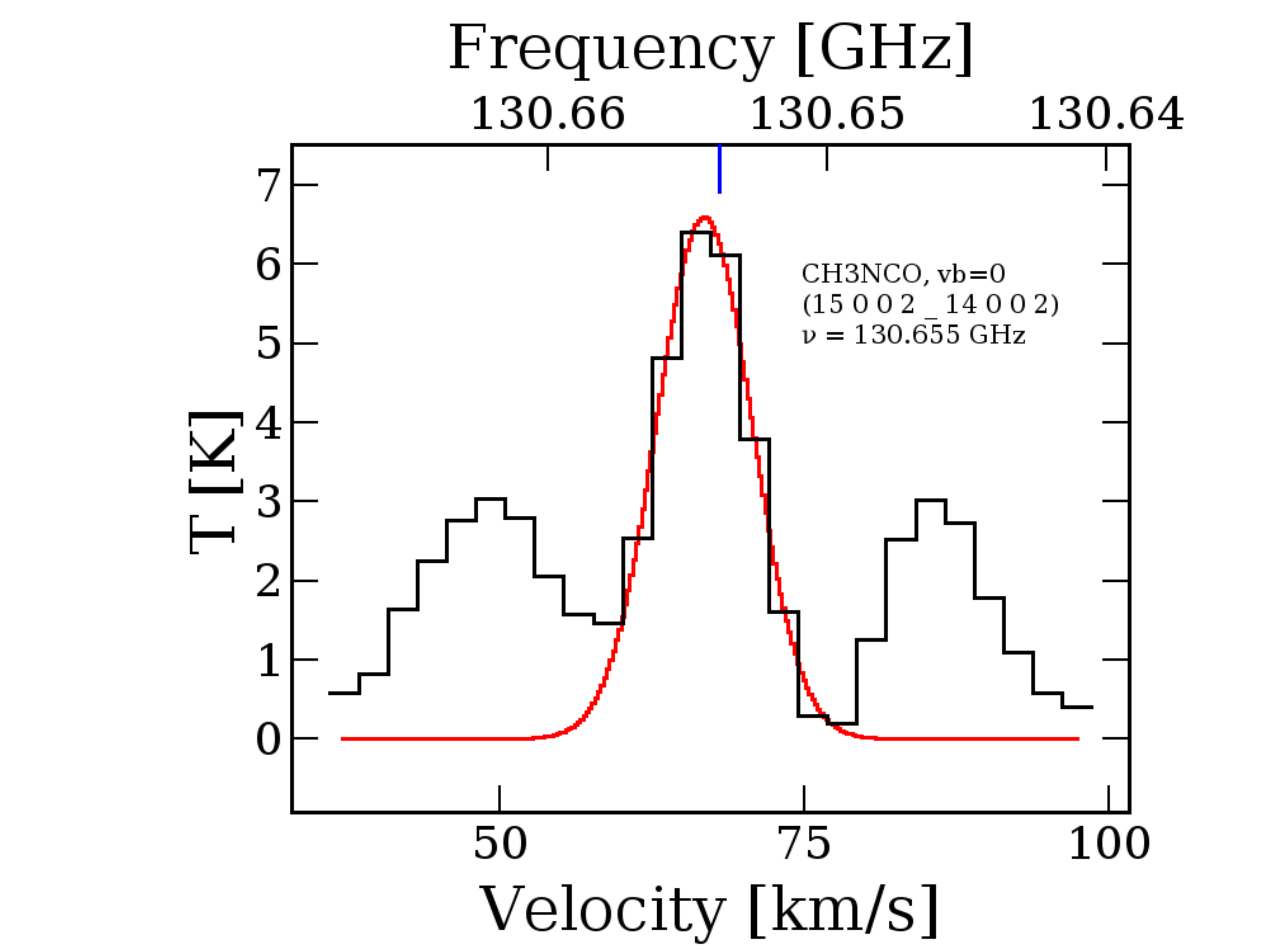}
\end{minipage}
\begin{minipage}{0.35\textwidth}
\includegraphics[width=\textwidth]{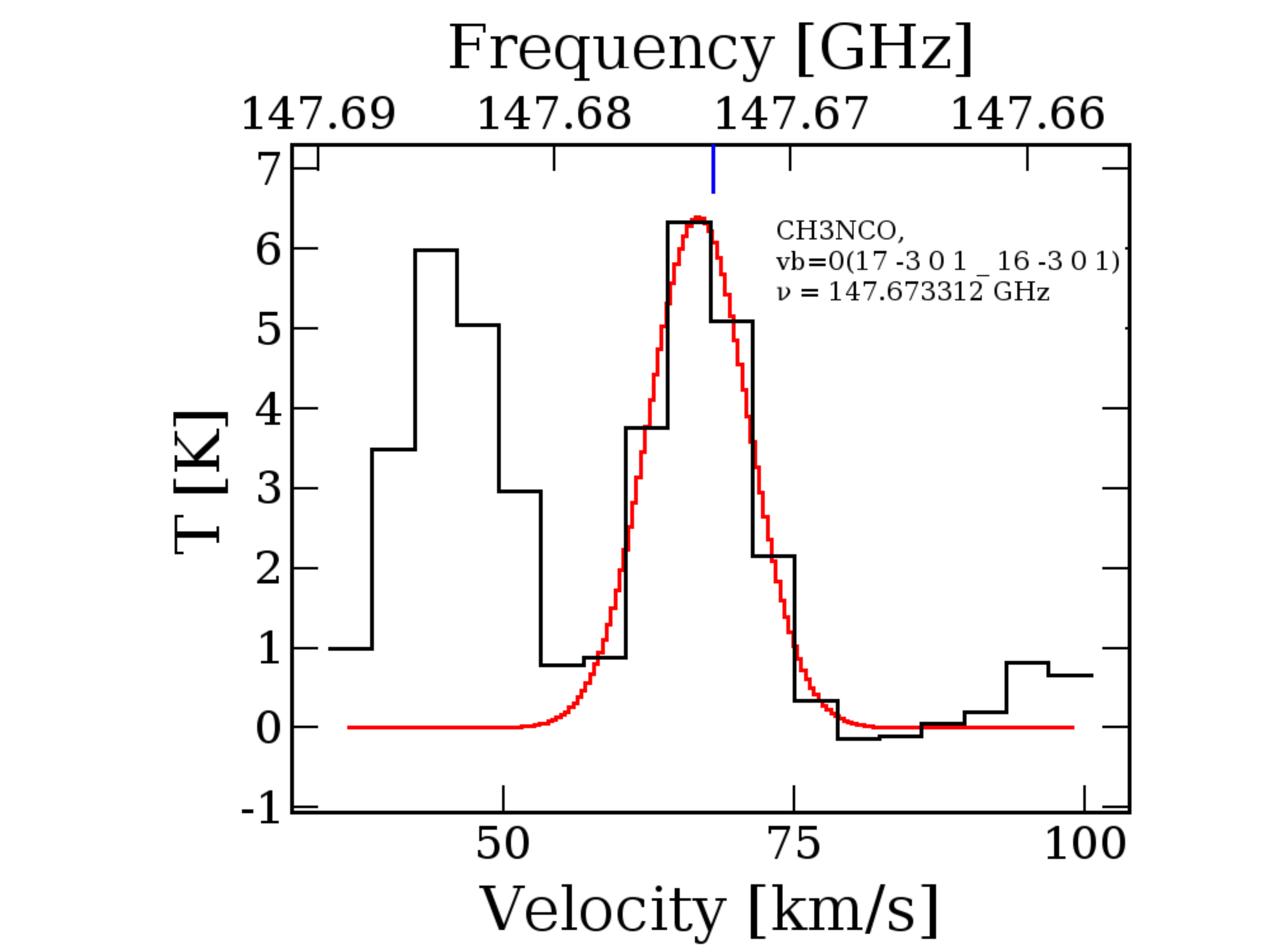}
\end{minipage}
\begin{minipage}{0.35\textwidth}
\includegraphics[width=\textwidth]{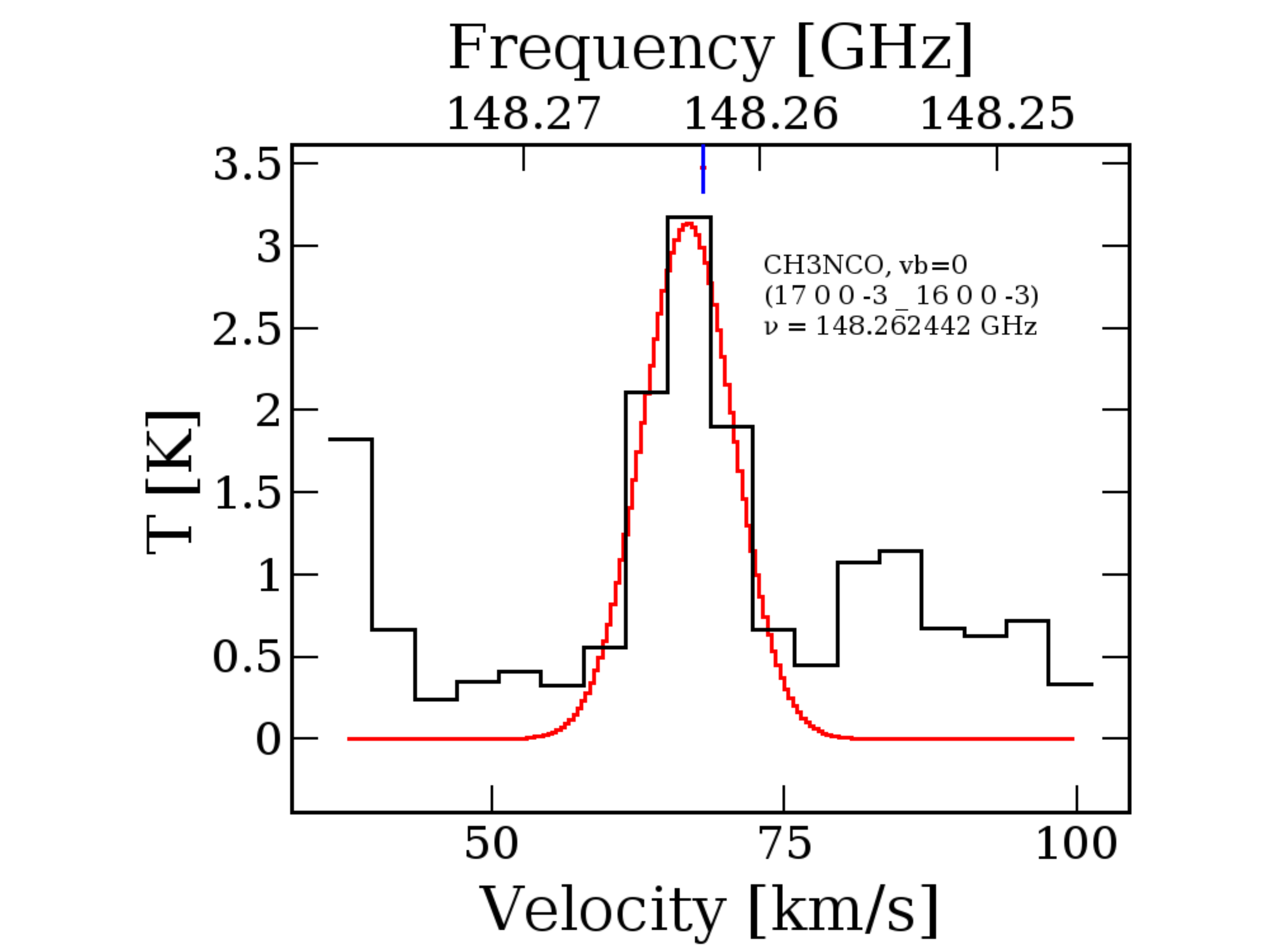}
\end{minipage}
\begin{minipage}{0.35\textwidth}
\includegraphics[width=\textwidth]{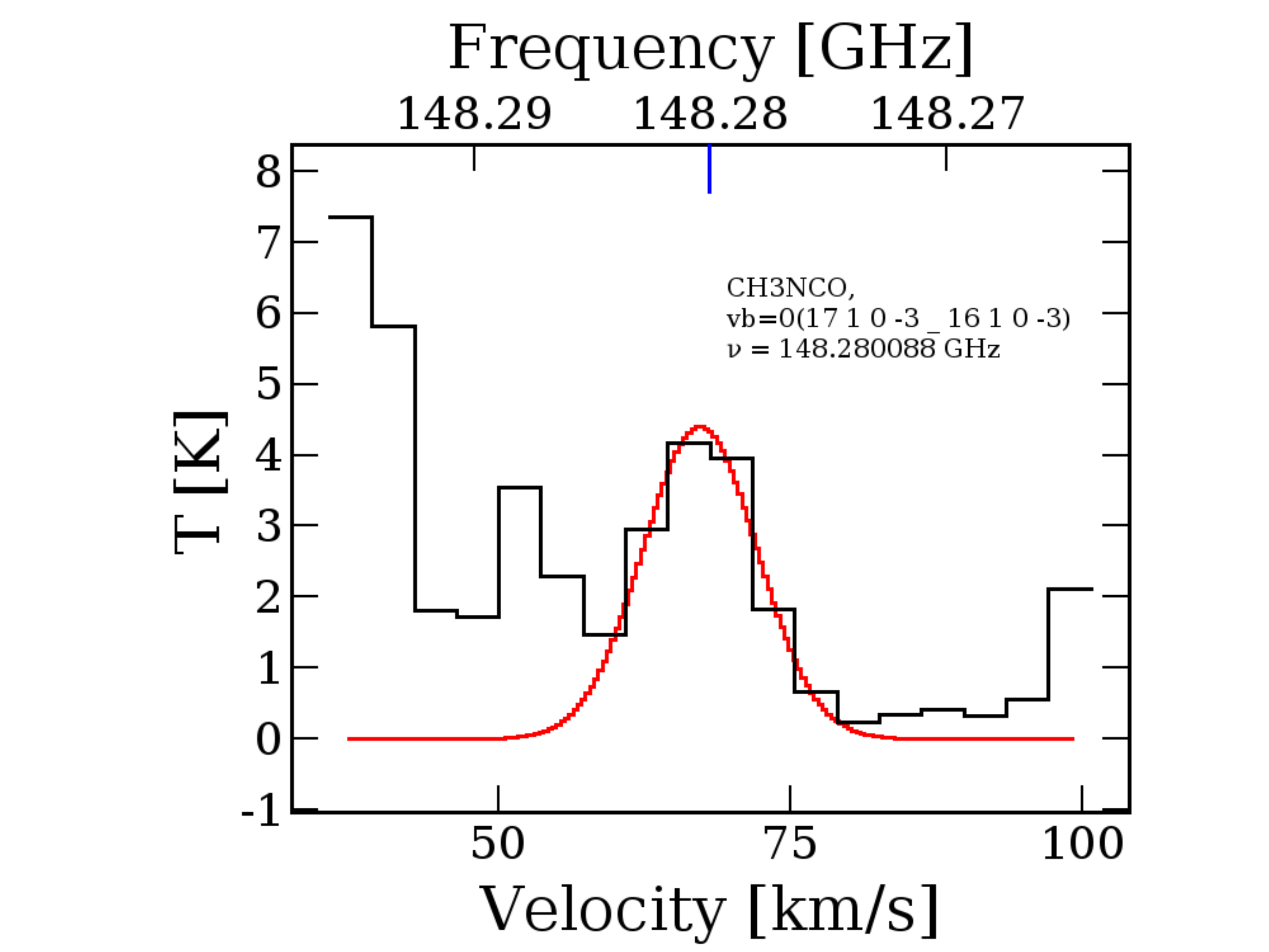}
\end{minipage}
\begin{minipage}{0.35\textwidth}
\includegraphics[width=\textwidth]{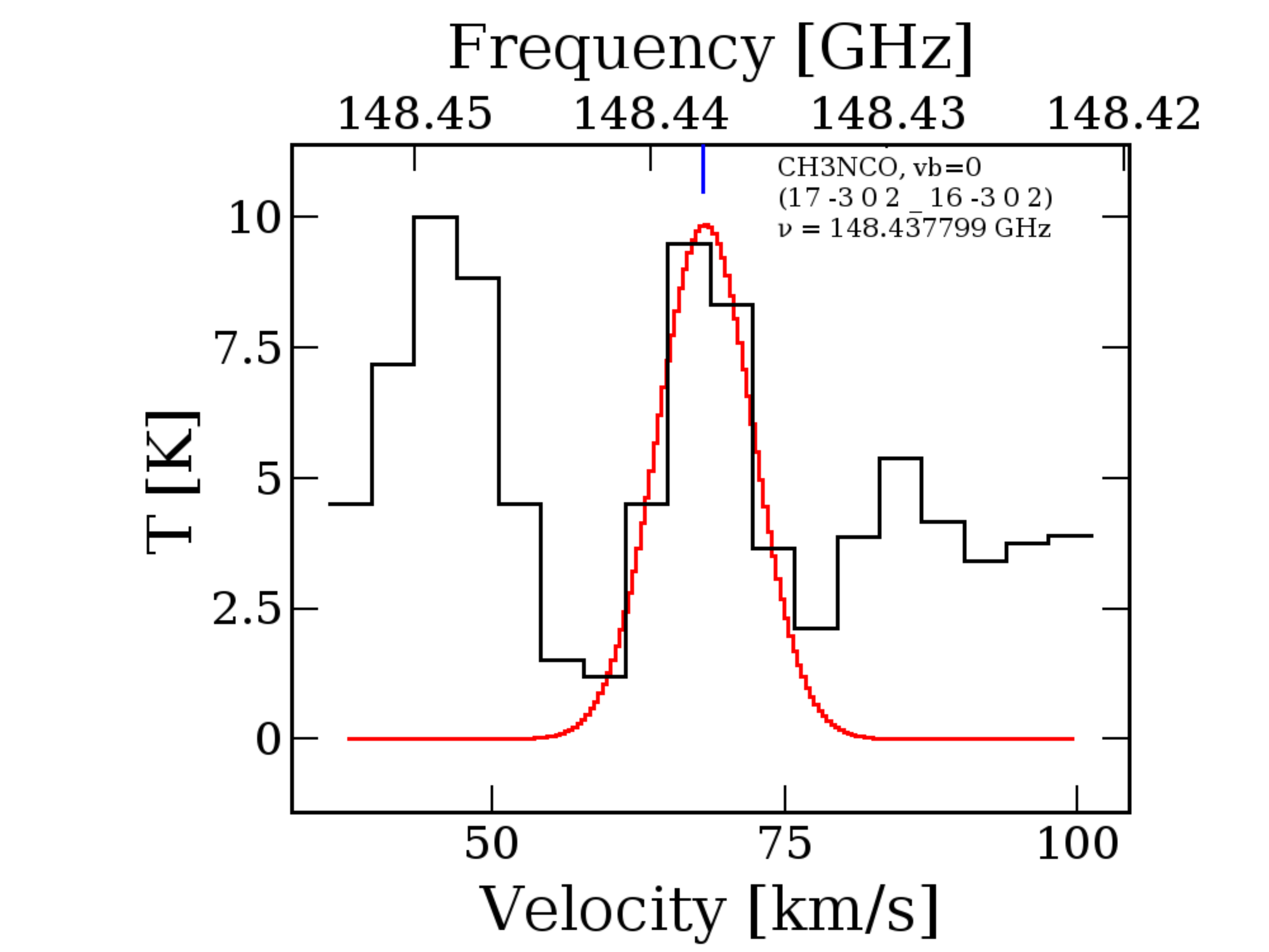}
\end{minipage}
\begin{minipage}{0.35\textwidth}
\includegraphics[width=\textwidth]{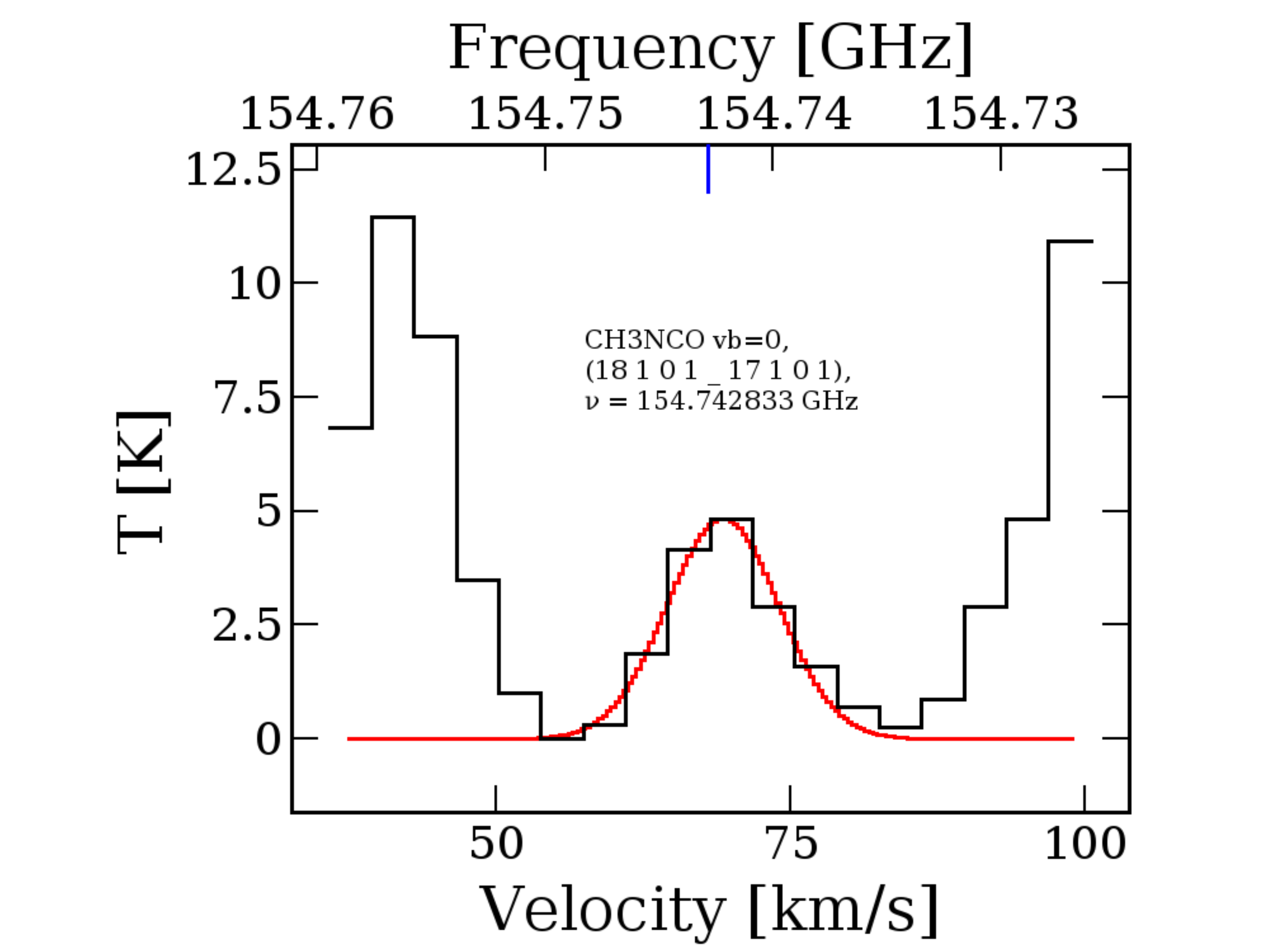}
\end{minipage}
\caption{{ Gaussian fitting of the observed emission spectra of CH$_3$NCO towards G10. Black line represents observed emission spectra and red line represents a Gaussian profile fitted to the observed 
spectra.}}
\label{Gfit-ch3nco}
\end{figure}
\end{document}